\documentclass[hyper]{JHEP3}
\usepackage{amsfonts} 
\usepackage{amsmath,epsfig,subfigure}

\title{Fermion Resonances on Multi-field Thick Branes}

\author{Yu-Xiao Liu$^1$,
        Hai-Tao Li$^1$\thanks{Corresponding author.},
        Zhen-Hua Zhao$^1$,
        Jing-Xin Li$^2$,
        Ji-Rong Ren$^1$\\
 $^1$Institute of Theoretical Physics,
     Lanzhou University, Lanzhou 730000, China\\
 $^2$College of Atmospheric Sciences,
     Lanzhou University, Lanzhou 730000, China\\
  E-mail: \email{liuyx@lzu.edu.cn},
          \email{liht07@lzu.cn},
          \email{zhaozhh02@gmail.com},
          \email{lijx08@lzu.cn},
          \email{renjr@lzu.edu.cn}
          }

\abstract{Motivated by the recent work on the fermion resonances on
scalar-constructed thick branes (arXiv:0901.3543 and
arXiv:0904.1785), we extend the idea to multi-scalar generated thick
branes and complete previous work. The fermion localization and
resonances on the three-field and two-field thick branes are
investigated. With the Numerov method, our numerical results show
that the resonance states also exist in the brane besides the
single-field thick branes and the two-field thick branes in the
former cases. This interesting phenomenon is related to the internal
structure of the brane and the coupling of fermions and scalars. We
find that the Kaluza-Klein chiral decomposition of massive fermion
resonances is the parity-chiral decomposition. For the couplings
$\eta\overline{\Psi}\phi^{k}\chi\rho\Psi$ and
$\eta\overline{\Psi}\phi^{k}\chi\rho\Psi$ for three-field and
two-field models with odd positive $k$, respectively, the number of
the resonant states decreases with $k$. This result is opposite to
the one obtained in the single-field de Sitter thick brane
(arXiv:0904.1785).}


\begin{document}

\section{Introduction}

A great deal of effort has been paid to the study of higher
dimensional space-time with large extra
dimensions~\cite{Akama:82a,Akama:83a,Rubakov:83a,Rubakov:83b,Visser:85a,Randjbar:86a,Antoniadis:90a,Arkani-Hamed:98a,Randall:99a,Randall:99b,Arkani-Hamed:00a,Lykken:00a,Kehagias:04a}.
Phenomenologically, the idea that our observed four-dimensional
world is a brane embedded in a higher-dimensional
space-time~\cite{Akama:83a,Rubakov:83a,Rubakov:83b,Randjbar:86a,Arkani-Hamed:98a,Randall:99a,Randall:99b,Lykken:00a,Kehagias:04a}
opens up a route towards resolving the mass hierarchy problem and
the cosmological constant problem. A first particle physics
application of this idea was put forward in
Refs.~\cite{Akama:82a,Rubakov:83a}. In the framework of
(3+1)-dimensional brane scenarios, gravity is free to propagate in
all dimensions, while all matter fields are confined to a $3$-brane.
In Ref.~\cite{Arkani-Hamed:98a}, Arkani-Hamed, Dimopoulos, and Dvali
proposed the large extra dimensions model, which lowers the energy
scale of quantum gravity to 1 TeV by localizing the standard model
fields to a $4$-brane in a higher dimensional space-time. In this
scenario, extra dimensions are compacted into a large volume that
effectively dilutes the strength of gravity from the fundamental
scale (the TeV scale) to the Planck scale. In
Ref.~\cite{Randall:99a}, Randall and Sundrum introduced a warped
branewolrd model, which provides a novel solution of the gauge
hierarchy problem in particle physics. Later developments suggested
that the warped metric could even provide an alternative to
compactification for the extra
dimensions~\cite{Randall:99b,Arkani-Hamed:00a}.

In recent years, an increasing interest has been focused on the
study of thick brane scenarios in higher dimensional space-time
\cite{DeWolfe:00a,Gremm:00a,Gremm:00b,Csaki:00a,Campos:02a,WangPRD2002,thickBranes,Guerrero:02a,Dzhunushaliev:08a,Bazeia:03a,Shtanov:09a}.
From a realistic point of view, a brane should have thickness. Thick
brane scenarios based on gravity coupled to scalars have been
constructed. It is known that the modulus is not stable in RS warped
scenario. While it can be stabilized by introducing scalar fields in
the bulk \cite{GW}. These bulk scalar fields also provide us with a
way of generating the brane as a domain wall (thick brane) in five
dimensions. An important feature of these models is that we can
obtain branes by a very natural way rather than by introducing them
artificially. One can refer to a recent review article on the
subject of the thick brane solutions~\cite{Dzhunushaliev:09a}.

In brane world scenarios, an important problem is localizing
gravity and various bulk fields on the branes by a natural
mechanism. In particular, the localization of spin $1/2$ fermions
is very interesting and rich. For localizing fermions on the
branes or domain walls, one needs to introduce other interactions
in addition to gravity. In the last few years, localization
mechanisms of fermions on a domain wall has been extensively
studied~\cite{Ringeval:02,Slatyer:07a,Liu:08b,Liu:08c,Liu:09a}.
Meanwhile, the same problem can also be discussed in other
contexts such as gauge field~\cite{Parameswaran:07a,Liu:07a},
supergravity~\cite{dePol:00a}, vortex
background~\cite{Liu:07b,Wang:05a,Rafael:08a,Starkman:02a} and
general spacetime background~\cite{Randjbar-Daemi:00a}.

Usually, one examines the localization problem in the content of
brane models constructed with one background scalar field. The
finite discrete Kaluza-Klein (KK) modes (bound states) and a
continuous gapless spectrum starting at a positive $m^{2}$ can be
obtained for example in
Refs.~\cite{Barbosa-Cendejas:08a,Liu:08a,Kodama:09a,20082009}. In
some brane models, the quasibound KK modes with a certain lifetime
will
appear~\cite{Csaki:00a,Bazeia:06a,Almeida:0901,Liu:09b,0801.0801}.
Especially, in recent work of Ref.~\cite{Almeida:0901}, the
authors investigated fermions on the Bloch brane \cite{Bazeia:04a}
constructed with two scalar fields $\phi$ and $\chi$. With the
simplest Yukawa coupling $\eta\overline{\Psi}\phi\chi\Psi$, the
localization problem of fermions was studied. Resonances for both
chiralities were found and their appearance is related to the
internal structure of the Bloch brane. In Ref.~\cite{Liu:09b}, the
localization and resonance spectrum of fermions on a one-scalar
generated dS thick brane were investigated. For a class of
scalar-fermion couplings $\eta\overline{\Psi}\phi^{k}\Psi$ with
positive odd integer $k$, some new nature about resonances were
obtained. On the other hand, thick branes with two or more scalar
background fields also have been
constructed~\cite{Bazeia:02,Bazeia:04a,Dzhunushaliev:06a}. In this
paper, our goal is to extend the idea of
Refs~\cite{Almeida:0901,Liu:09b} to multi-field generated branes
for refreshing the understanding of the localization and
resonances of fermions in these brane models. We will consider
fermions in the three-field constructed thick brane model as an
explicit example. We also give a further analysis about resonances
on the Bloch brane and compare the results obtained in different
models.

The structure of this paper is as follows. In Sec. \ref{secModels},
we review the multi-field thick brane models for spin $0$ scalar
fields and spin $1/2$ fermion fields. In Sec. \ref{sec3Fields} we
provide a complete investigation of the fermion localization and
resonances on a three-scalar generated thick brane in detail. In
Sec. \ref{sec2Fields}, we re-analysis the resonances in a Bloch
brane model. Finally, our comments and conclusion are presented in
Sec. \ref{secConclusion}.

\section{The multi-field thick brane models}
\label{secModels}

In order to maintain the continuity of the whole work and make the
background of this paper clear, in this section, we review the
multi-field thick brane models. Followed the similar procedure in
Refs. \cite{Almeida:0901,Liu:09b,Bazeia:02}, the field equations for
scalars and fermions are obtained, especially we get the first-order
equations by the superpotential.

\subsection{Spin 0 scalar fields}

Let us consider the 4+1 dimensional multi-field thick brane
models. Specifically, we introduce $n$ scalar background fields.
The action for such a system is given by
\begin{eqnarray}
 S = \int d^{4}x dy \sqrt{-g}
  \left[ \frac{1}{4}R -\frac{1}{2}\partial_{M}\phi\partial^{M}\phi
                      -\frac{1}{2}\partial_{M}\chi\partial^{M}\chi - \cdots
                      -\frac{1}{2}\partial_{M}\rho\partial^{M}\rho
  -V(\phi,\chi,\cdots,\rho) \right] \nonumber  \\
 \label{scalarAction}
\end{eqnarray}
and the line-element is assumed as
\begin{eqnarray}
 ds^{2} =g_{MN}dx^{M}dx^{N}
        =e^{2A}\eta_{\mu\nu}dx^{\mu}dx^{\nu}-dy^{2}, \label{metric}
\end{eqnarray}
where $g = \det(g_{MN})$, $M,N = 0,1,2,3,4$, $e^{2A}$ is the warp
factor, and $y$ stands for the extra coordinate. The warp factor
and the scalar fields are considered to be functions of $y$ only,
i.e., $A = A(y)$, $\phi = \phi(y)$, $\chi = \chi(y)$, $\cdots$,
and $\rho = \rho(y)$.

The field equations generated from the action (\ref{scalarAction})
with the ansatz (\ref{metric}) reduce to the following coupled
nonlinear differential equations
\begin{eqnarray}
 \phi^{\prime\prime} &=& \frac{\partial
 V(\phi,\chi,\cdots,\rho)}{\partial\phi}
     -4A^{\prime}\phi^{\prime}\,,  \nonumber\\
 \chi^{\prime\prime}
   &=& \frac{\partial V(\phi,\chi,\cdots,\rho)}{\partial\chi}
   -4A^{\prime}\chi^{\prime}\,, \nonumber\\
 \vdots  \label{2ndOrderEqs} \\
 \rho^{\prime\prime} &=& \frac{\partial
    V(\phi,\chi,\cdots,\rho)}{\partial\rho}
   -4A^{\prime}\rho^{\prime}\,, \nonumber\\
A^{\prime\prime} &=&
-\frac{2}{3}(\phi^{\prime2}+\chi^{\prime2}+\cdots+\rho^{\prime2})\,,
\nonumber\\
A^{\prime2} &=&
\frac{1}{6}(\phi^{\prime2}+\chi^{\prime2}+\cdots+\rho^{\prime2})
 -\frac{1}{3}V(\phi,\chi,\cdots,\rho)\,. \nonumber
\end{eqnarray}
We suppose that the $n$-field models are defined by the potential
\begin{eqnarray}
V(\phi,\chi,\cdots,\rho) = \frac{1}{8}\Biggl[\Biggl(\frac{\partial
W}{\partial\phi}\Biggl)^{2}+\Biggl(\frac{\partial
W}{\partial\chi}\Biggl)^{2}+\cdots+\Biggl(\frac{\partial
W}{\partial\rho}\Biggl)^{2}\Biggl]-\frac{1}{3}W^{2}\,,
 \label{PotentialV1}
\end{eqnarray}
where $W=W(\phi,\chi,\cdots,\rho)$ is the superpotential, which
leads to the following first-order equations that solve the second
order equations (\ref{2ndOrderEqs}):
\begin{eqnarray}
 \phi^{\prime} = \frac{1}{2}\frac{\partial W}{\partial\phi}\,,~~
 \chi^{\prime} =\frac{1}{2}\frac{\partial W}{\partial\chi}\,,~~
 \cdots,~~
 \rho^{\prime} = \frac{1}{2}\frac{\partial W}{\partial\rho}\,,~~
 A^{\prime} = -\frac{1}{3}W\,. \label{1stEqs}
\end{eqnarray}

\subsection{Spin 1/2 fermion fields}

In this subsection, we are ready to investigate whether spin half
fermions can be localized on the brane. Let us consider $n$ real
scalars coupled to a massless bulk fermion by means of a general
Yukawa coupling in five dimension. The starting action reads
\begin{eqnarray}
 S_{1/2} &=& \int d^{5}x\sqrt{-g}\bigg[\bar{\Psi}\Gamma^{M}D_{M}\Psi
    -\eta\bar{\Psi}F(\phi,\chi,\cdots,\rho)\Psi\bigg],
    \label{fermion field action}
\end{eqnarray}
where the spin connection $\omega^{\bar{M}\bar{N}}_{M}$ in the
covariant derivative
\begin{eqnarray}
D_{M}\Psi &=& (\partial_{M}+\omega_{M})\Psi
 = \left(\partial_{M}+\frac{1}{4}\omega^{\bar{M}\bar{N}}_{M}
         \Gamma_{\bar{M}}\Gamma_{\bar{N}}\right)\Psi
\label{covariant derivative}
\end{eqnarray}
is defined as
\begin{eqnarray}
\omega^{\bar{M}\bar{N}}_{M}
 &=&\frac{1}{2}E^{N\bar{M}}(\partial_{M}E^{\bar{N}}_{N}
         -\partial_{N}E^{\bar{N}}_{M})  
-\frac{1}{2}E^{N\bar{N}}(\partial_{M}E^{\bar{M}}_{N}
         -\partial_{N}E^{\bar{M}}_{M})\nonumber\\
 &-& \frac{1}{2}E^{P\bar{M}}E^{Q\bar{N}} E^{\bar{R}}_{M}
   (\partial_{P}E_{Q\bar{R}}-\partial_{Q}E_{P\bar{R}})\,.
   \label{SpinConnection}
\end{eqnarray}
In five dimension, fermions are four component spinors and their
Dirac structure is determined by
$\Gamma^{M}=E^{M}_{\bar{M}}\Gamma^{\bar{M}}$ with the
$E^{M}_{\bar{M}}$ being the vielbein and
$\{\Gamma^{M},\Gamma^{N}\}=2g^{MN}$. The indices of
five-dimensional spacetime coordinates and the local lorentz
indices are labelled with capital Latin letters $M,N,\ldots$ and
$\bar{M},\bar{N},\ldots$, respectively. $\Gamma^{M}$ are the
curved gamma matrices and $\gamma^{\bar{M}}$ are the flat ones.

Following, we will derive the Schr\"{o}dinger equation for the KK
modes of fermions. In order to get the corresponding
mass-independent potential, we perform the coordinate
transformation
\begin{equation}
dz=e^{-A(y)}dy \label{coordinateTransformation}
\end{equation}
to get a conformally flat metric
\begin{equation}
ds^{2}_{5}=e^{2A(z)}(\eta_{\mu\nu}dx^{\mu}dx^{\nu}+dz^{2}) \,.
\label{conformallyFlatMetric}
\end{equation}
With the metric (\ref{conformallyFlatMetric}), the nonvanishing
components of the spin connection $\omega_{M}$ are calculated as
\begin{eqnarray}
\omega_{\mu}=\frac{1}{2}(\partial_{z}A)\gamma_{\mu}\gamma_{5}\,.
\label{SpinConnectionOmegaMu}
\end{eqnarray}
Then the five-dimensional Dirac equation is read as
\begin{eqnarray}
[\gamma^{\mu}\partial_{\mu}+\gamma^{5}(\partial_{z}+2\partial_{z}A)-\eta
e^{A}F(\phi,\chi,\rho)]\Psi=0\,,
 \label{DiracEq1}
\end{eqnarray}
where $\gamma^{\mu}\partial_{\mu}$ is the Dirac operator on the
brane.

We are now ready to study the above Dirac equation for
five-dimensional fluctuations, and write the fermion field $\Psi$
in terms of four-dimensional effective fields. Following
\cite{Liu:09a}, we have the general chiral decomposition
\begin{eqnarray}
\Psi(x,z)=e^{-2A}\biggl(\sum_{n}\psi_{Ln}(x)f_{Ln}(z)+\sum_{n}\psi_{Rn}(x)f_{Rn}(z)\biggl)\,,
\label{the general chiral decomposition}
\end{eqnarray}
where $\psi_{Ln}(x)=-\gamma^{5}\psi_{Ln}(x)$ and
$\psi_{Rn}(x)=\gamma^{5}\psi_{Rn}(x)$ are the left-handed and
right-handed components of a four-dimensional Dirac fermion
fields, respectively, they satisfy the four-dimensional massive
Dirac equations
$\gamma^{\mu}\partial_{\mu}\psi_{Ln}(x)=m_{n}\psi_{Rn}(x)$ and
$\gamma^{\mu}\partial_{\mu}\psi_{Rn}(x)=m_{n}\psi_{Ln}(x)$. The KK
modes $f_{Ln}(z)$ and $f_{Rn}(z)$ of the chiral decomposition of
the spinor satisfy the following coupled equations
\begin{eqnarray}
 \big[\partial_{z} +{\eta} e^{A}F(\phi,\chi,\rho)\big]
  f_{Ln}(z) &=& \,\,\,\,\,m_{n}f_{Rn}(z)\,,\label{spinorCoupledEqs1} \\
 \big[\partial_{z}-\eta e^{A}F(\phi,\chi,\rho)\big]
  f_{Rn}(z) &=& -m_{n}f_{Ln}(z)\,. \label{spinorCoupledEqs2}
\end{eqnarray}
In order to obtain the standard four-dimensional action for the
massive chiral fermions, we need the following orthonormality
conditions for $f_{Ln}(z)$ and $f_{Rn}(z)$:
\begin{eqnarray}
 \int^{+\infty}_{-\infty}f_{Lm}f_{Ln}dz=
 \int^{+\infty}_{-\infty}f_{Rm}f_{Rn}dz=\delta_{mn}\,,~~
 \int^{+\infty}_{-\infty}f_{Lm}f_{Rn}dz=0\,.
  \label{OrthonormalityConditions}
\end{eqnarray}
With Eqs.~(\ref{spinorCoupledEqs1}) and (\ref{spinorCoupledEqs2}),
we have
\begin{eqnarray}
 \biggl(\frac{d}{dz}-\eta e^{A}F(\phi,\chi,\rho)\biggl)
       \biggl(\frac{d}{dz}+\eta e^{A}F(\phi,\chi,\rho)\biggl)f_{L}
    =-m^{2}_{n}f_{L}\,,\\
 \biggl(\frac{d}{dz}+\eta e^{A}F(\phi,\chi,\rho)\biggl)
       \biggl(\frac{d}{dz}-\eta e^{A}F(\phi,\chi,\rho)\biggl)f_{R}
    =-m^{2}_{n}f_{R} \,.
\end{eqnarray}
Hence, we get the Schr\"{o}dinger-like equations for the left- and
right-chiral fermions \cite{Almeida:0901,Liu:09b}
\begin{eqnarray}
H_{L}f_{L}(z)=m^{2}f_{L}(z)\,,\\
H_{R}f_{R}(z)=m^{2}f_{R}(z)\,,
\end{eqnarray}
with the corresponding Hamiltonians
\begin{eqnarray}
 H_{L}=\biggl(-\frac{d}{dz}+\eta e^{A}F(\phi,\chi,\rho)\biggl)
  \biggl(\frac{d}{dz}+\eta e^{A}F(\phi,\chi,\rho)\biggl)\,,\\
          \label{HamiltonianLeft}
 H_{R}=\biggl(-\frac{d}{dz}-\eta e^{A}F(\phi,\chi,\rho)\biggl)
 \biggl(\frac{d}{dz}-\eta e^{A}F(\phi,\chi,\rho)\biggl)\,.
          \label{HamiltonianRight}
\end{eqnarray}
The Schr\"{o}dinger equations can be expressed as
\begin{subequations}\label{Scheq}
\begin{eqnarray}
 [-\partial_{z}^{2}+V_{L}(z)]f_{L} &=& m^{2}f_{L} \,,
    \\ \label{ScheqLeft}
 [-\partial_{z}^{2}+V_{R}(z)]f_{R} &=& m^{2}f_{R} \,,
       \label{ScheqRight}
\end{eqnarray}
\end{subequations}
where the effective Schr\"{o}dinger-like potentials for the KK
modes $f_{L,R}$ are
\begin{subequations}\label{Vz}
\begin{eqnarray}
V_{L}(z)=\eta^{2} e^{2A}F^{2}(\phi,\chi,\rho)-\eta
e^{A}\partial_{z}F(\phi,\chi,\rho)-\eta
e^{A}(\partial_{z}A)F(\phi,\chi,\rho)\,,  \label{VzL}  \\
V_{R}(z)=\eta^{2} e^{2A}F^{2}(\phi,\chi,\rho)+\eta
e^{A}\partial_{z}F(\phi,\chi,\rho)+\eta
e^{A}(\partial_{z}A)F(\phi,\chi,\rho)\,.   \label{VzR}
\end{eqnarray}
\end{subequations}

\section{Fermion localization and resonances on a three-field thick brane}
\label{sec3Fields}

\begin{figure}[htb]
\begin{center}
\includegraphics[width=7cm,height=5cm]{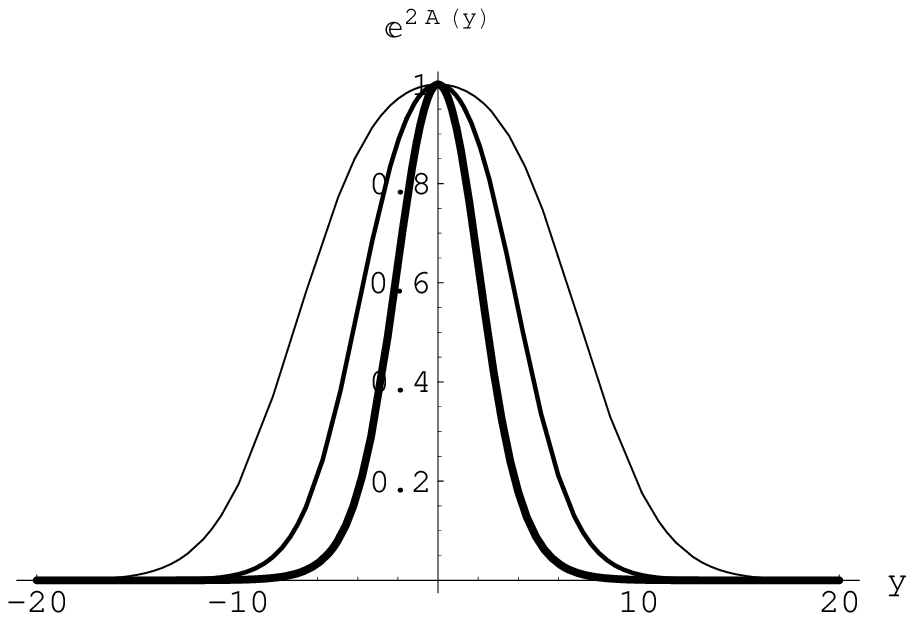}
\includegraphics[width=7cm,height=5cm]{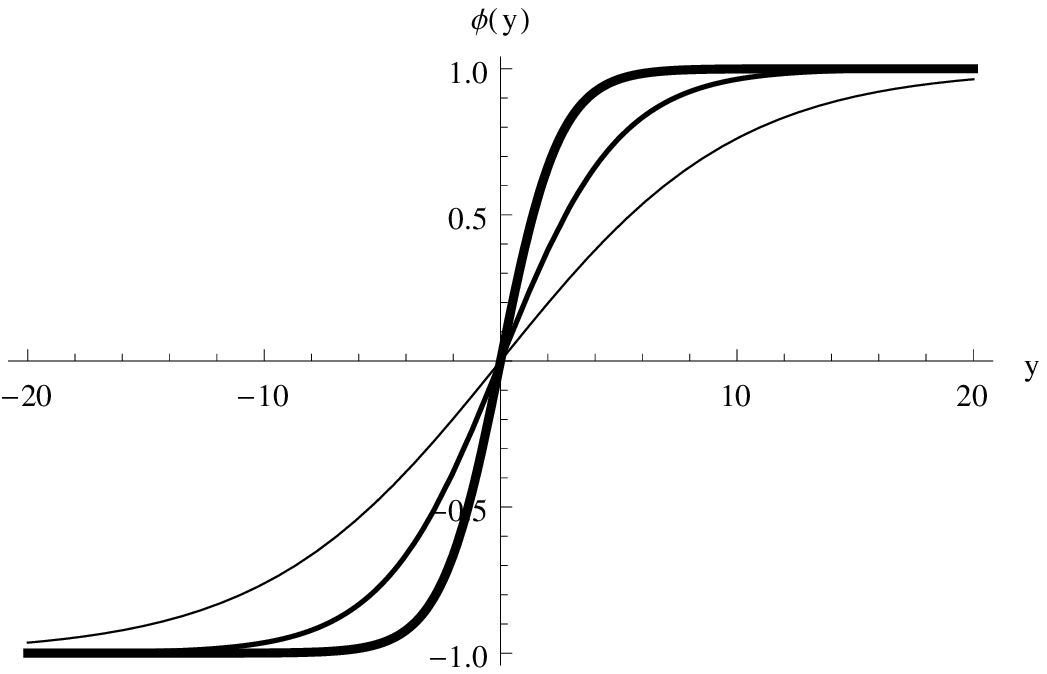}
\includegraphics[width=7cm,height=5cm]{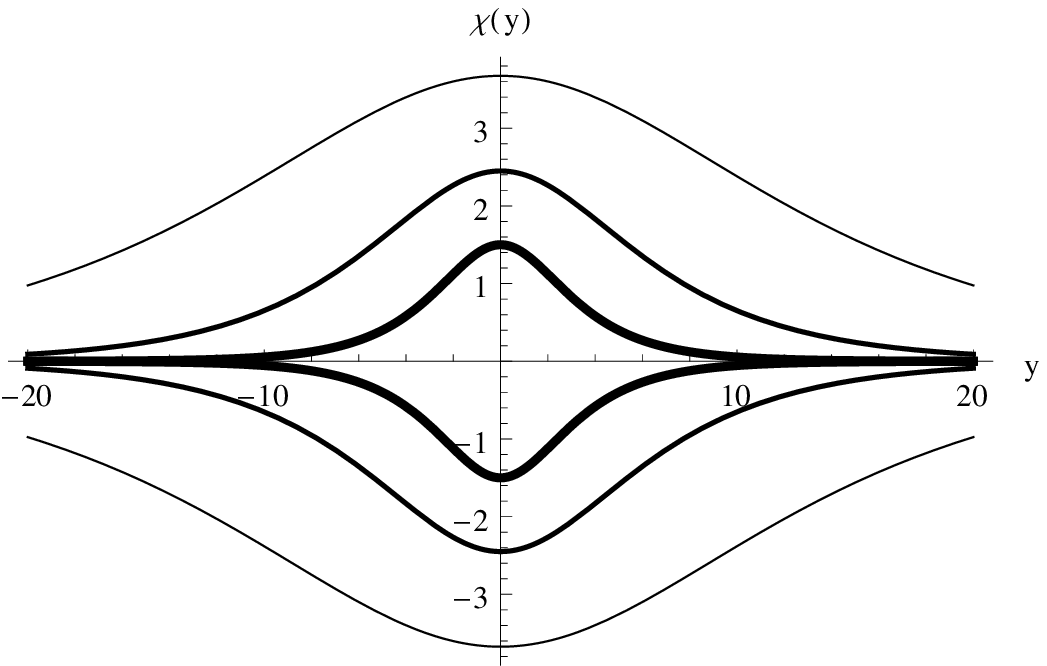}
\includegraphics[width=7cm,height=5cm]{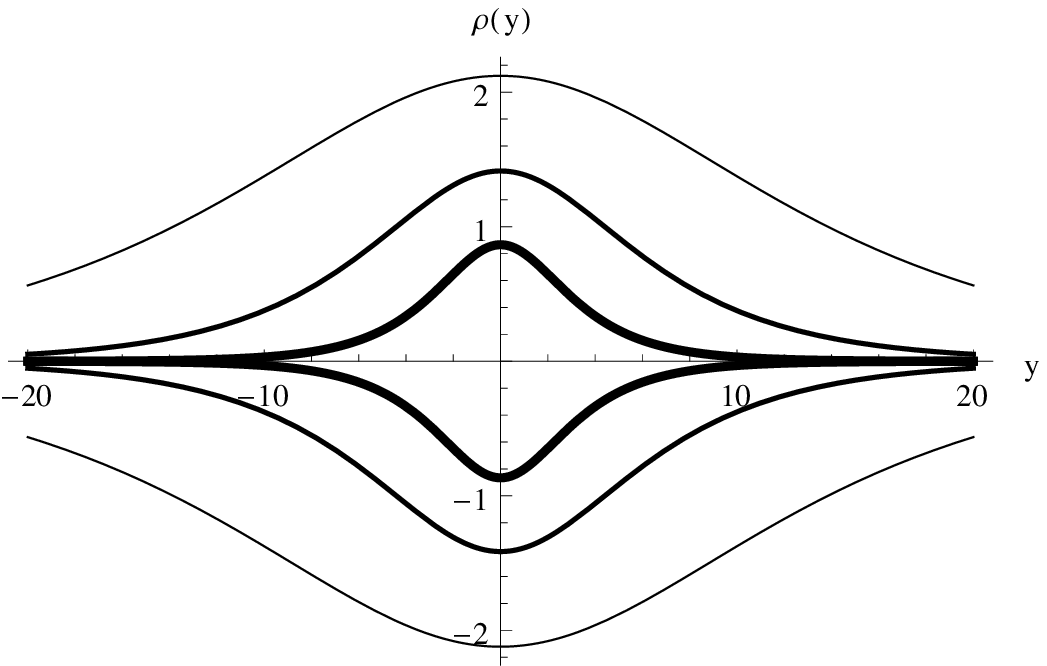}
\end{center}\vskip -2mm
\caption{Warp factor $e^{2A(y)}$, scalar fields  $\phi(y)$,
$\chi(y)$ and $\rho(y)$. The parameters are set to $\theta=\pi/6$,
$a=0.05$ (thin trace), $0.1$, $0.2$ (thick trace). The thickness of
lines is increases with $a$.}
 \label{WarpFactorScalars}
\end{figure}

Let us now investigate the three-field brane models. As a natural
extension of the single-field and two-field scenarios, the
superpotential with three-fields could be
\begin{eqnarray}
W(\phi,\chi,\rho)=2\left(\phi-\frac{1}{3}\phi^{3}-a\phi(\chi^{2}+\rho^{2})\right)\,,
\label{superpotentialW}
\end{eqnarray}
which was considered in Refs \cite{Bazeia:02,Izquierdo:02}. Here $a$
is a real parameter. In this case, the corresponding potential
(\ref{PotentialV1}) is
\begin{eqnarray}
 V= \frac{1}{2}
  \bigg[4a^2\phi^{2}(\rho^{2}+\chi^{2})
        +\big[1-\phi^{2}+a(\rho^{2}+\chi^{2})\big]^{2}
  \bigg]
  -\frac{4}{3}\bigg[\phi -\frac{1}{3}\phi^{3}
     -a\phi(\chi^{2}+\rho^{2})\bigg]^{2}\,.
 \label{PotentialV2}
\end{eqnarray}
The first order equations are
\begin{subequations}\label{1stOrderEqsWithW}
\begin{eqnarray}
\frac{d\phi}{dy} &=& 1-\phi^{2}-a(\chi^{2}+\rho^{2})\,,\\
\frac{d\chi}{dy} &=& -2a\phi\chi\,,\\
\frac{d\rho}{dy} &=& -2a\phi\rho\,,\\
\frac{d A}{dy} &=& -\frac{2}{3}\Biggl(\phi-\frac{1}{3}\phi^{3}
 -a\phi(\chi^{2}+\rho^{2})\Biggl)\,.
\end{eqnarray}\end{subequations}
The solution is given by
\begin{subequations}\label{SolutionI}
\begin{eqnarray}
 \phi(y)&=&\tanh(2ay), \\
 \chi(y) &=& \pm\sqrt{\frac{1}{a}-2}\;\cos\theta \;
      \text{sech}(2ay)\,,\label{SolutionIa}\\
 \rho(y) &=& \pm\sqrt{\frac{1}{a}-2}\;\sin\theta \;
      \text{sech}(2ay)\,,\label{SolutionIb}\\
 A(y) &=& \frac{1}{9a}\bigg[
      (1-3a)\tanh^{2}(2ay)
      -2\ln \cosh(2ay)\bigg]\,, \label{SolutionIc}
\end{eqnarray}\end{subequations}
where $0<a<1/2$ and $\theta\in[0,2\pi)$. The shapes of the solution
are plotted in Fig. \ref{WarpFactorScalars}. It can be seen that the
scalar $\phi(y)$ is a kink while $\chi(y)$ and $\rho(y)$ are
campanulate. The warp factor has a normal shape and the brane
thickness decreases with $a$. We note that the warp factor here is
the same with the one in two-field thick brane case in Refs.
\cite{Bazeia:04a,Almeida:0901} but different from the single-field
one in Refs. \cite{WangPRD2002,Liu:09b}. In this paper, we take
$\theta=\pi/6$ and consider the positive solutions of $\chi(y)$ and
$\rho(y)$.

\begin{figure}[htb]
\begin{center}
\includegraphics[width=7cm,height=5cm]{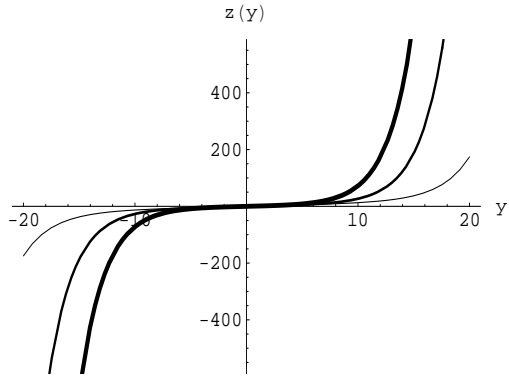}
\end{center} \caption{Plot of the function $z(y)$
 for $a=0.05$, 0.1 and 0.2.
 The thickness of lines is increases with $a$.}
 \label{fig:zy}
\end{figure}

Now, we turn to the $z$-coordinate according to
Eq.~(\ref{coordinateTransformation}). For general value of $a$,
the expression of $A(y)$ (\ref{SolutionIc}) cannot be integrated
in a known explicit form. Even for the particular case $a=1/3$,
for which we have an analytical expression of $z(y)$, it is hard
to get an expression for the inverse $y(z)$ in an explicit form in
terms of some analytical functions, so some numerical methods are
needed. The numerical result for $z(y)$ is depicted in Fig.
\ref{fig:zy}. The derivatives with respect to $z$, can be
calculated as a function of $y$ by using the coordinate
transformation (\ref{coordinateTransformation})
\begin{eqnarray}
 \frac{d A(z)}{dz} &=& \frac{d A(y)}{dy}\frac{dy}{dz}
         =\frac{d A}{dy}e^{A(y)}\,,\\
 \frac{d F(z)}{dz} &=& \frac{d F(y)}{dy}\frac{dy}{dz}
         =\frac{d F(y)}{dy}e^{A(y)}\,.
\end{eqnarray}
With the above expressions, we can rewrite the potentials
(\ref{Vz}) as a function of $y$:
\begin{subequations}\label{Vy}
\begin{eqnarray}
V_{L}(z(y))&=&\eta e^{2A}\bigg[\eta F^{2}(\phi,\chi,\rho)-
            \partial_{y}F(\phi,\chi,\rho)
            -F(\phi,\chi,\rho)\partial_{y}A(y)\bigg],  \label{VyL}  \\
V_{R}(z(y))&=&V_{L}(z(y))|_{\eta\rightarrow-\eta}.   \label{VyR}
\end{eqnarray}
\end{subequations}
It can be seen that, for the left- or right-chiral fermion
localization, there must be some kind of scalar-fermion coupling. In
addition, for the kink configuration of the scalar $\phi(y)$ and the
lump configurations of $\chi(y)$ and $\rho(y)$, $F(\phi,\chi,\rho)$
should be an odd function of $\phi(y)$ when one demands that
$V_L(z(y))$ or $V_R(z(y))$ is $Z_2$-even with respect to $y$. Here,
we would like to consider two cases: the simplest Yukawa coupling
$F(\phi,\chi,\rho)=\phi\chi\rho$ and the general coupling
$F(\phi)=\phi^k\chi\rho$ with odd positive $k$. Surely, other more
complex cases can be investigated.

\subsection{Case I: $F(\phi,\chi,\rho)=\phi^{k}\chi\rho$ with $k=1$}

Firstly, we investigate the simplest Yukawa coupling
$F(\phi,\chi,\rho)=\phi\chi\rho$ for the three-scalar generated
thick branes. For simplicity, we consider the brane solution
(\ref{SolutionI}) only. The explicit forms of the potentials
(\ref{Vy}) are given by
\begin{subequations}\label{VyCaseI}
\begin{eqnarray}
 V_{L}(y)&=& \frac{1}{288a^{2}}\biggl[\eta(2a-1)
             \exp\left(\frac{2(1-3a)}{9a}\tanh^{2}(2ay)\right)
             \cosh^{-6-4/(9a)}(2ay)\biggl]\nonumber\\
        &&\times  \bigg[3\big(52\sqrt{3}a^{2}+9\eta-2a(2\sqrt{3}+9\eta)\big)
                         -4\sqrt{3}(1+9a)\cosh(8ay) \label{VyCaseIL}\\
        &&~~~~+\big(16\sqrt{3}a+24\sqrt{3}a^{2}-27\eta+54a\eta\big)\cosh(4ay)
                  \bigg], \nonumber\\
 V_{R}(y)&=&V_{L}(y)|_{\eta\rightarrow-\eta}. \label{VyCaseIR}
\end{eqnarray}
\end{subequations}
Here, we note that $y$ cannot be expressed in an explicit form in
terms of $z$, so the potentials are expressed with the variable
$y$. From Fig.~\ref{fig:zy}, we can easily obtain the asymptotic
behavior of the potentials (\ref{VyCaseI}). The values of the
potentials at $z=y=0$ and $y$ or $z\rightarrow\pm\infty$ are given
by
\begin{eqnarray}
&& V_{L}(0) =-\frac{\sqrt{3}}{2}(1-2a)\eta= -V_{R}(0),\label{VL0}\\
&& V_{L}(\pm\infty)=0=V_{R}(\pm\infty).\label{VLRinfty}
\end{eqnarray}
It can be seen that both potentials have the same asymptotic
behavior when $z \rightarrow \pm\infty$, but opposite behavior at
the origin $z=0$, which results in the well-known conclusion: only
one of the massless left- and right-chiral fermions could be
localized on the brane.

\begin{figure}[htb]
\begin{center}
\subfigure[$V_L$, $\eta=1$]{\label{fig:VLRCaseIa}
\includegraphics[width=7cm,height=4.5cm]{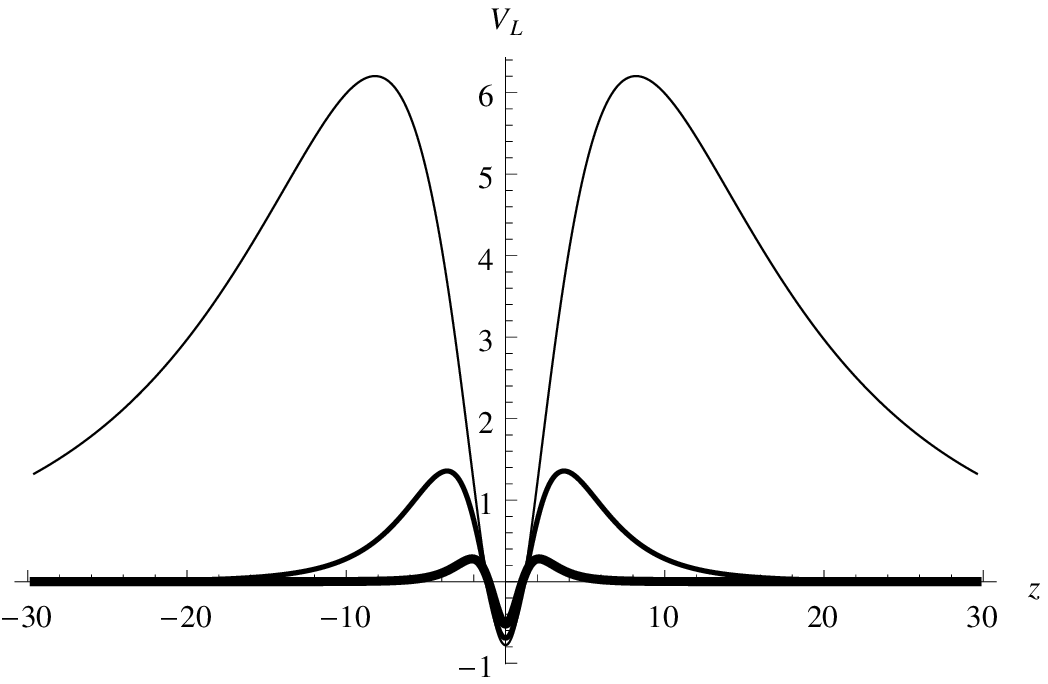}}
\subfigure[$V_R$, $\eta=1$]{\label{fig:VLRCaseIb}
\includegraphics[width=7cm,height=4.5cm]{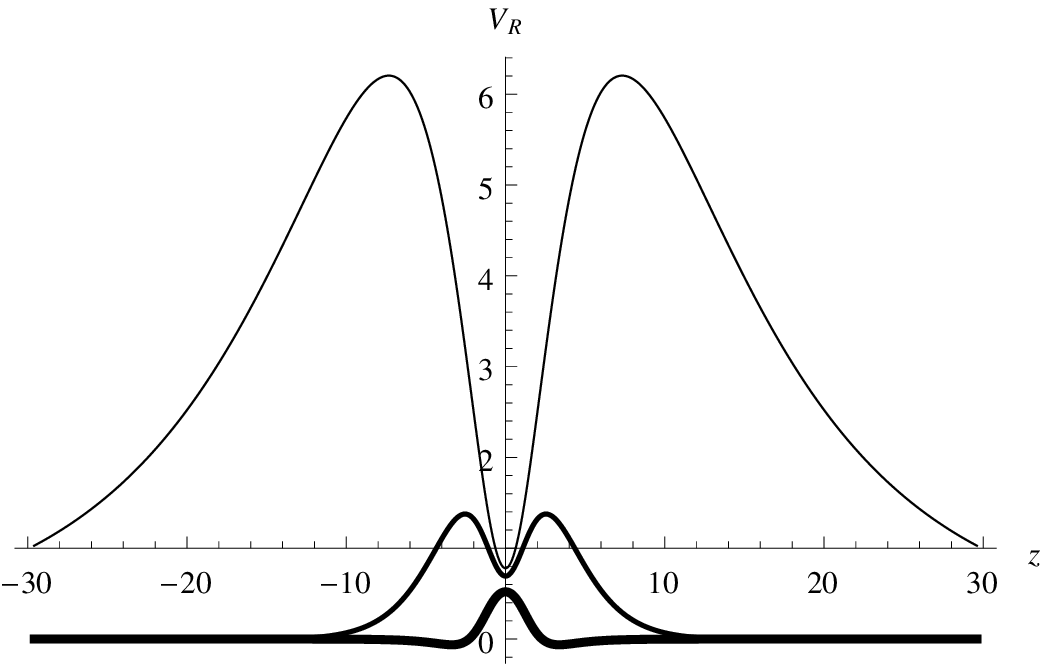}}
\subfigure[$V_R$, $\eta=3$]{\label{fig:VLRCaseIc}
\includegraphics[width=7cm,height=4.5cm]{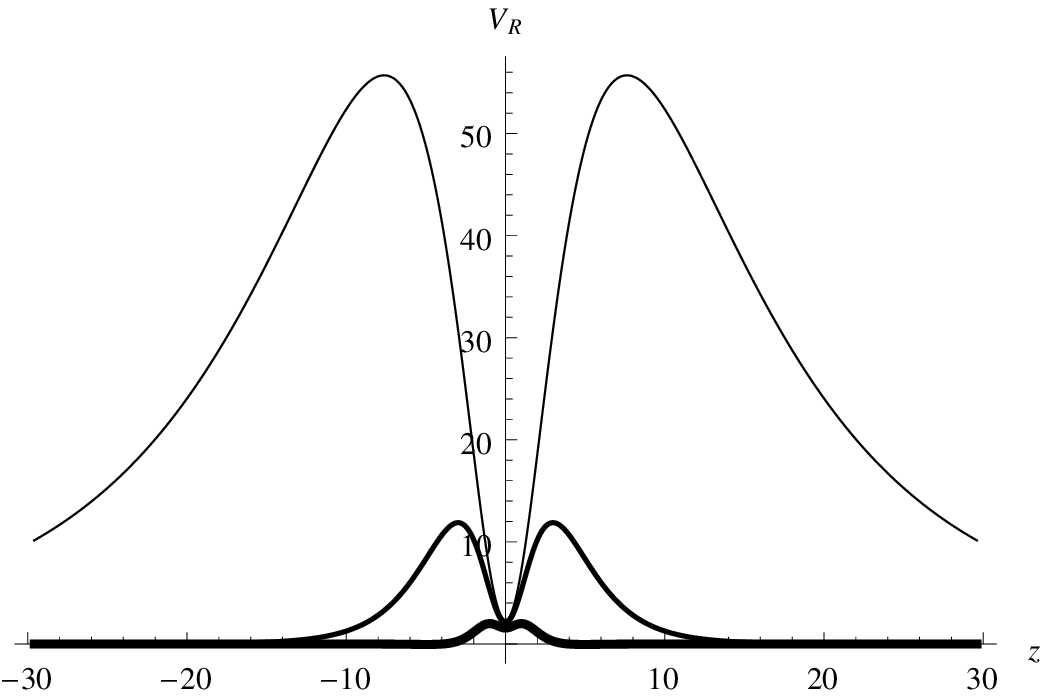}}
\subfigure[$V_R$, $\eta=7$]{\label{fig:VLRCaseId}
\includegraphics[width=7cm,height=4.5cm]{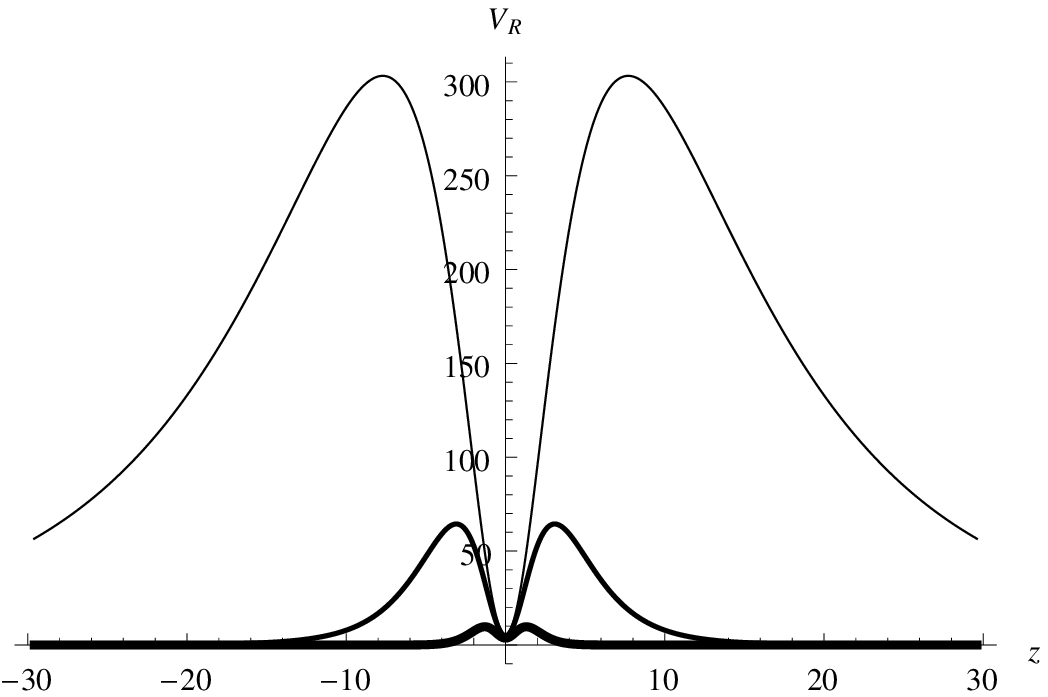}}
\end{center} \vskip -5mm
\caption{Potentials $V_{L}(z)$ and $V_{R}(z)$ for left- and
right-chiral fermions with $F(\phi,\chi,\rho)=\phi\chi\rho$. The
parameter $a$ is set to $0.05$ (thin trace), $0.1$, $0.2$ (thick
trace).}
 \label{fig:VLRCaseI}
\end{figure}

As mentioned above, with the numerical methods, we can get the
shape of potential function graphics as shown in Fig.
\ref{fig:VLRCaseI}. Clearly, for any value of $0<a<1/2$ and
$\eta>0$, $V_{L}(z)$ is a volcano type of potential. Therefore,
the potential of the left-handed fermions does not provide mass
gap between the zero-mode and KK excitation modes, and there is a
continuous gapless spectrum of KK excitation modes. The zero mode
of the left-handed fermions is turned out to be
\begin{eqnarray}
 f_{L0}(z)&\propto&\exp\bigg(-\eta\int_{0}^{z}d\bar{z}e^{A(\bar{z})}
       \phi(\bar{z})\chi(\bar{z})\rho(\bar{z})\bigg) \nonumber \\
 &=& \exp\bigg( -\eta\int_{0}^{y} d\bar{y}
       \phi(\bar{y})\chi(\bar{y})\rho(\bar{y})  \bigg)
       \label{fL0CaseI}
\end{eqnarray}
and the normalization condition
\begin{eqnarray}
 \int_{-\infty}^{\infty}f^2_{L0}(z) dz
 &=& \int_{-\infty}^{\infty}f^2_{L0}(y) e^{-A(y)} dy \nonumber \\
 &\propto& \int_{-\infty}^{\infty}  \exp\bigg( -A(y)-2\eta\int_{0}^{y} d\bar{y}
       \phi(\bar{y})\chi(\bar{y})\rho(\bar{y})  \bigg)dy \nonumber \\
 &=& \int_{-\infty}^{\infty}  \exp\bigg( -A(y)-2\eta\frac{3(1-2a)}{16a^2}\tanh^2(2ay)  \bigg)dy
 <\infty
       \label{fL0CaseI}
\end{eqnarray}
cannot be satisfied because $A(y)\rightarrow-\frac{4}{9}y$ while
$\tanh^2(2ay)\rightarrow 1$ when $y\rightarrow\infty$. Hence, the
zero mode (\ref{fL0CaseI}) is nonnormalizable. The reason is that
the coupling $F(\phi,\chi,\rho)$ contains the lump configurations
$\chi$ and $\rho$. In order get localized zero mode for
left-handed fermions, we can only consider the coupling of
fermions and the kink $\phi$, i.e., $F(\phi,\chi,\rho)=\phi$, for
which the zero mode is normalizable provided $\eta>2/9$. In this
paper, we would not consider the special case. On the other hand,
the potential $V_{R}(y)$ at the brane location is always positive,
and when far away from the brane it gradually becomes zero. We
know that this type of potential cannot trap any bound state
fermions with right-chirality and there exists no zero mode of
right-handed fermions. As is well known this result is consistent
with the previous well-known conclusion that massless fermions
must be single-handed chirality in the brane world models
\cite{Ringeval:02,Liu:09b}.

However, the structure of the potential $V_R$ is determined by the
coupling constant $\eta$ and parameter $a$ jointly. For a given
$a$, as $\eta$ increases, there will be a potential well and the
depth of the well will be deeper and deeper. On the other hand,
for a given $\eta$, the smaller the parameter $a$ is, the deeper
the depth of the potential well. There is a fine-tuning
relationship between $a$ and $\eta$. This means that a competition
between $a$ and $\eta$ exists. We consider the situation of the
extreme ones. As $a$ range from $0$ to $1/2$, for a value $a$
which is close to $1/2$, in order to produce a potential well, we
need very large $\eta$, i.e., the scalar and fermion have a strong
coupling. All we can do is to find the essence of the phenomenon
behind. This is the most interesting things and phenomena.

\subsubsection{Left-handed fermions}

We will find the following fact that the emergence of the
potential wells are closely related to the resonance states, which
are massive fermions with finite lifetime. A similar phenomenon
for left- and right-chiral fermions can be found in
Refs.~\cite{Liu:09b,Almeida:0901}, the former is in the context of
two-scalar constructed thick brane with internal structure, the
latter is in the background of the single-scalar generated $dS$
thick brane with a class of scalar-fermion coupling. Here, we
extend this point of view and method to a three-scalar generated
thick brane obtained in Ref.~\cite{Bazeia:02}. We follow
Refs.~\cite{Liu:09b,Almeida:0901} through solving numerically the
equations (\ref{Scheq}) with numerical Schr\"{o}dinger potentials
in (\ref{VyCaseI}) to study the massive modes of fermions. In
particular, in order to obtain the probability of massive modes of
the fermions on the brane, we adopt the new method of calculating
the probability in Ref.~\cite{Liu:09b}. In addition, our results
are consistent with the previous two models, and some new
properties about resonances are discovered.

In order to get the solutions of the KK modes $f_{L,R}(z)$ from
the second order ordinary differential equations (\ref{Scheq}), we
need additional two types of initial conditions:
\begin{eqnarray}
f(0)=c_{0},\,\, f'(0)=0 \,, \label{evencondition}
\end{eqnarray}
and
\begin{eqnarray}
f(0)=0,\,\, f'(0)=c_{1} \,. \label{oddcondition}
\end{eqnarray}
From Fig. \ref{fig:VLRCaseI}, we can see that the potentials which
we will consider have even-parity. According to the knowledge of
quantum mechanics, we can know that the wave functions of a
Schr\"{o}dinger equation with a finite smooth potential are
continuous at any position. The above two types of initial
conditions will lead to even-parity and odd-parity KK modes,
respectively. It is worth noting that, according to a specific
numerical procedure, the form of the initial conditions of
equivalent deformation will be used in our followed numerical
calculations. The constants $c_{0}$ and $c_{1}$ for the nonbound
massive KK modes are arbitrary, and in accordance with specific
conditions of numerical calculations.

\begin{figure}[htb]
\begin{center}
 \subfigure[$m^{2}=2.8848$]{\label{fig_fL_Eigenvalue_a}
  \includegraphics[width=4.5cm,height=3.5cm]{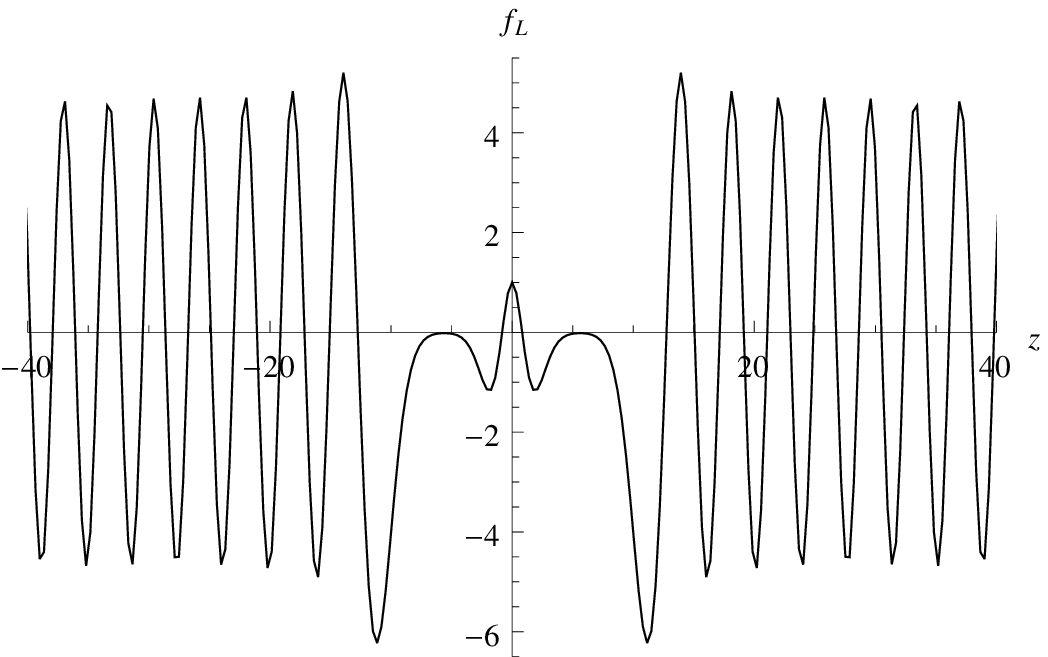}}
 \subfigure[$m^{2}=2.884901$]{\label{fig_fL_Eigenvalue_b}
  \includegraphics[width=4.5cm,height=3.5cm]{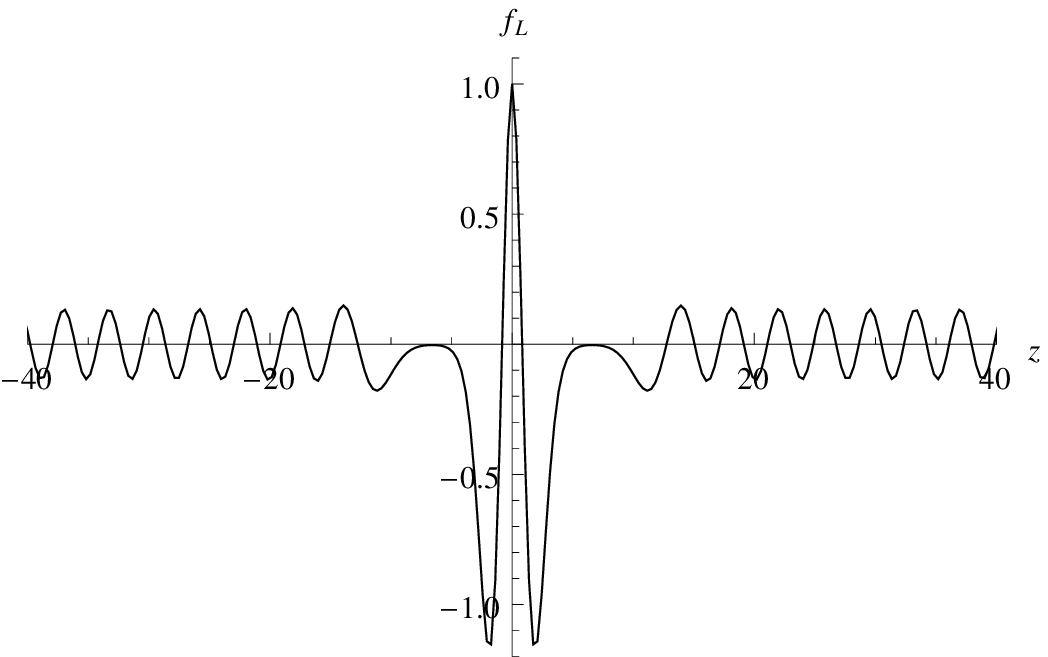}}
 \subfigure[$m^{2}=2.8849035$]{\label{fig_fL_Eigenvalue_f}
  \includegraphics[width=4.5cm,height=3.5cm]{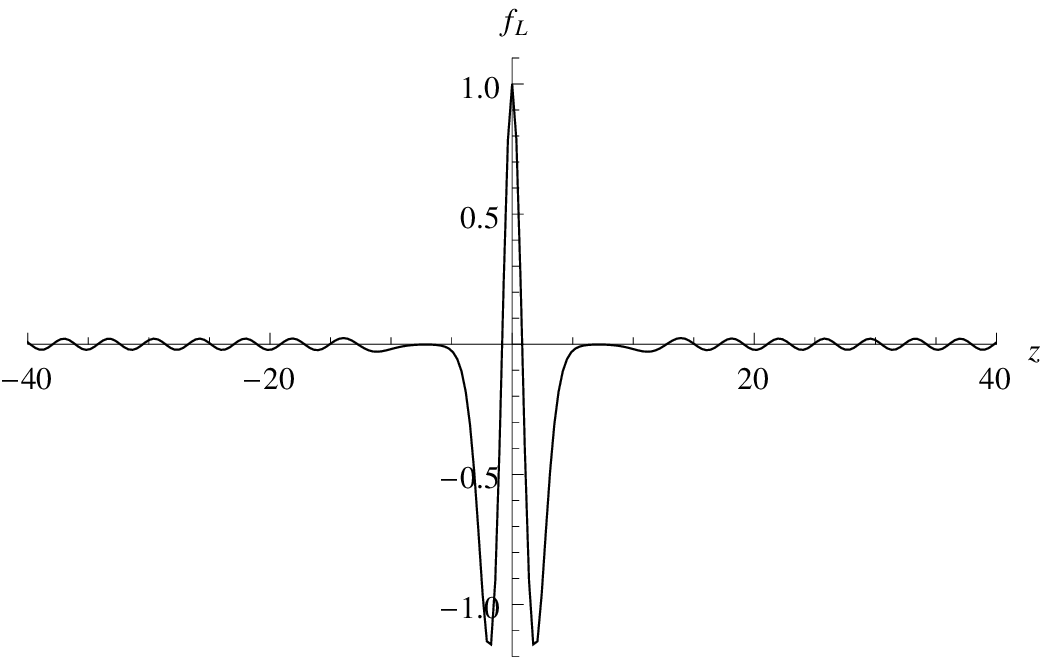}}
 \subfigure[$m^{2}=2.884904$]{\label{fig_fL_Eigenvalue_c}
  \includegraphics[width=4.5cm,height=3.5cm]{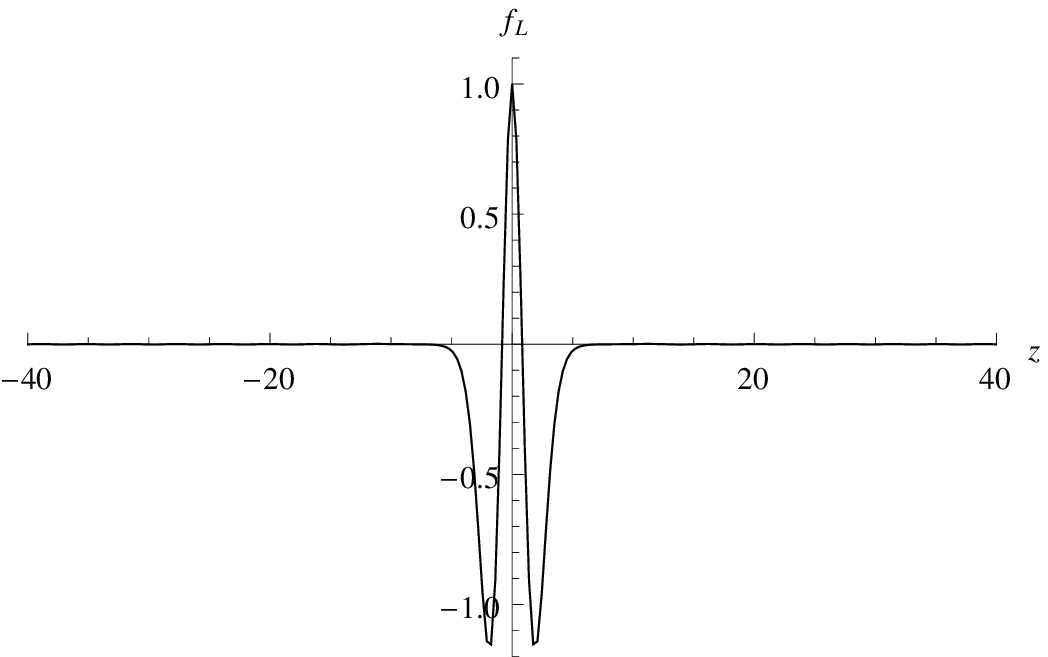}}
 \subfigure[$m^{2}=2.884907$]{\label{fig_fL_Eigenvalue_d}
  \includegraphics[width=4.5cm,height=3.5cm]{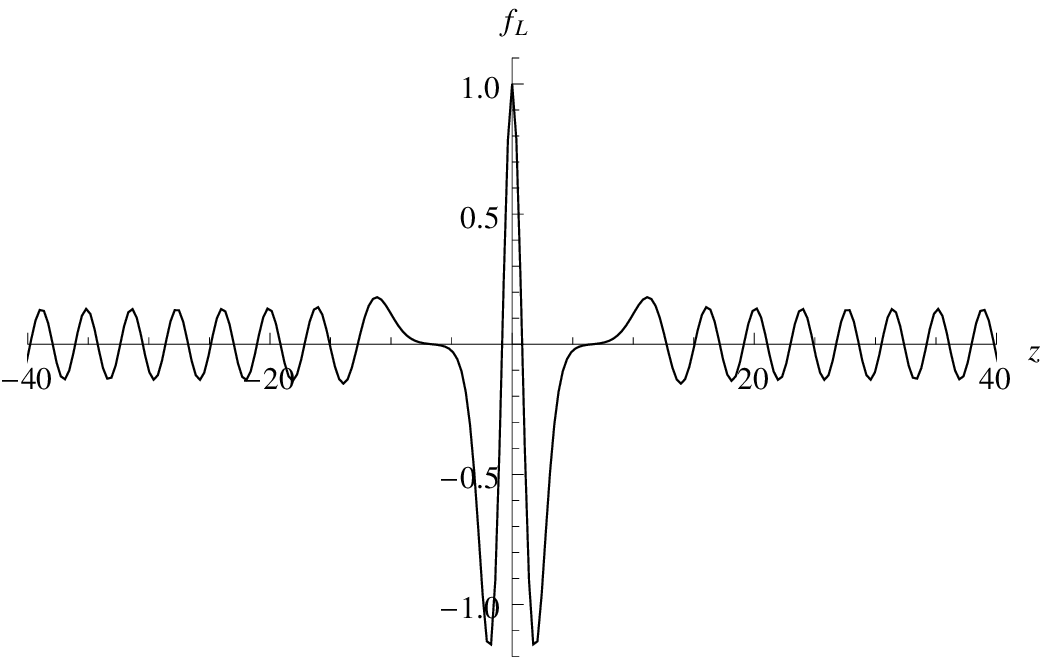}}
 \subfigure[$m^{2}=2.885004$]{\label{fig_fL_Eigenvalue_e}
  \includegraphics[width=4.5cm,height=3.5cm]{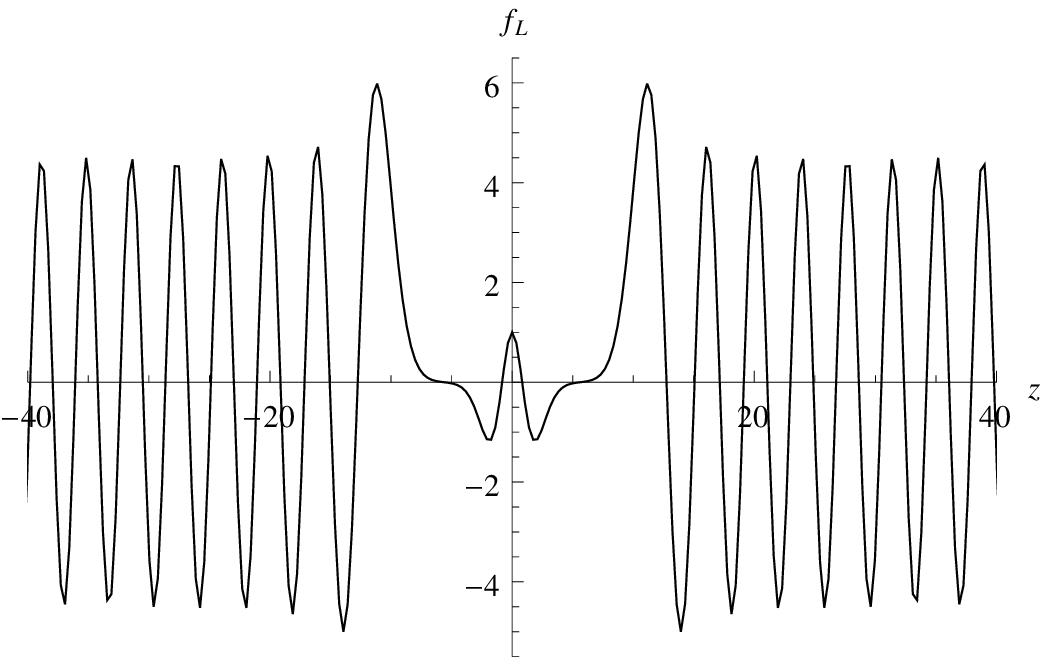}}
\end{center}
\caption{The shapes of the even parity massive KK modes $f_L(z)$ of
left-chiral fermions for the coupling
$F(\phi,\chi,\rho)=\phi\chi\rho$ with different $m^{2}$. The
parameters are set to $z_{max}=210$, $a=0.05$ and $\eta=1$ .}
 \label{fig_fL_Eigenvalue}
\end{figure}

From the point of view of quantum mechanics, as the general
effects of the quantum mechanics, there will be tunneling effect
when the massive KK modes experience the potential barrier near
the brane. By means of Numerov method \cite{Numerov,Bazeia:07},
for a given potential function combined with the Schr\"{o}dinger
equations, the numerical results show that the corresponding KK
modes with a series of masses and lifetimes will be obtained. The
KK modes with finite lifetime are also called resonances, i.e.,
quasibound states or metastable states. Some even-parity massive
KK modes of left-handed chiral fermions for the coupling
$F(\phi,\chi,\rho)=\phi\chi\rho$ with different $m^{2}$ are
depicted in Fig.~\ref{fig_fL_Eigenvalue}. These graphics indicate
that there are some resonance states when the mass accesses to
certain specific values. The results we get here are consistent
with that given in Ref.~\cite{Almeida:0901,Liu:09b}. However,
there are some new characteristics. With different models and
coupling mechanisms, some new properties of resonances and
physical meaning may exist. In what follows, we will carefully
discuss the resonance problem in the background of the
three-scalar constructed thick brane.

From Fig. \ref{fig_fL_Eigenvalue}, we see that, to obtain a very
clear resonance, one need a very good accuracy for the parameter
$m^2$. If we want to search for a resonance directly through the
eigenfunction and energy eigenvalue, we will encounter great
difficulties and workload. According to the formal theory of
resonance of quantum wave functions, we can study the probability
of finding the massive KK modes around the vicinity of the brane
location within a relatively large region
\cite{Almeida:0901,Liu:09b}.

\begin{figure}[htb]
\begin{center}
 \subfigure[$m=1.2257$]{\label{fig_fL_Probability_a}
  \includegraphics[width=4.5cm,height=3.5cm]{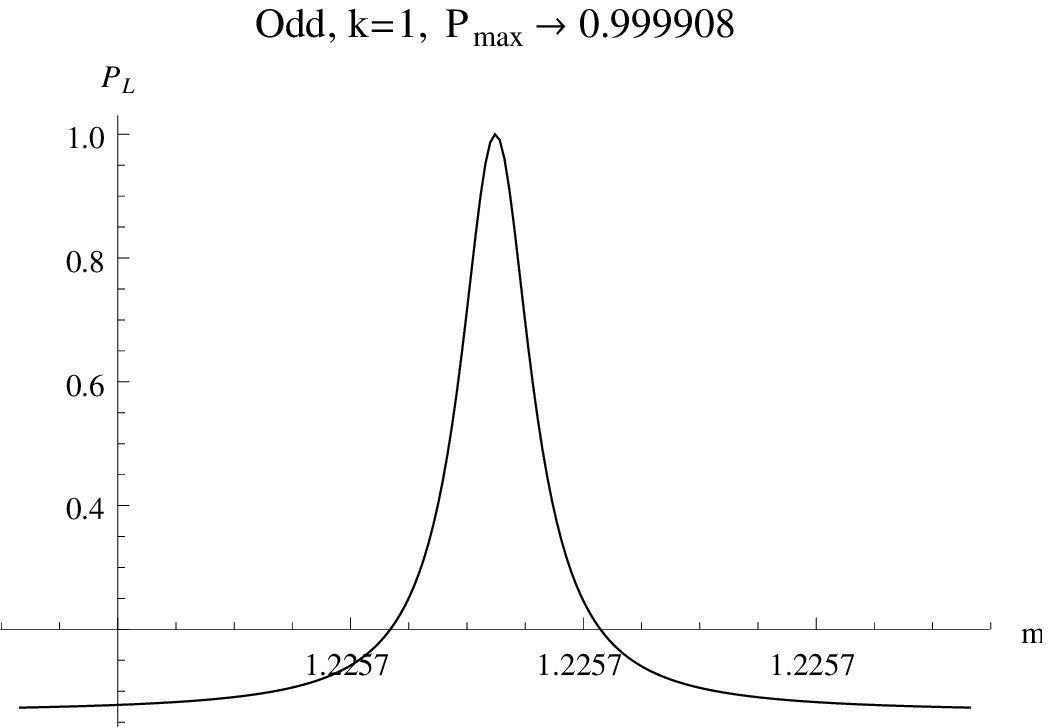}}
 \subfigure[$m=1.6985$]{\label{fig_fL_Probability_b}
  \includegraphics[width=4.5cm,height=3.5cm]{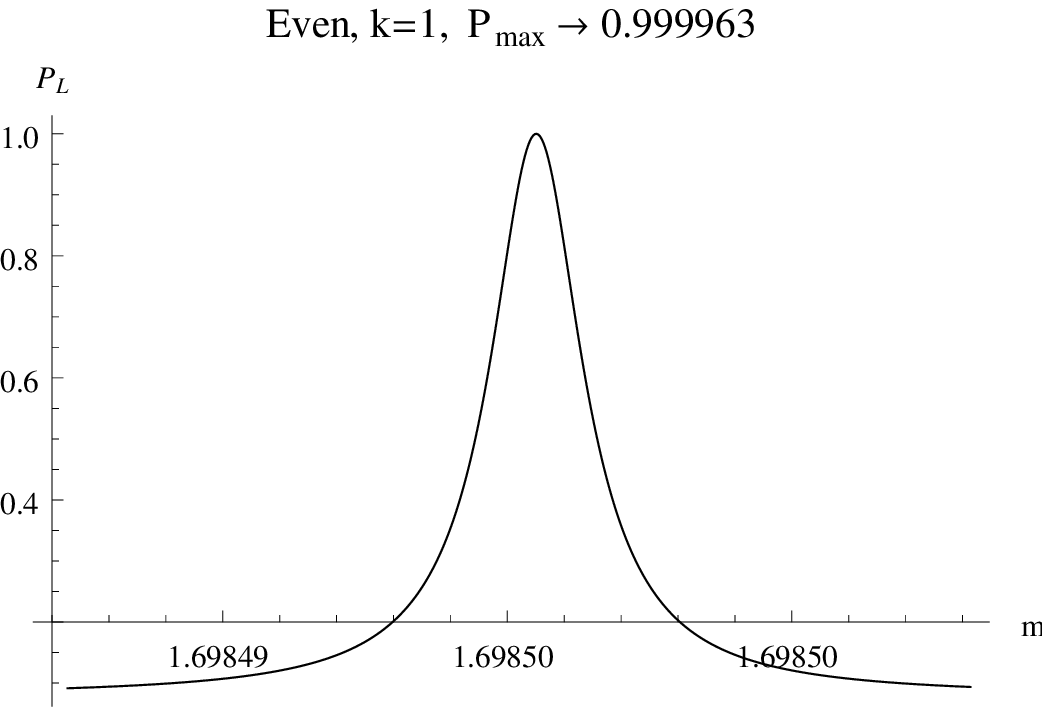}}
 \subfigure[$m=2.03213$]{\label{fig_fL_Probability_c}
  \includegraphics[width=4.5cm,height=3.5cm]{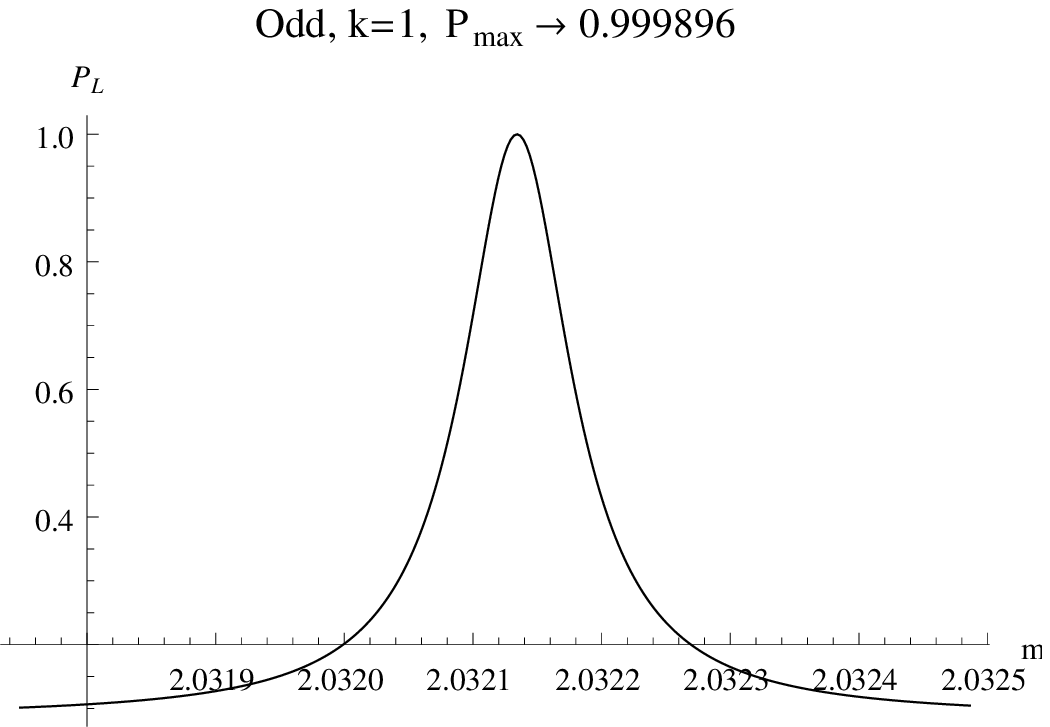}}
 \subfigure[$m=2.28112$]{\label{fig_fL_Probability_d}
  \includegraphics[width=4.5cm,height=3.5cm]{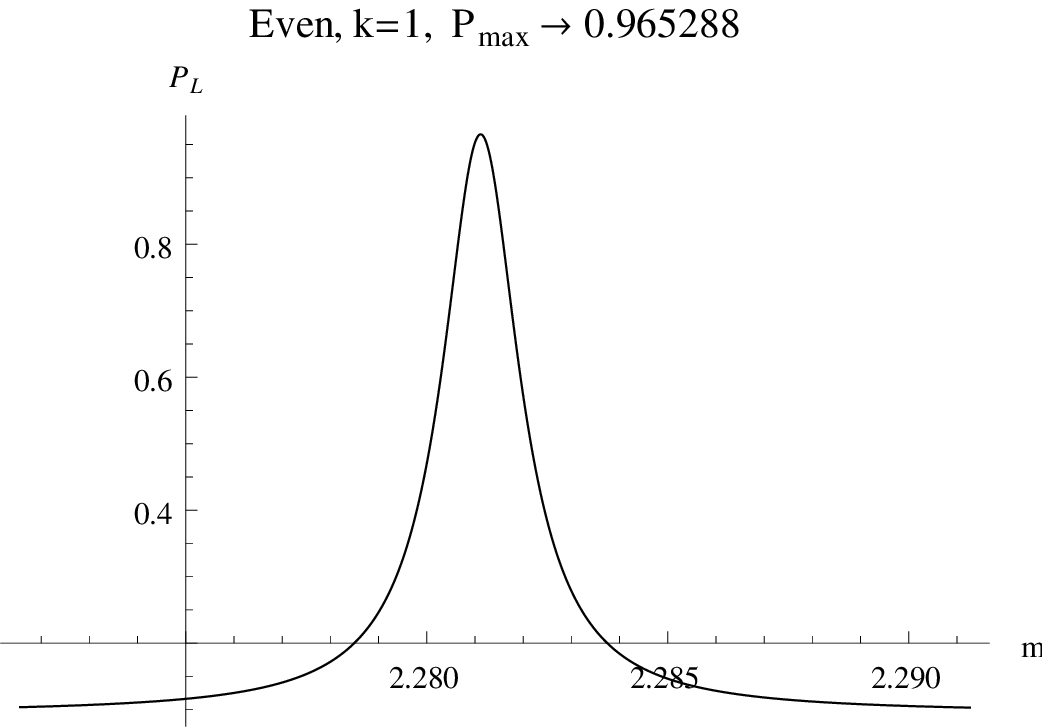}}
 \subfigure[$m=2.45601$]{\label{fig_fL_Probability_e}
  \includegraphics[width=4.5cm,height=3.5cm]{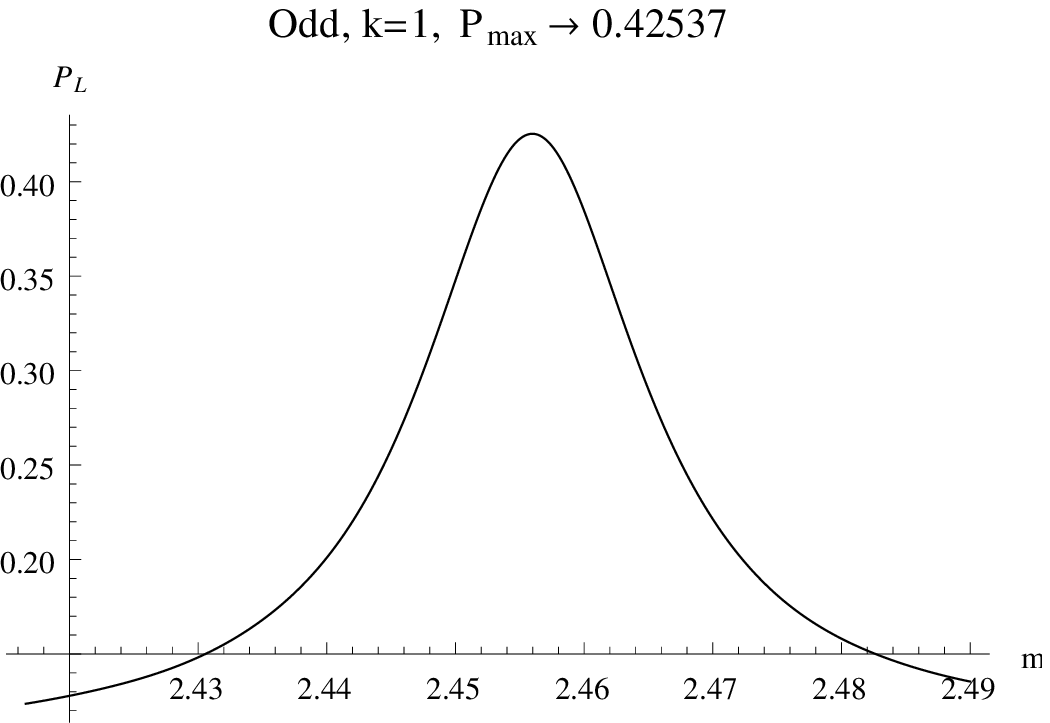}}
\end{center}\vskip -5mm
\caption{The probability $P_{L}$ (as a function of $m$) for finding
the even parity (under three resonant peaks) and odd parity (upper
three resonant peaks) massive KK modes of left-chiral fermions
around the brane location for the coupling
$F(\phi,\chi,\rho)=\phi\chi\rho$. The parameters are set to
$z_{max}=210$, $a=0.05$ and $\eta=1$.}
 \label{fig_fL_Probability}
\end{figure}

\begin{figure}[htb]
\begin{center}
 \subfigure[$m^{2}=1.50234155$]{\label{fig_fL_Eigenvalue_1a}
  \includegraphics[width=4.5cm,height=3.5cm]{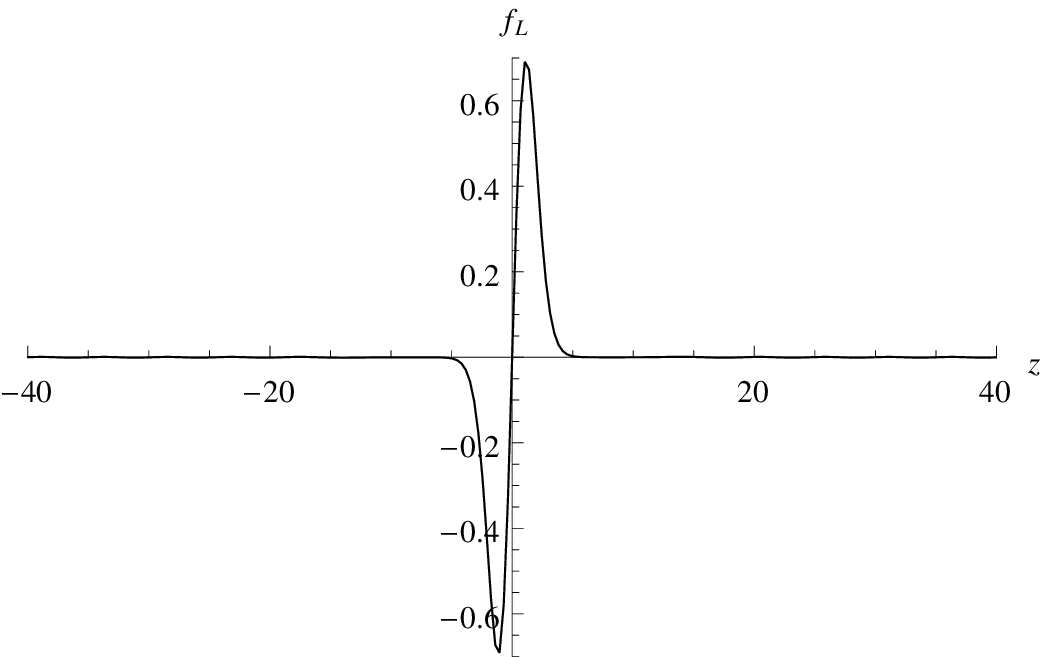}}
 \subfigure[$m^{2}=2.884904$]  {\label{fig_fL_Eigenvalue_1b}
  \includegraphics[width=4.5cm,height=3.5cm]{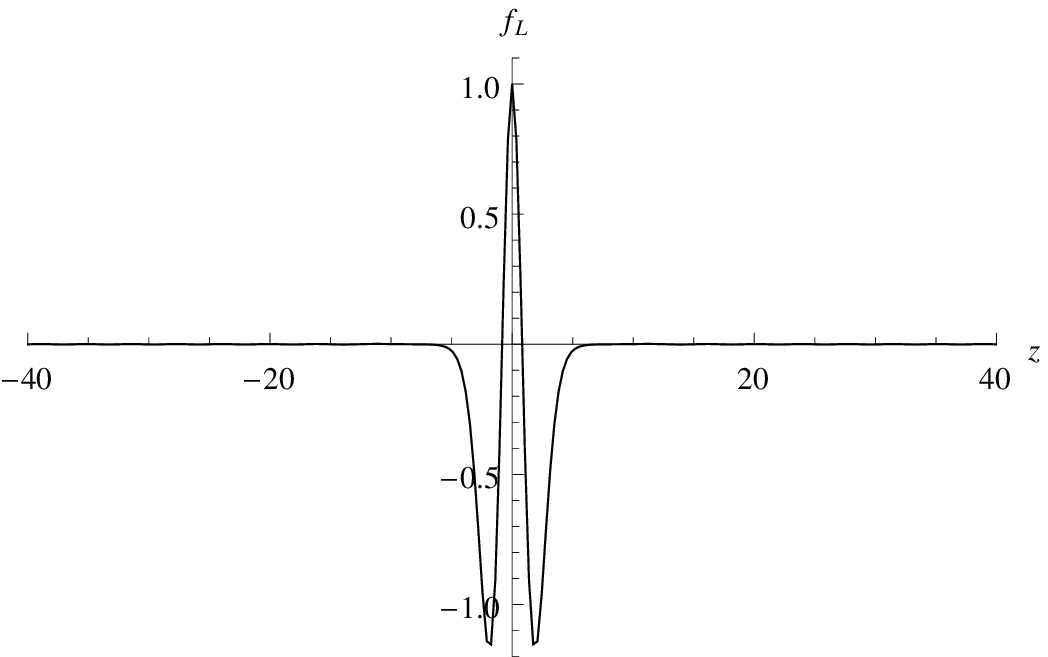}}
 \subfigure[$m^{2}=4.12957$]   {\label{fig_fL_Eigenvalue_1c}
  \includegraphics[width=4.5cm,height=3.5cm]{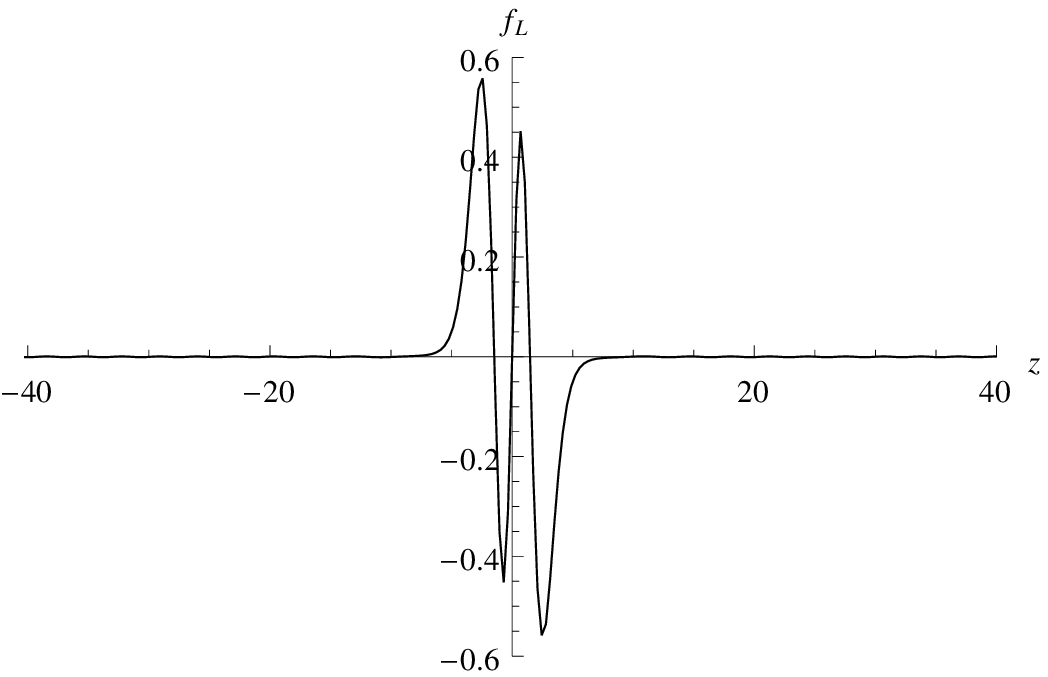}}
 \subfigure[$m^{2}=5.2035$]    {\label{fig_fL_Eigenvalue_1d}
  \includegraphics[width=4.5cm,height=3.5cm]{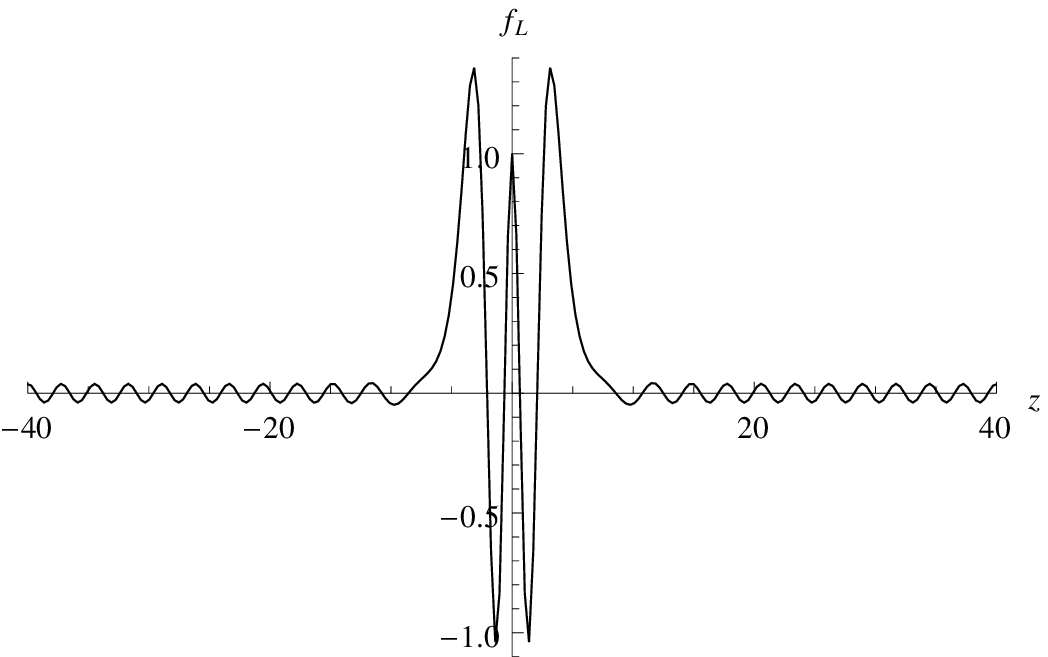}}
 \subfigure[$m^{2}=6.032$]     {\label{fig_fL_Eigenvalue_1e}
  \includegraphics[width=4.5cm,height=3.5cm]{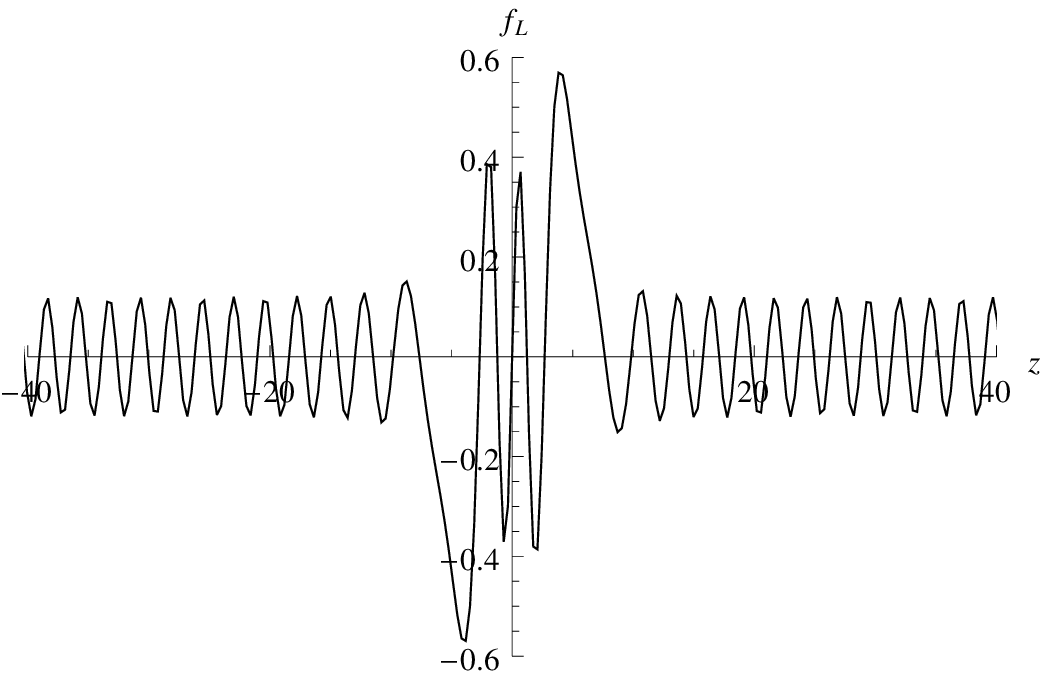}}
\end{center} \vskip -5mm
\caption{The shapes of the even parity and odd parity massive KK
modes $f_L(z)$ of left-chiral fermions for the coupling
$F(\phi,\chi,\rho)=\phi\chi\rho$ with different $m^{2}$. The
parameters are set to $z_{max}=210$, $a=0.05$ and $\eta=1$ .}
 \label{fig_fL_Eigenvalue1}
\end{figure}

Because the formula (\ref{Scheq}) can be re-written in the form
$\mathcal {O}^{\dag}_{L,R}\mathcal
{O}_{L,R}f_{L,R}(z)=m^{2}f_{L,R}(z)$, so $|f_{L,R}(z)|^{2}$ can be
interpreted as the probability for finding the massive KK modes at
the position $z$ along extra dimension. In Ref.~\cite{Almeida:0901},
the authors suggested that large peaks in the distribution of
$f_{L,R}(0)$ as a function of $m$ would reveal the existence of
resonance states. Here, we follow the procedures of the extended
idea in Ref.~\cite{Liu:09b} for the two types of initial conditions
(\ref{evencondition}) and (\ref{oddcondition}). For a given
eigenvalue $m^{2}$, the corresponding relative probability is
defined as \cite{Liu:09b}:
\begin{eqnarray}
P_{L,R}(m)=\frac{\int^{z_{b}}_{-z_{b}}|f_{L,R}(z)|^{2}dz}{\int^{z_{max}}_{-z_{max}}|f_{L,R}(z)|^{2}dz}\,,
\label{relative probability}
\end{eqnarray}
where the relation of $z_{b}$ and $z_{max}$ is selected as
$z_{max}/z_{b}=10$, for which, the probability for a plane wave
mode with mass $m$  is $0.1$. For the KK modes with eigenvalue
$m^{2}$ much larger than the maximum of the corresponding
potential function, they will have a very good plane wave
approximation, and the corresponding probability $P(m)$ will tend
to $0.1$. For the parameters $k=1$, $a=0.05$, $\eta=1$, the
potential wells distribute in the range
$[-z_{b},z_{b}]$($z_{b}=21$) along extra dimension, so we choose
$z_{max}=210$, and the corresponding graphics of the resonance
spectra are depicted in Fig.~\ref{fig_fL_Probability}. For
left-handed fermions, we find a total of five clear resonant peaks
with approximate eigenvalues $m^{2}=1.50234155$, $2.884904$,
$4.12957$, $5.2035$, $6.032$. The corresponding left-handed
eigenfunctions are shown in Fig.~\ref{fig_fL_Eigenvalue1}. It can
be seen that the configurations of
Figs.~\ref{fig_fL_Eigenvalue_1a}, ~\ref{fig_fL_Eigenvalue_1c} and
~\ref{fig_fL_Eigenvalue_1e} are odd-parity eigenfunctions, and the
other two are even-parity ones. From Figs.
\ref{fig_fL_Probability} and \ref{fig_fL_Eigenvalue1}, we can see
that odd-parity eigenfunctions and even-parity eigenfunctions are
placed at the five resonant peaks with irregular intervals of mass
eigenvalues.

As mentioned above, it is usual to describe the resonance by its
width $\Gamma=\Delta m$ at half-maximum of the corresponding
resonant peak \cite{Almeida:0901}. In this way, we can calculate
the lifetime $\tau$ of a resonance state with the width $\Gamma$.
This means that the massive fermions disappeared along the
direction of extra dimension after time
$\tau=1/\Gamma$~\cite{Gregory:00}. The eigenvalues $m^2$, mass
$m$, width $\Gamma$ and lifetime $\tau$ for resonances of
left-handed fermions corresponding to the resonant peaks shown in
Fig. \ref{fig_fL_Probability} are listed in Table
\ref{TableSpectra3Fields}. We can see that the resonance states
with relatively larger eigenvalues have a relatively smaller
lifetimes.

\begin{figure}[htb]
\begin{center}
 \subfigure[$m_{Step}=10^{-11}$]{\label{fig_Fluctuationeven_a}
  \includegraphics[width=7cm,height=5cm]{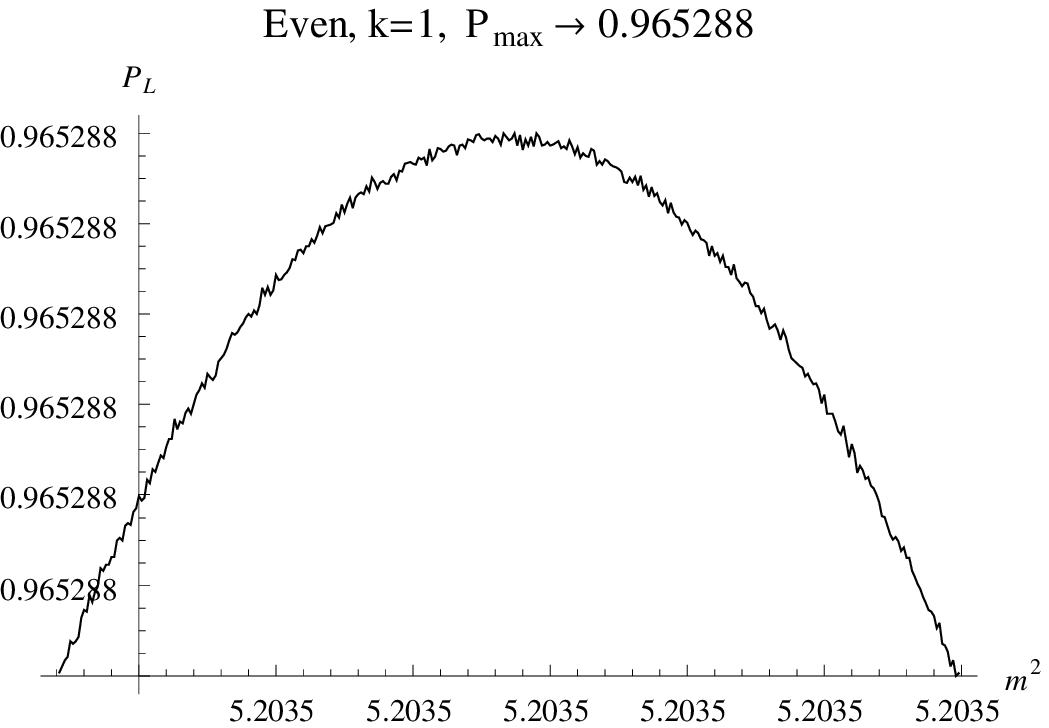}}
 \subfigure[$m_{Step}=10^{-12}$]
  {\label{fig_Fluctuationeven_b}
  \includegraphics[width=7cm,height=5cm]{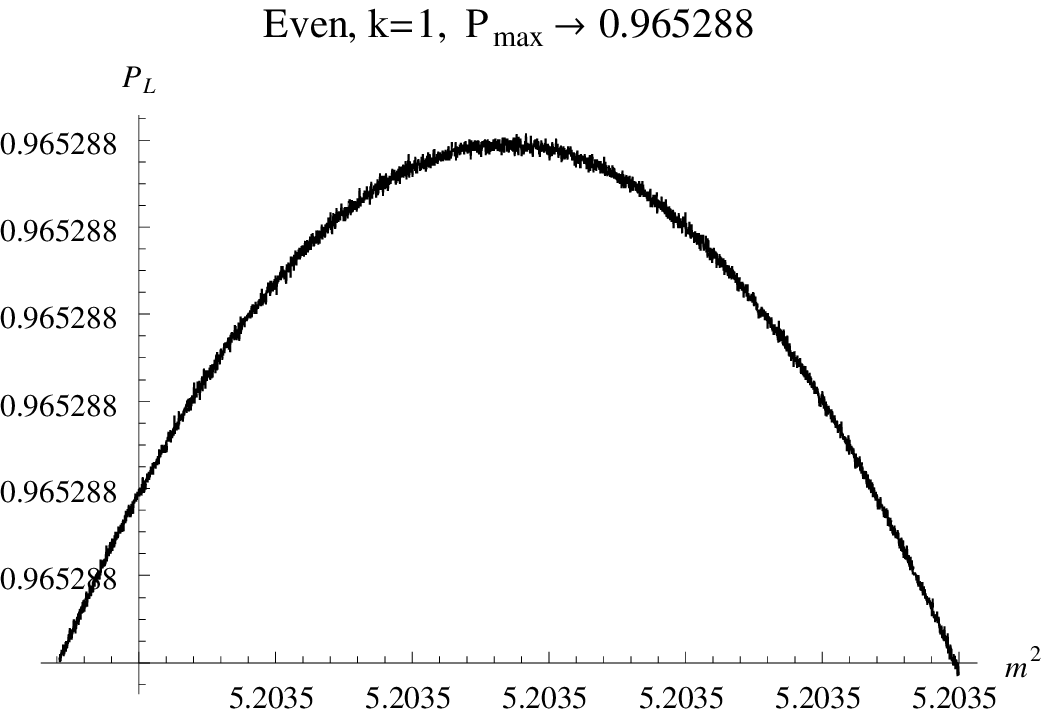}}
\end{center} \vskip -5mm
\caption{The fluctuation of the probability $P_{L}$ (as a function
of $m$) for finding the even parity massive KK modes of left-chiral
fermions around the brane location for the coupling
$F(\phi,\chi,\rho)=\phi\chi\rho$ with different step accuracy
$m_{Step}$ in numerical method. The parameters are set to $a=0.05$
and $\eta=1$.}
 \label{fig_Fluctuationeven}
\end{figure}

In addition, it should be noted that the solutions we obtain with
numerical methods are some approximate solutions which meet a
certain accuracy. For example, for the even-parity resonance states
shown in Fig.~\ref{fig_fL_Probability_d}, when the step accuracy of
the mass $m$ accesses to a certain value, the fluctuation of the
probability likes a hedgehog would emerge around the corresponding
resonant peak (see Fig. \ref{fig_Fluctuationeven}). The overall
profile of the probability fluctuation still corresponds to the
resonant peak.

\begin{table}[h]
\begin{center}
\begin{tabular}{||c|c|c|c|c|c|c|c||}
 \hline
 $k$ & $\mathcal{C}$ & $\mathcal {P}$ & $m^{2}_{n}$ & $m_{n}$ & $\Gamma$& $\tau$ &  $P_{max}$   \\
 \hline\hline

 & & odd & 1.50234155      & 1.2257    & $7.92943$$\times$$10^{-9}$   & $1.26112$$\times$$10^{8}$  & 0.999908 \\ \cline{3-8}
 &  &  even    & 2.884904         & 1.6985     & $2.14302$$\times$$10^{-6}$    & 466631  &  0.999963  \\ \cline{3-8}
 & $\mathcal{L}$ & odd  & 4.12957 & 2.03213 & 0.00011232 & 8903.15&  0.999896 \\ \cline{3-8}
  & & even  & 5.2035  &  2.28112      & 0.00217082     & 460.655 & 0.965288
  \\ \cline{3-8}
  & & odd  & 6.03  & 2.45601 & 0.0296346 & 33.7443 & 0.424972 \\
  \cline{2-8}
  \raisebox{2.3ex}[0pt]{1}&  & even & 1.502620398 & 1.22581 & $7.98459$$\times$$10^{-9}$ & $1.25241$$\times$$10^{8}$&0.998721
  \\ \cline{3-8} \cline{3-8}
  &&odd  & 2.885601     & 1.69871     & $2.14401$$\times$$10^{-6}$   & 466416  &  0.999967 \\ \cline{3-8}
  & $\mathcal{R}$ &even      & 4.13079  & 2.03243      & 0.000112761    & 8868.28  &  0.999897  \\ \cline{3-8}
  & &odd& 5.2051 & 2.28147 & 0.00218277 & 458.134  & 0.964972  \\ \cline{3-8}
   & &even      & 6.033  & 2.45622      & 0.0300386    & 33.2905  &  0.42376  \\ \hline \hline

 & & odd & 0.136755      & 0.369804   & 0.0000263936   & 37888  & 0.999961 \\ \cline{3-8}
 & $\mathcal{L}$ &  even    & 0.39192         & 0.626035      & 0.00123317    & 810.918  & 0.965227   \\ \cline{3-8}
 & & odd  & 0.607 & 0.779102 & 0.0219376 & 45.5838 & 0.365007 \\
  \cline{2-8}\cline{2-8}
  \raisebox{2.3ex}[0pt]{3}&& even & 0.136763 & 0.369815 & 0.0000263447 & 37958.2 & 0.99861
  \\ \cline{3-8} \cline{3-8}
  & $\mathcal{R}$ &odd  & 0.39195      & 0.626059     & 0.00123482   & 809.836  & 0.965131  \\ \cline{3-8}
  &  &even      & 0.607  & 0.779102      & 0.0221197     & 45.2086  &  0.364746  \\ \hline\hline

 & & odd & 0.04936      & 0.222171    & 0.000811223   & 1232.71  & 0.965989 \\ \cline{3-8}
 & \raisebox{2.3ex}[0pt]{$\mathcal{L}$} &  even    & 0.1419         & 0.376696      & 0.040258    & 24.8398  & 0.245507   \\  \cline{2-8}
  \raisebox{2.3ex}[0pt]{5}&  & even & 0.04937 & 0.222194  & 0.000810324 & 1234.07 & 0.966584
  \\ \cline{3-7} \cline{3-8}
  & \raisebox{2.3ex}[0pt]{$\mathcal{R}$} &odd  & 0.1421      & 0.376962     & 0.0439204   & 22.7684 & 0.241676 \\  \hline\hline

 & $\mathcal {L}$ & odd & 0.02423      & 0.15566    & 0.00524879   & 190.52 &  0.565059 \\ \cline{2-8}
  \raisebox{2.3ex}[0pt]{7}& $\mathcal{R}$ & even & 0.02425 & 0.155724  & 0.00516514  & 193.606 & 0.567736 \\ \hline\hline

 & $\mathcal {L}$ & odd & 0.01331     & 0.115369    & 0.0193406   & 51.7047 & 0.313287 \\ \cline{2-8}
  \raisebox{2.3ex}[0pt]{9}& $\mathcal{R}$ & even & 0.01325 & 0.115109 & 0.0171465 & 58.3209 & 0.322988 \\ \hline
\end{tabular}\\
\caption{The eigenvalues $m^2$, mass, width and lifetime of left-
and right-chiral fermions with odd-parity and even-parity
solutions for the coupling $F(\phi,\chi,\rho)=\phi^{k}\chi\rho$.
$\mathcal {C}$ and $\mathcal {P}$ stand for chirality and parity,
respectively. $\mathcal {L}$ and $\mathcal {R}$ are short for
left-handed and right-handed, respectively. The parameters are
$k=\{1$, $3$, $5$, $7$, $9\}$, $a=0.05$ and $\eta=1$.}
\label{TableSpectra3Fields}
\end{center}
\end{table}

\subsubsection{Right-handed fermions}

\begin{figure}[htb]
\begin{center}
 \subfigure[$m=1.22581$]{\label{fig_fR_Probability_a}
  \includegraphics[width=4.5cm,height=3.5cm]{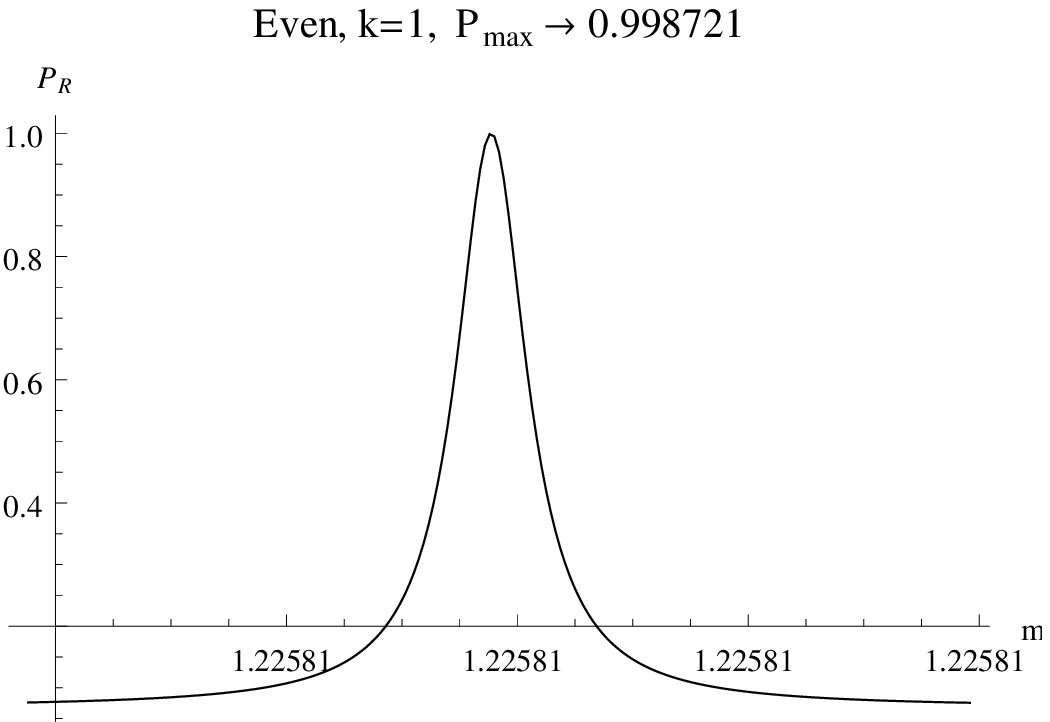}}
 \subfigure[$m=1.69871$]{\label{fig_fR_Probability_b}
  \includegraphics[width=4.5cm,height=3.5cm]{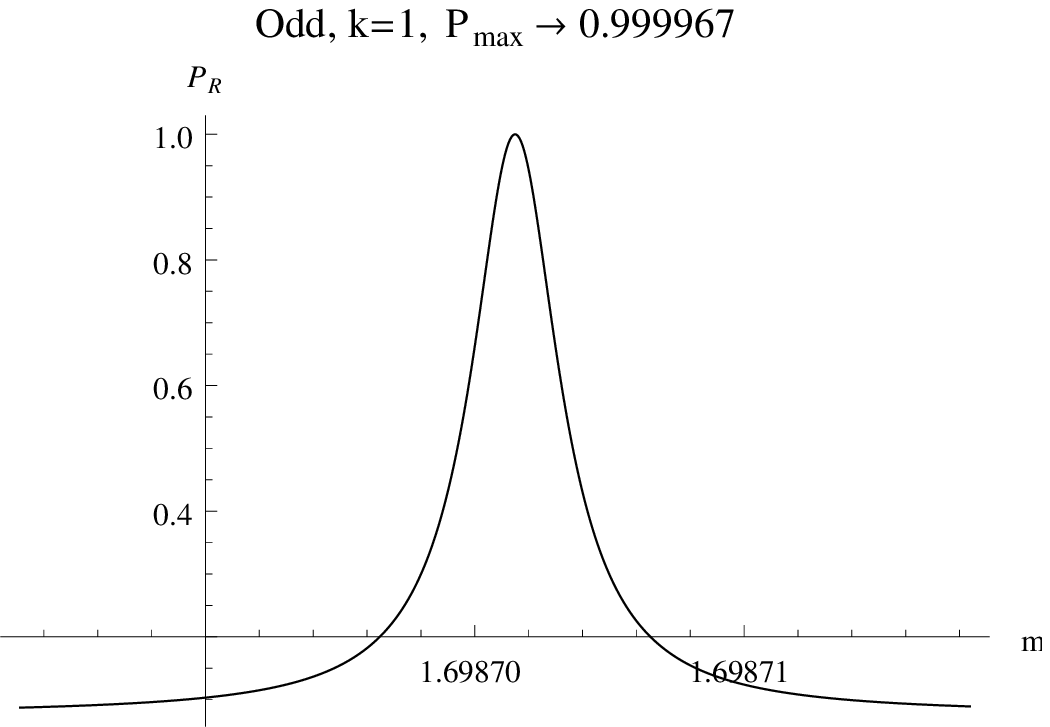}}
 \subfigure[$m=2.03243$]{\label{fig_fR_Probability_c}
  \includegraphics[width=4.5cm,height=3.5cm]{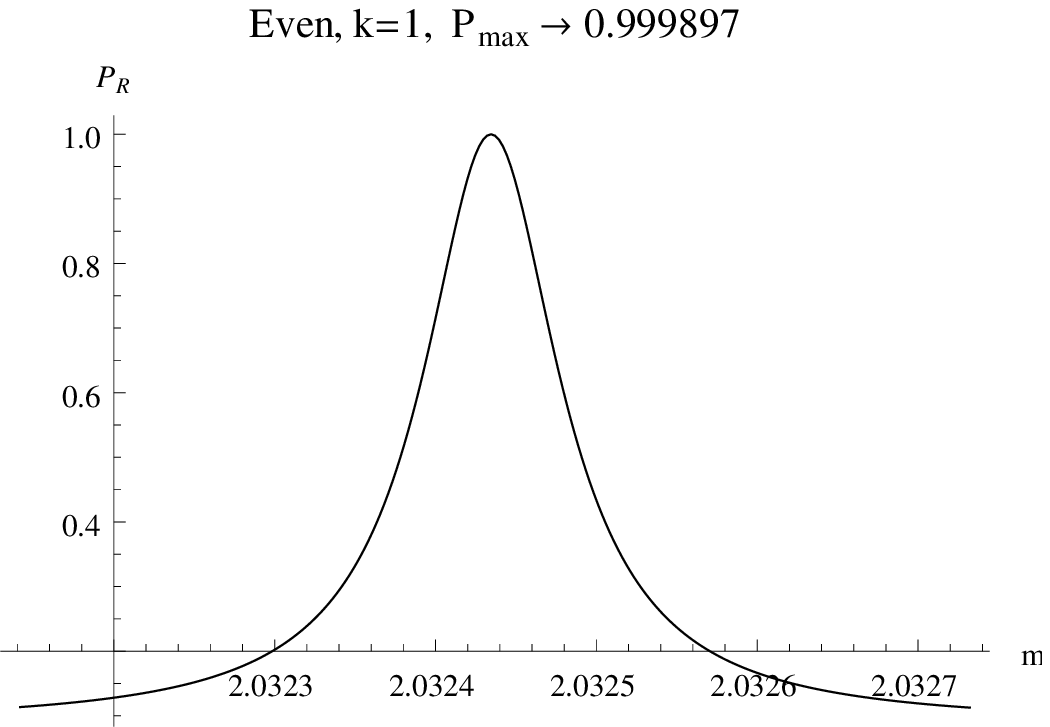}}
   \subfigure[$m=2.28147$]{\label{fig_fR Probability_d}
  \includegraphics[width=4.5cm,height=3.5cm]{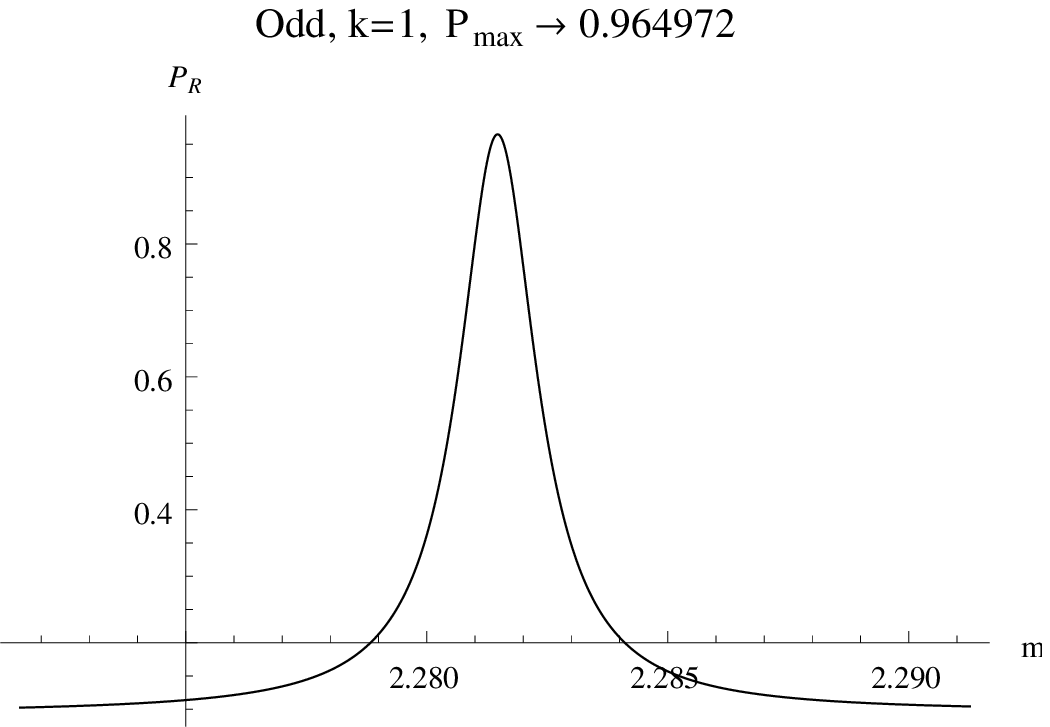}}
 \subfigure[$m=2.45622$] {\label{fig_fR_Probability_e}
  \includegraphics[width=4.5cm,height=3.5cm]{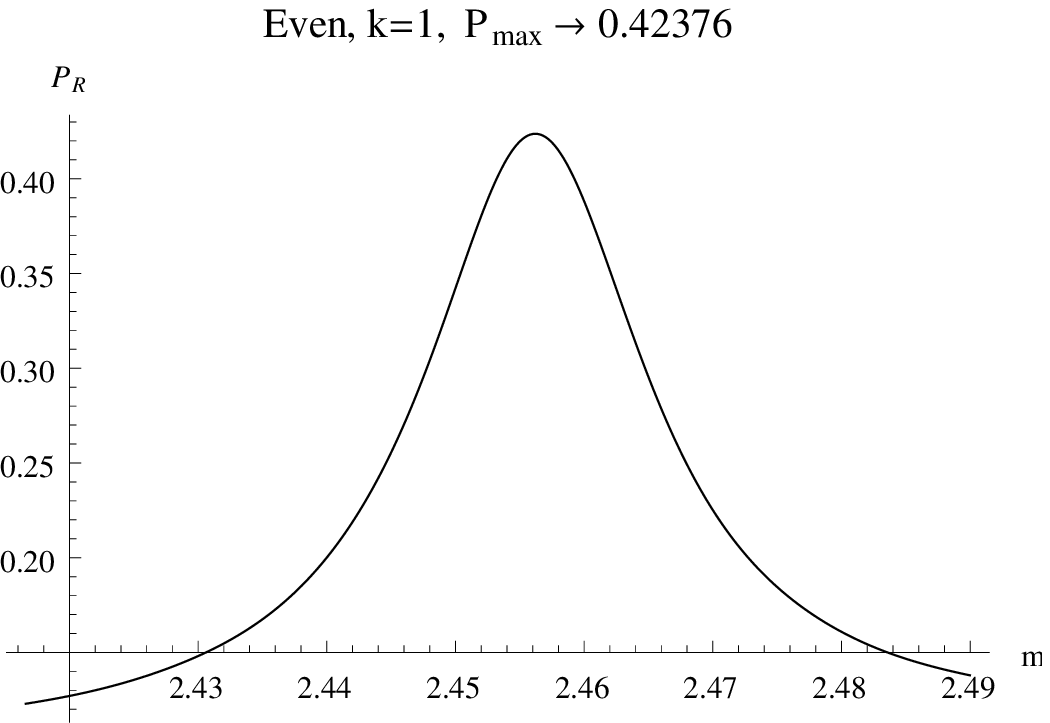}}
\end{center} \vskip -5mm
\caption{The probability $P_{R}$ (as a function of $m$) for finding
the even parity (upper three resonant peaks) and odd parity (under
three resonant peaks) massive KK modes of the right-chiral fermions
around the brane location for the coupling
$F(\phi,\chi,\rho)=\phi\chi\rho$. The parameters are set to
$z_{max}=210$, $a=0.05$ and $\eta=1$.}
 \label{fig_fR_Probability}
\end{figure}

\begin{figure}[htb]
\begin{center}
 \subfigure[$m^{2}=1.502620398$]{\label{fig_fR_Eigenvalue_1a}
  \includegraphics[width=4.5cm,height=3.5cm]{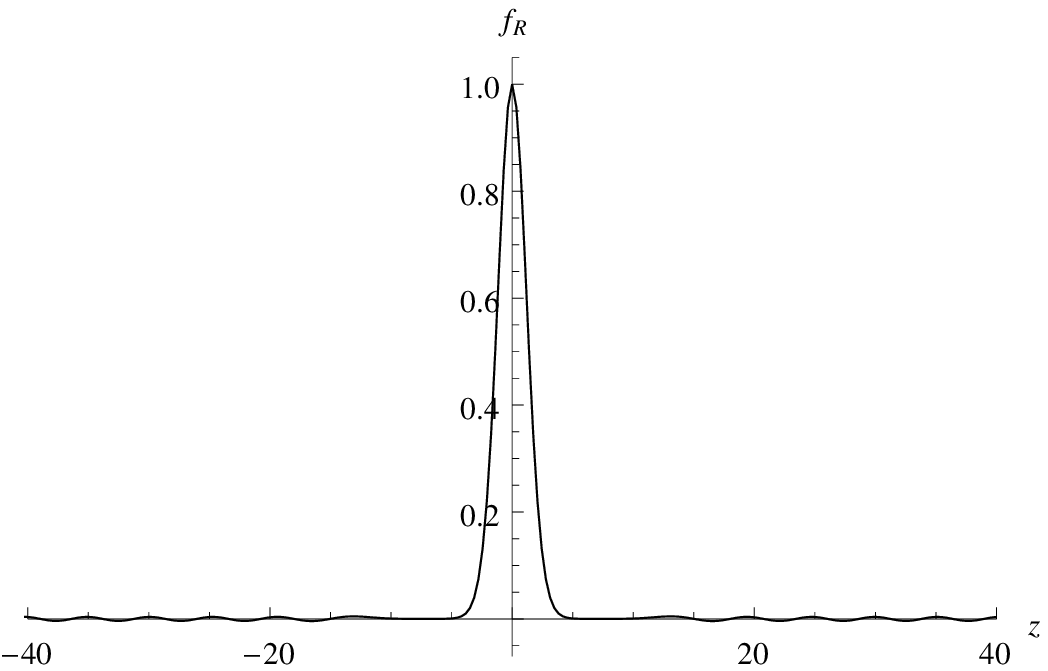}}
 \subfigure[$m^{2}=2.885601$]  {\label{fig_fR_Eigenvalue_1b}
  \includegraphics[width=4.5cm,height=3.5cm]{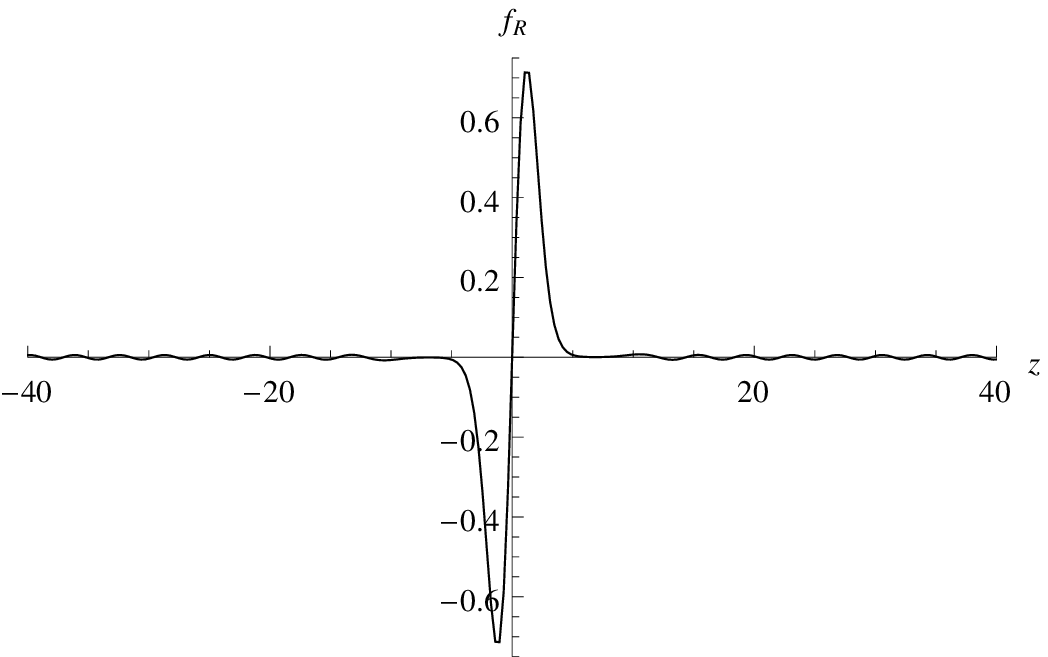}}
 \subfigure[$m^{2}=4.13079$]  {\label{fig_fR_Eigenvalue_1c}
  \includegraphics[width=4.5cm,height=3.5cm]{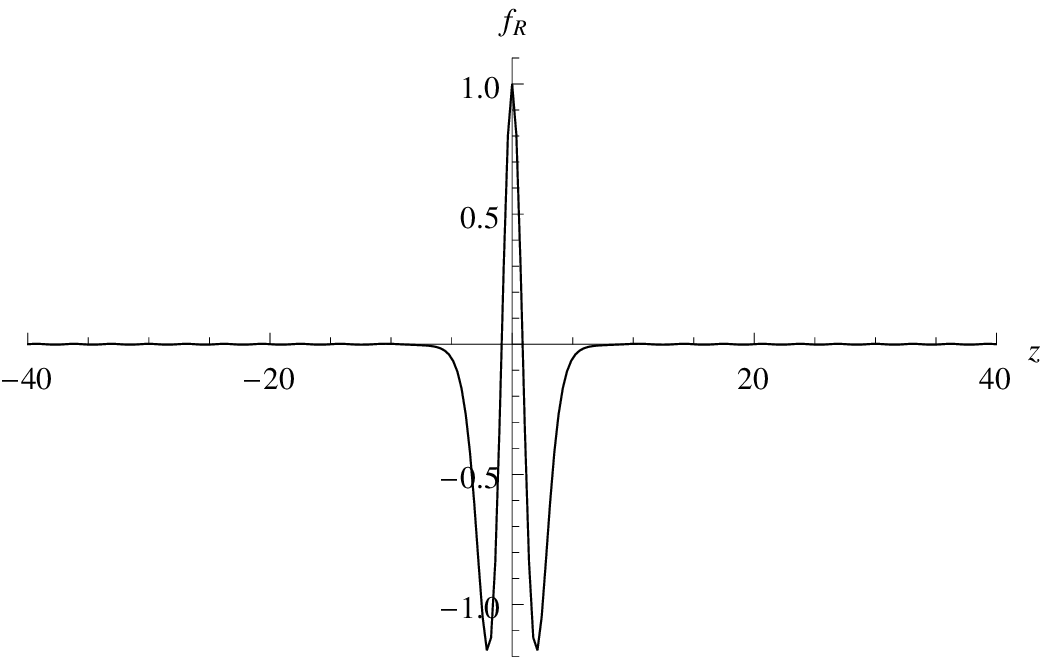}}
   \subfigure[$m^{2}=5.2051$]{\label{fig_fR_Eigenvalue_1d}
  \includegraphics[width=4.5cm,height=3.5cm]{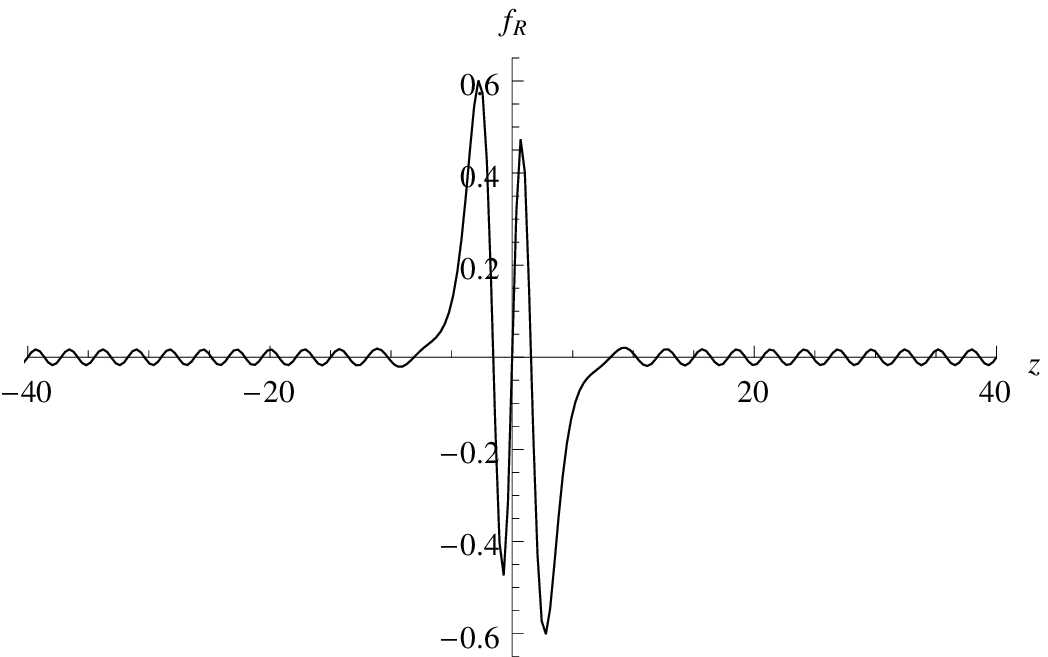}}
 \subfigure[$m^{2}=6.033$]  {\label{fig_fR_Eigenvalue_1e}
  \includegraphics[width=4.5cm,height=3.5cm]{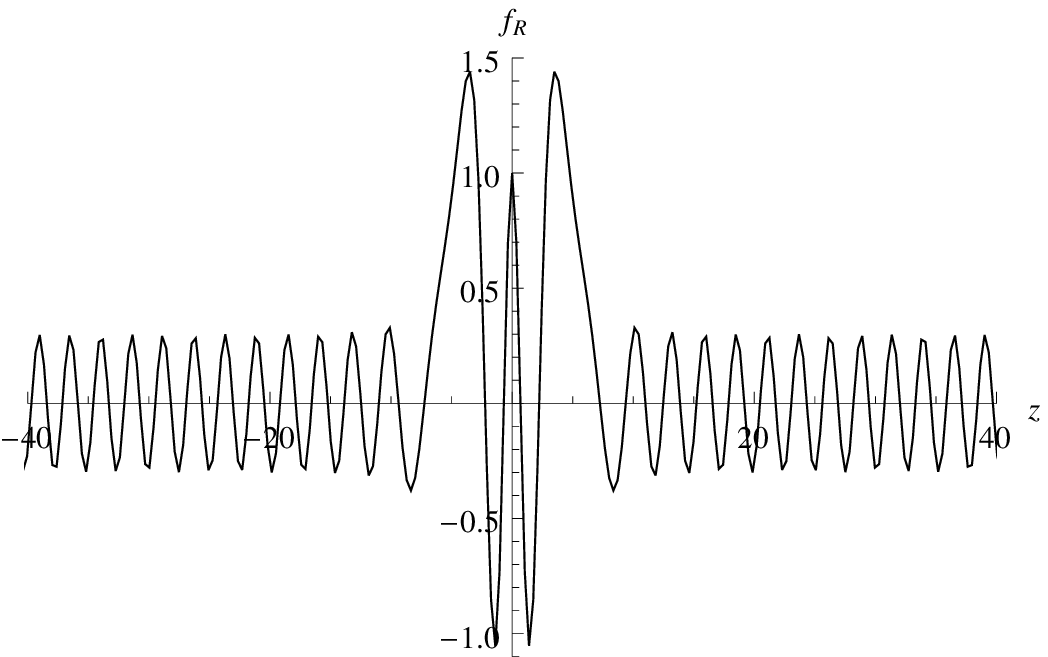}}
\end{center}  \vskip -5mm
\caption{The shapes of the even parity (upper three) and odd parity
(under three) resonance states $f_R(z)$ of right-chiral fermions for
the coupling $F(\phi,\chi,\rho)=\phi\chi\rho$ with different
$m^{2}$. The parameters are set to $z_{max}=210$, $a=0.05$ and
$\eta=1$ .}
 \label{fig_fR_Eigenvalue1}
\end{figure}

For right-handed fermions, as we mentioned earlier, there is no
normalized zero-mode. Repeating the analysis of the previous
subsection, we can get the similar results about resonances. And
the resonance states are closely related to the emergence of the
potential well shown in Fig. \ref{fig:VLRCaseI}. For the
parameters $a=0.05$ and $\eta=1$, the corresponding graphics of
the resonance spectra are depicted in
Fig.~\ref{fig_fR_Probability}. For right-handed fermions, we also
find a total of five clear resonant peaks with approximate
eigenvalues $m^{2}=1.502620398$, $2.885601$, $4.13079$, $5.2051$,
$6.033$. The corresponding right-handed eigenfunctions are shown
in Fig.~\ref{fig_fR_Eigenvalue1}. The configurations of
Figs.~\ref{fig_fR_Eigenvalue_1a}, ~\ref{fig_fR_Eigenvalue_1c} and
~\ref{fig_fR_Eigenvalue_1e} are even-parity eigenfunctions, and
the other two are odd-parity ones. This situation is the reverse
of that with left-handed fermions. From the Figs.
\ref{fig_fR_Probability} and \ref{fig_fR_Eigenvalue1}, we can also
see that odd-parity eigenfunctions and even-parity eigenfunctions
are placed at the five resonant peaks with irregular intervals of
$m$. The difference is that the first resonance state is known as
even-parity.

Comparing Figs. \ref{fig_fL_Probability} and
\ref{fig_fL_Eigenvalue1} with Figs. \ref{fig_fR_Probability} and
\ref{fig_fR_Eigenvalue1}, respectively, we find that the $n$th
massive resonance with left-chirality and the $n$th  one with
right-chirality have the same mass, i.e., the spectra of massive
odd (even) left-handed and even (odd) right-handed fermionic
resonances are the same. At the same time, excluded from the error
in the numerical calculations, their lifetimes listed in Table
\ref{TableSpectra3Fields} are in general of the same order of
magnitude. This demonstrates that it is possible to compose a
Dirac fermion from a left-handed fermion with odd-parity and a
right-handed one with even-parity, and vice versa. This means that
the general chiral decomposition expression Eq.~(\ref{the general
chiral decomposition}) becomes two explicit forms
\begin{subequations}\label{the general chiral decomposition explicit}
\begin{eqnarray}
\Psi(x,z)&=&e^{-2A}\biggl(\sum_{n}\psi_{Ln}(x)f^{(E)}_{Ln}(z)+\sum_{n}\psi_{Rn}(x)f^{(O)}_{Rn}(z)\biggl)\,,
\label{the general chiral decomposition explicit a}\\
\Psi(x,z)&=&e^{-2A}\biggl(\sum_{n}\psi_{Ln}(x)f^{(O)}_{Ln}(z)+\sum_{n}\psi_{Rn}(x)f^{(E)}_{Rn}(z)\biggl)\,,
\label{the general chiral decomposition explicit b}
\end{eqnarray}
\end{subequations}
where the superscripts $E$ and $O$ stands for even-parity and
odd-parity, respectively. We call this kind of decomposition the
parity-chiral decomposition instead of the general chiral
decomposition. In fact, this is not difficult to understand,
because in mathematics, any function can be decomposed into an
even function and an odd one. This implies that the parity and the
chirality of massive fermions are conserved in a sense. We can
also see that the resonance states with relatively smaller mass
have a relatively longer lifetime. This conclusion is consistent
with the results obtained in the single-scalar constructed dS
thick brane \cite{Liu:09b}.

It is worth noting that, the key point here is that we have
considered two types of initial conditions (\ref{evencondition}) and
(\ref{oddcondition}). These two types of initial conditions lead to
odd- and even-parity solutions. In fact, as mentioned in the earlier
work, we found in Refs.~\cite{Liu:08c,Liu:09a} that the spectra of
the bound massive KK modes of left- and right-chiral fermions are
also the same, where the effective potentials for KK modes of
fermions are modified P\"{o}schl-Teller potentials.

\subsection{Case II: $F(\phi,\chi,\rho)=\phi^{k}\chi\rho$ with odd $k>1$}

\begin{figure}[htb]
\begin{center}
 \includegraphics[width=7cm,height=5cm]{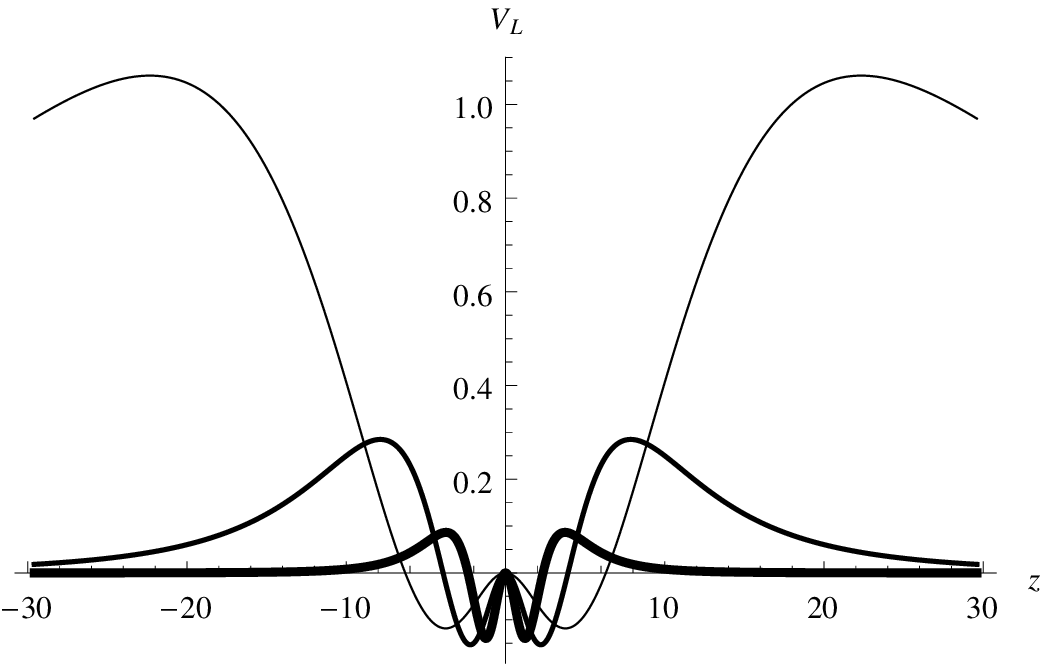}
 \includegraphics[width=7cm,height=5cm]{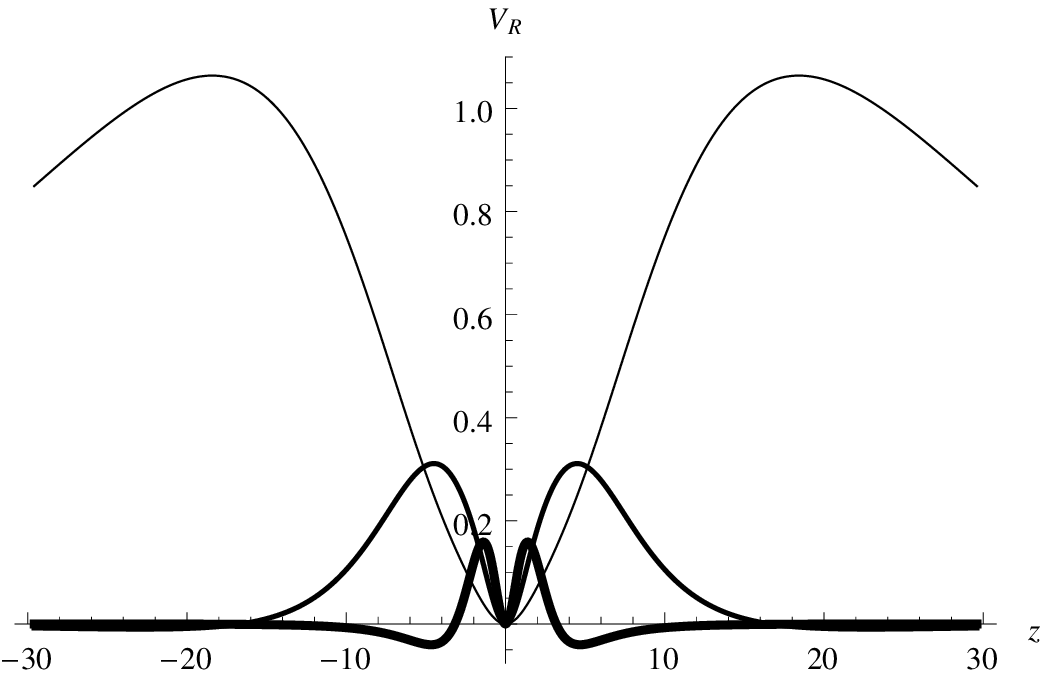}
\end{center}  \vskip -5mm
\caption{Potentials $V_{L}(z)$ and $V_{R}(z)$ for left- and
right-handed  chiral fermions with
$F(\phi,\chi,\rho)=\phi^{3}\chi\rho$.
 The parameters are set to  $\eta=1$, and $a=0.05$ (thin trace),
 $0.1$, $0.2$ (thick trace).}%
    \label{Vfermionsphi3}
\end{figure}

\begin{figure}[htb]
\begin{center}
 \includegraphics[width=7cm,height=5cm]{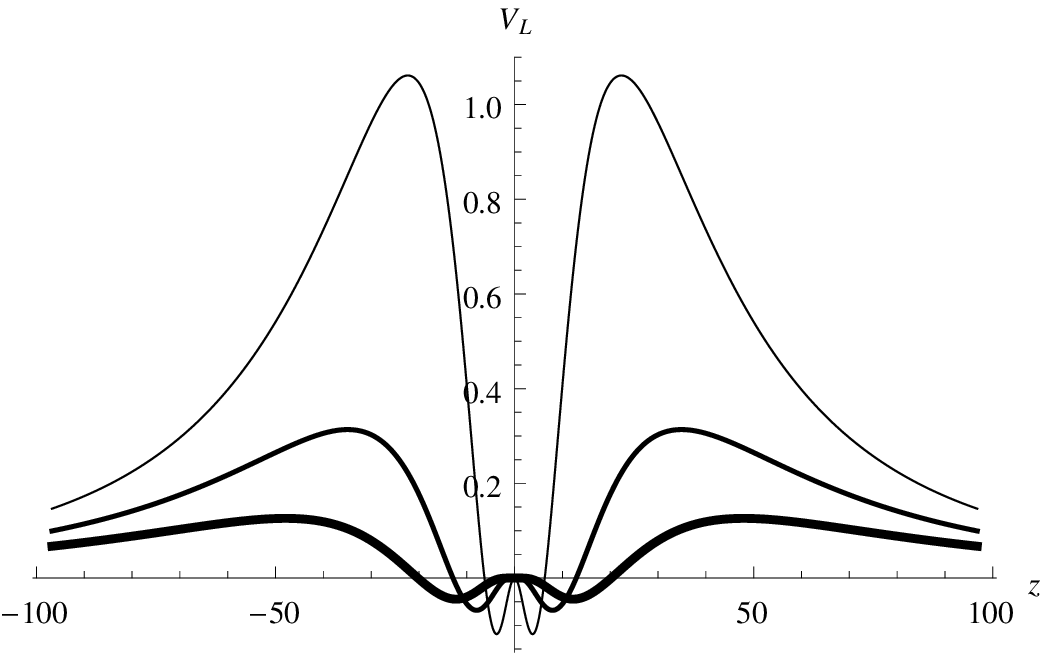}
 \includegraphics[width=7cm,height=5cm]{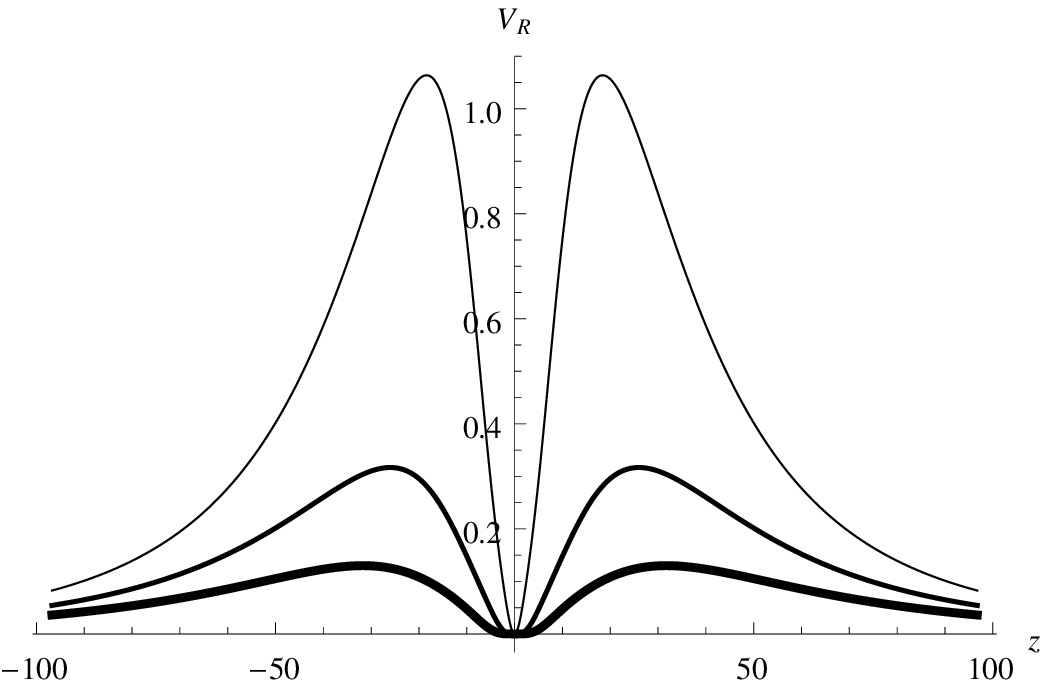}
\end{center}  \vskip -5mm
\caption{Potentials $V_{L}(z)$ and $V_{R}(z)$ for left- and
right-handed
 chiral fermions with $F(\phi,\chi,\rho)=\phi^{k}\chi\rho$.
 The parameters are set to $a=0.05$, $\eta=1$, and $k=3$(thin trace),
 $5$, $7$(thick trace).}
    \label{Vfermionsphi357}
\end{figure}

Next, we consider a natural generalization of the simplest Yukawa
coupling: $F(\phi,\chi,\rho)=\phi^{k}\chi\rho$, where $k$ is a
positive odd integer, and $k>1$. The similar generalized couplings
have been studied in Refs. \cite{Liu:09b,Liu:09c}. For this
coupling, the potentials for the KK modes of left- and right-handed
fermions (\ref{Vy}) become the form below
\begin{subequations}\label{VyCaseII}
\begin{eqnarray}
 V_{L}(y) &=& \frac{2a-1}{144a^{2}} \eta
      \exp\left(\frac{2(1-3a)}{9a}\tanh^{2}(2ay)\right)
      \cosh^{-4-\frac{4}{9a}}(2ay)\tanh^{k-1}(2ay)\nonumber\\
  &&\times\bigg\{27(2a-1)\eta\tanh^{k+1}(2ay)
           -16\sqrt{3}a(1+9a) \text{sinh}^2(2ay) \nonumber\\
  && ~~~   +8\sqrt{3}a\bigg[9ak-(6a-2)\tanh^{2}(2ay)\bigg]\bigg\}\,,
  \label{VyCaseIIa}\\
 V_{R}(y) &=& V_{L}(y)|_{\eta\rightarrow-\eta}\,,
\label{VyCaseIIb}
\end{eqnarray}
\end{subequations}
It is easy to see that both the two potentials have a simple
asymptotic behavior:
\begin{eqnarray}
V_{L,R}(\pm\infty)=V_{L,R}(0)=0\,.\label{VLR1infty}
\end{eqnarray}
The potentials cannot be expressed as explicit functions with the
variable $z$. But by means of numerical methods, we can get the
relationship between the potentials and the variable $z$. The
graphics of the numerical potentials are depicted in Figs.
\ref{Vfermionsphi3} and \ref{Vfermionsphi357} for different values
of $a$, $\eta$ and $k$. As mentioned in Ref.~\cite{Liu:09b}, the
potential well for left-handed fermions becomes a double-well,
while, for any positive $\eta$, there is a potential well for
right-handed fermions. The lowest point of the potential well for
right-handed fermions sits on the origin of the extra coordinate
$z$. For given $\eta$ and $k$, as $a$ increases, the depth of the
potential wells for left- and right-handed fermions will be
shallower and shallower. On the other hand, for given $\eta$ and
$a$, the smaller the parameter $k$ is, the deeper the depth of the
potential wells. The third situation is that, for given $a$ and
$k$, the greater the absolute value of $\eta$ is, the deeper the
depth of the potential wells. The graphics are not drawn here.

As we know, there is a continuous gapless spectrum of KK modes for
both left- and right-chiral fermions. The zero modes of left-handed
fermions
\begin{eqnarray}
 f_{L0}(z)\propto\exp\bigg(-\eta\int_{0}^{z}d\bar{z}e^{A(\bar{z})}
  \phi^{k}(\bar{z})\chi(\bar{z})\rho(\bar{z})\bigg)\,
  \label{ZeroModeCaseII}
\end{eqnarray}
can also not be normalized, while the following one for $F=\phi^k$
with $\eta>2/9$ is normalizable:
\begin{eqnarray}
 f_{L0}(z)\propto\exp\bigg(-\eta\int_{0}^{z}d\bar{z}e^{A(\bar{z})}
  \phi^{k}(\bar{z})\bigg).
  \label{ZeroModeCaseIInew}
\end{eqnarray}
The emergence of potential wells shown in Figs. \ref{Vfermionsphi3}
and \ref{Vfermionsphi357} are also related to resonances. From
Figs.~\ref{Vfermionsphi3} and ~\ref{Vfermionsphi357}, we see that
some characteristics of these potentials are very different from
those in Ref.~\cite{Liu:09b}. Here, for given $\eta$ and $a$, the
larger the parameter $k$ is, the lower the depth of the potential
wells. Therefore, we speculate that the number of resonance states
will be reduced with the increase of $k$, rather than the case in
Ref.~\cite{Liu:09b}. Next, we will prove this conjecture.

\begin{figure}[htb]
\begin{center}
 \subfigure[$m^{2}=0.136755$]{\label{fig_fR_Eigenvalue_12a}
  \includegraphics[width=4.5cm,height=3.5cm]{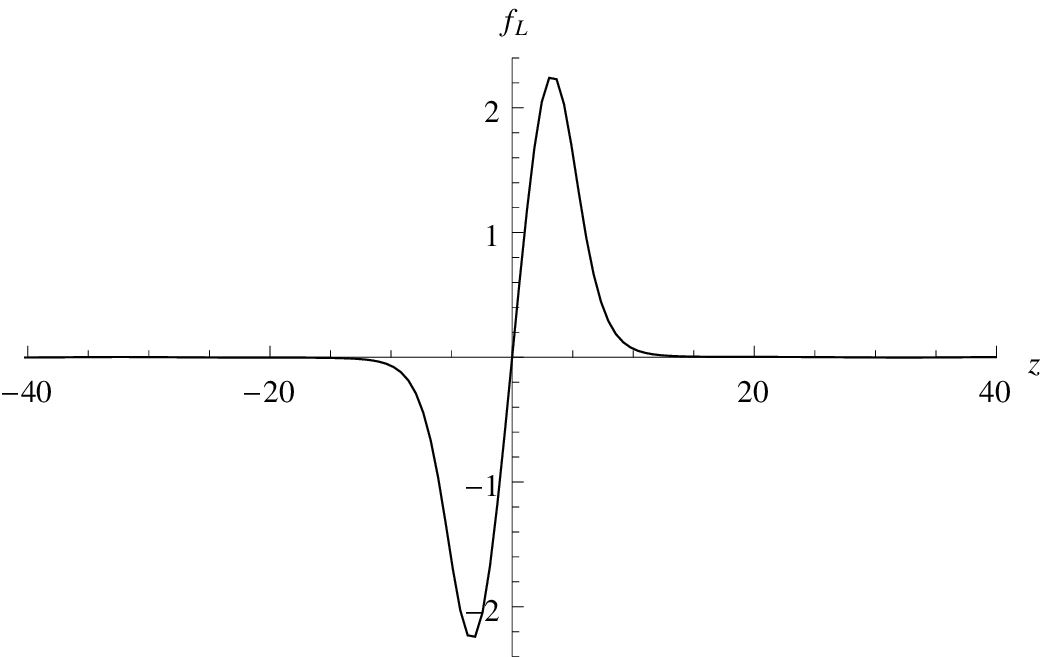}}
 \subfigure[$m^{2}=0.39192$]  {\label{fig_fR_Eigenvalue_12b}
  \includegraphics[width=4.5cm,height=3.5cm]{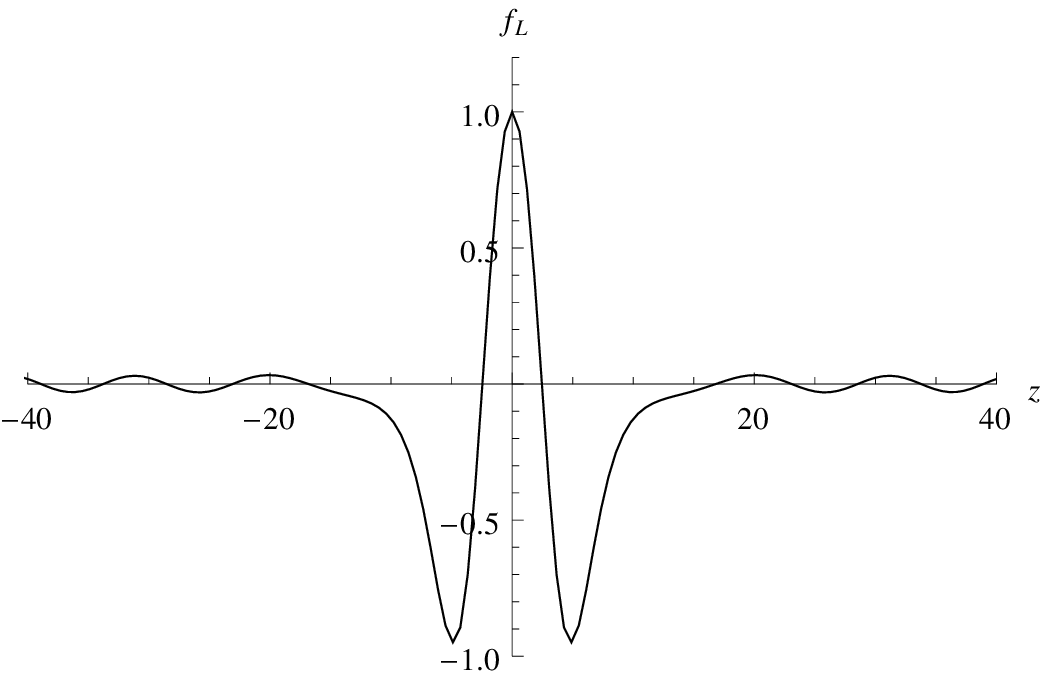}}
 \subfigure[$m^{2}=0.607$]  {\label{fig_fR_Eigenvalue_12c}
  \includegraphics[width=4.5cm,height=3.5cm]{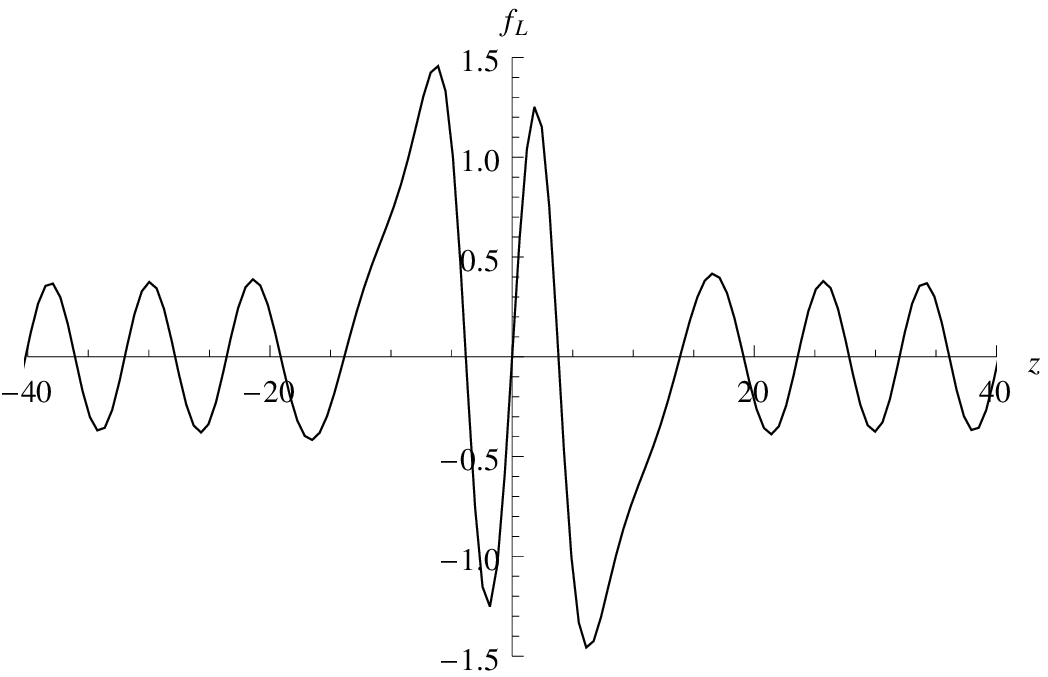}}
 \subfigure[$m^{2}=0.136763$]{\label{fig_fR_Eigenvalue_12d}
  \includegraphics[width=4.5cm,height=3.5cm]{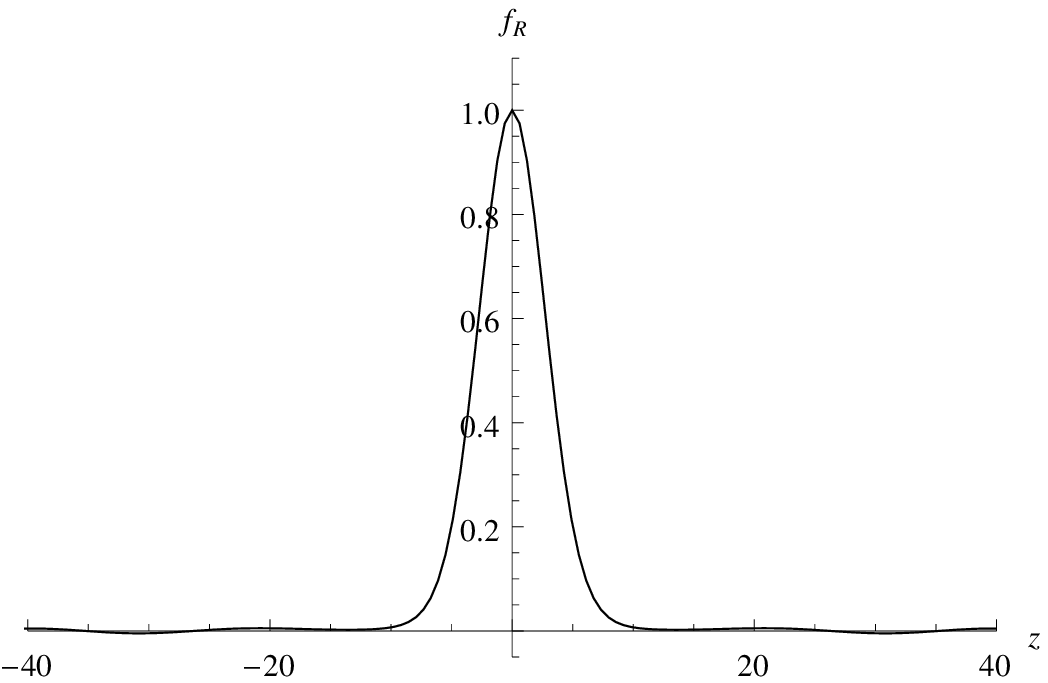}}
 \subfigure[$m^{2}=0.39195$]  {\label{fig_fR_Eigenvalue_12e}
  \includegraphics[width=4.5cm,height=3.5cm]{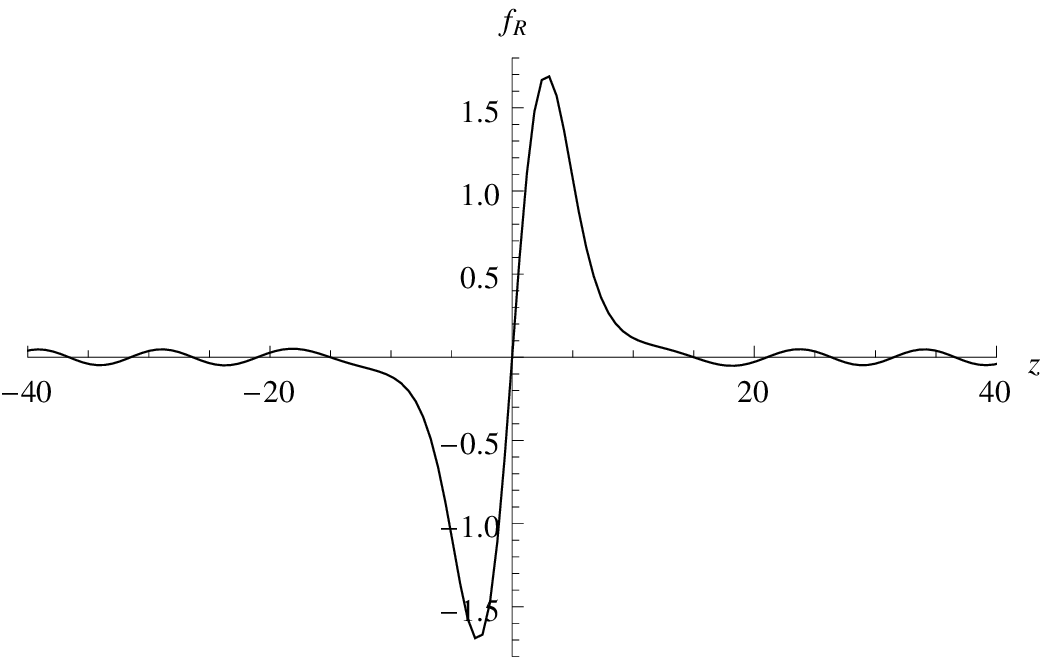}}
 \subfigure[$m^{2}=0.607$]  {\label{fig_fR_Eigenvalue_12f}
  \includegraphics[width=4.5cm,height=3.5cm]{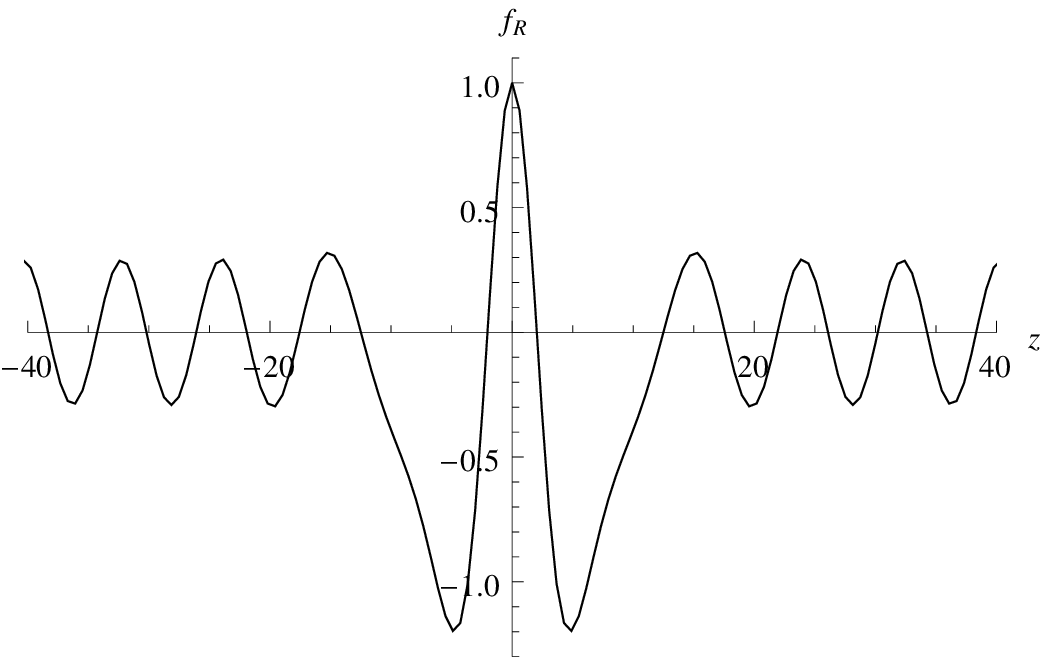}}
\end{center}  \vskip -5mm
\caption{The shapes of massive KK modes of left-handed (upper) and
right-handed (under) fermions with the even parity and odd parity
for the coupling $F(\phi,\chi,\rho)=\phi^{k}\chi\rho$ with different
$m^{2}$. The parameters are set to $k=3$, $z_{max}=370$, $a=0.05$,
and $\eta=1$.}
 \label{fig_fR_Eigenvalue12}
\end{figure}

\begin{figure}[htb]
\begin{center}
 \includegraphics[width=7cm,height=5cm]{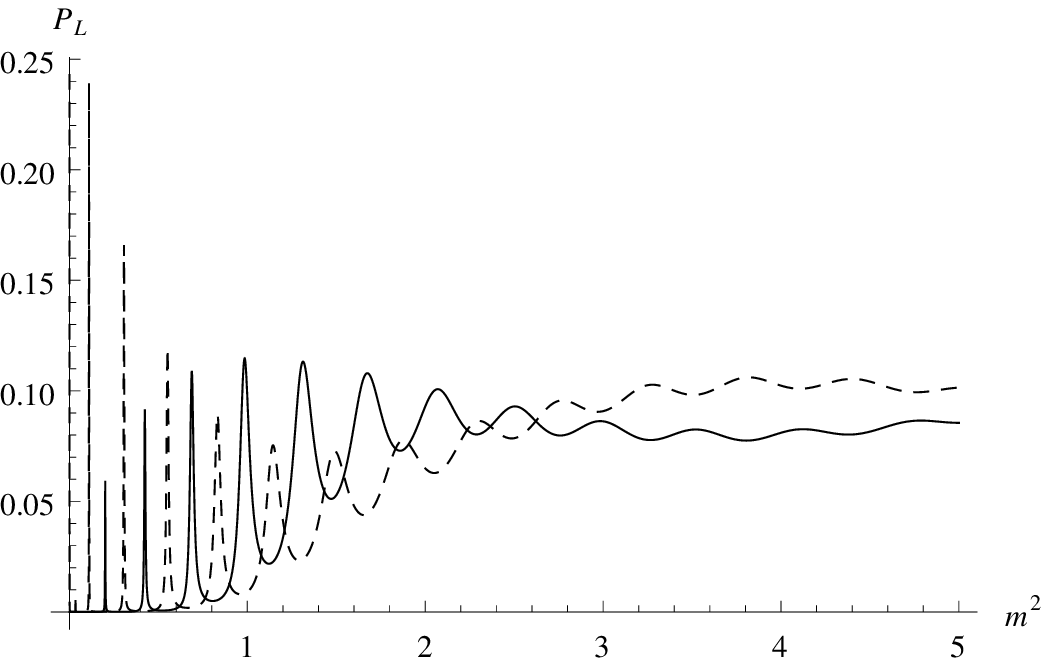}
 \includegraphics[width=7cm,height=5cm]{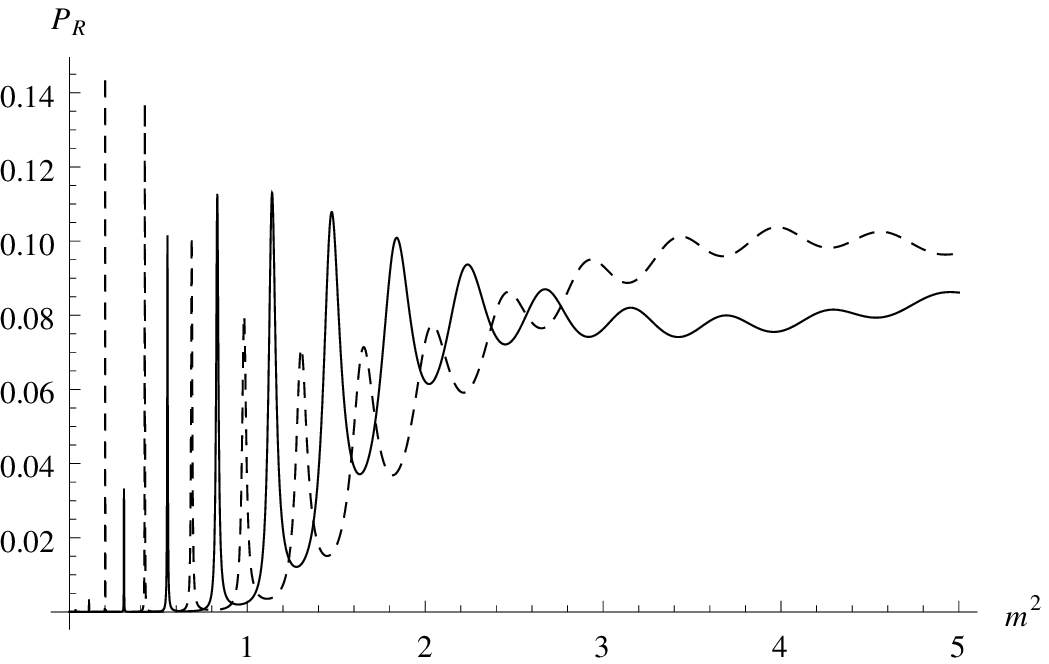}
\end{center} \vskip -5mm
\caption{The probability $P_{L,R}$ (as a function of $m^{2}$)
 for finding massive KK modes of left- and right-chiral
 fermions with mass $m^{2}$ around the brane location
 for the coupling $F(\phi,\chi,\rho)=\phi^{3}\chi\rho$.
 Solid lines and dashed lines are plotted for the odd-parity
 and even-parity massive fermions, respectively.
 The parameters are $a=0.01$ and $\eta=1$.}
 \label{figPk3a0.01}
\end{figure}

In order to study the resonance states, we consider the massive KK
modes now. Just the same as the previous subsection, using the
Numerov method, we will find the numerical solutions of
Schr\"{o}dinger equations with the purely numerical potentials
(\ref{VyCaseII}) under the two types of the initial conditions
(\ref{evencondition}) and (\ref{oddcondition}). We first consider
the case $F(\phi,\chi,\rho)=\phi^{3}\chi\rho$ to see what will
happen. For $a=0.05$ and $\eta=1$, we choose $z_{max}=370$. The
eigenfunctions of the massive KK modes of left-handed and
right-handed fermions with the even parity and odd parity are
obtained with different $m^{2}$. The results of the numerical
calculations are plotted in Fig. \ref{fig_fR_Eigenvalue12}.
Comparing Figs.~\ref{fig_fL_Eigenvalue1} and
~\ref{fig_fR_Eigenvalue1} with Fig.~\ref{fig_fR_Eigenvalue12}, we
see that the number of resonances of the case
$F(\phi,\chi,\rho)=\phi^{3}\chi\rho$ is less than that of
$F(\phi,\chi,\rho)=\phi\chi\rho$ for the same set of the parameters.
This proves our previous conjecture about the number of resonances.
However, if we reduce the value of a, the number of resonance states
will increase. For $a=0.01$ and $\eta=1$, the probability for
finding the massive KK modes of the left- and right-chiral fermions
around the brane location are shown in Fig.~\ref{figPk3a0.01}. Here,
we mainly discuss the number of resonance states. We can see that
there have been a lot of increases of the number of the resonance
states. However, because of the step size of the eigenvalue is not
small enough, the probability of the resonant peaks we can see in
Fig.~\ref{figPk3a0.01} is not true size, and the actual probability
is  generally larger than that shown in the graphics.

In addition, we can investigate the other cases when $k$ takes the
following values $\{3$, $5$, $7$, $9\}$. The eigenvalues
$m^{2}_{n}$, mass $m_{n}$, width $\Gamma$ and lifetime $\tau$ of
left- and right-chiral fermions for the coupling
$F(\phi,\chi,\rho)=\phi^{k}\chi\rho$ are also listed in Table
\ref{TableSpectra3Fields}, here, $n$ represents the
$n$th-resonance state. The results show that, when $k$ increases,
the number of the resonance states gradually reduces, and when
$k=11$, the number decreases to $0$. This means that, if
$k\geq11$, there would be no resonance state at all. And at the
same time, there exists a finite number of resonance states for
arbitrary $k$. These results are very different from that obtained
in Ref.~\cite{Liu:09b}, where the number of resonance states
increases with $k$, and it is boundless for $k\rightarrow\infty$.
The reason is that, for the coupling
$F(\phi,\chi,\rho)=\phi^{k}\chi\rho$, the absolute value of the
scalar field at finite $z$ in our current case is smaller than
$1$. Hence, for given $\eta$ and $a$, the larger the parameter $k$
is, the weaker the kink-fermion coupling and so smaller the depth
of the potential wells. For large enough $k$, there will be no
resonance states because the potential wells are not deep enough.

\begin{figure}[htb]
\begin{center}
 \includegraphics[width=13cm,height=16cm]{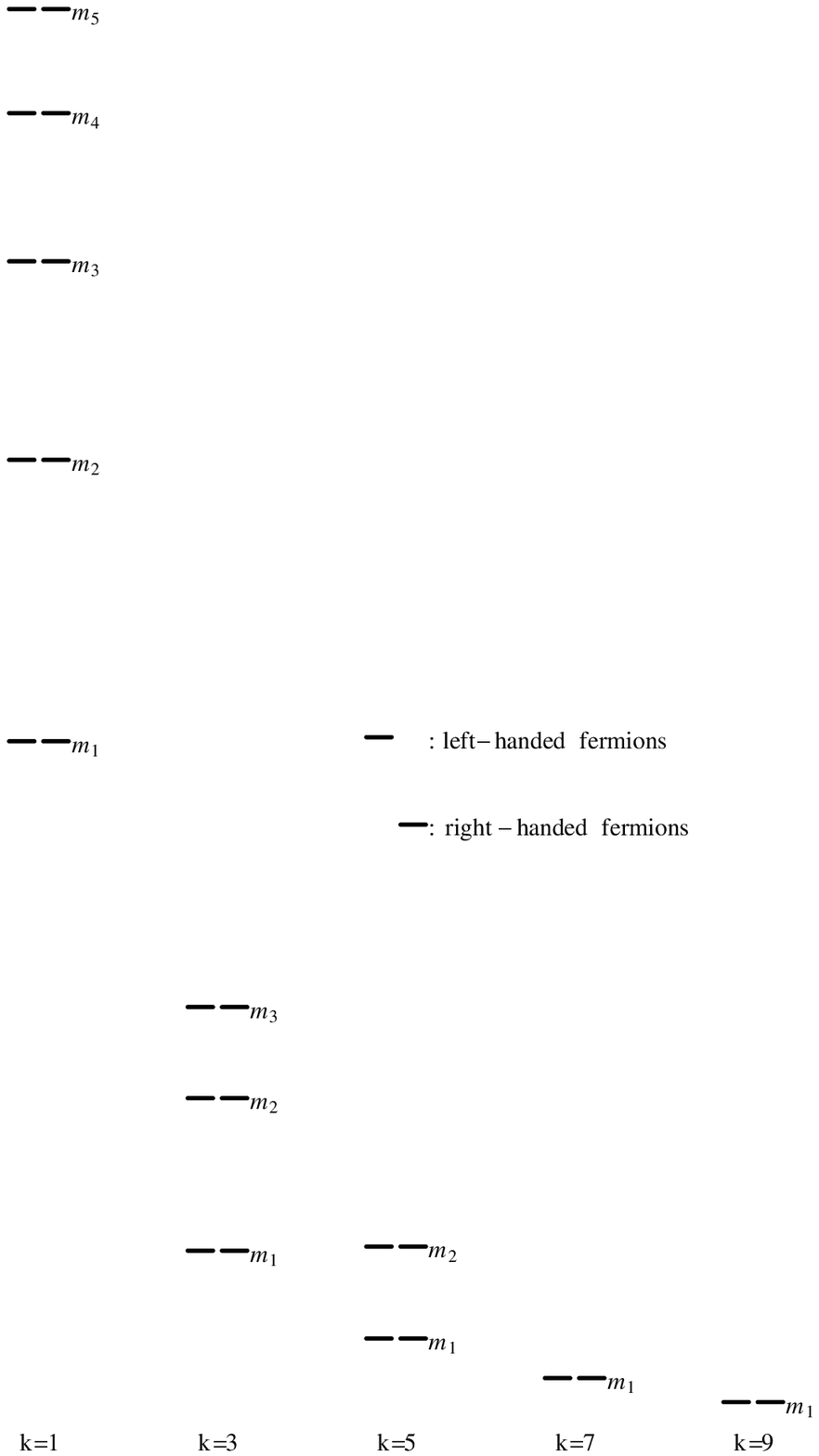}
 \end{center}  \vskip -5mm
\caption{Mass spectra of resonances for the coupling
 $F(\phi,\chi,\rho)=\phi^{k}\chi\rho$.
 The parameters are set to $k=\{1$, $3$, $5$, $7$, $9\}$,
 $a=0.05$ and $\eta=1$.}
 \label{threefieldspectrum}
\end{figure}

It can be seen that, for a given $k\geq3$, the resonances with
lower mass would have longer lifetime. This is consistent with the
previous result with $k=1$. In addition, it is worth noting that,
for $3\leq k\leq 9$, all of the resonance states satisfy the
parity-chiral decompositions (\ref{the general chiral
decomposition explicit}) in a given range of error. In order to
visualize the structure of resonances, we give all the mass
spectrum of resonances on the thick brane with the coupling
$F(\phi,\chi,\rho)=\phi^{k}\chi\rho$ for $k=1, 3, 5, 7, 9$ in
Fig.~\ref{threefieldspectrum}. It turns out that, for each
eigenvalue $m^2_{n}$, we get a pair of resonance states. They
always appear at the same time, and have opposite parity and
chiral. Remarkably, the first state of the resonance spectrum with
lower $k$ would have a relatively larger mass $m_{1}$. For a given
$k$, as $n$ increases, the mass gap $\Delta m=m_{n}-m_{n-1}$ of
the resonances is getting smaller and smaller.

\section{Resonances on a Bloch brane model}
\label{sec2Fields}

In this section, we re-analysis the problems of resonances in a
Bloch brane model \cite{Bazeia:04a} with the coupling
$F(\phi,\chi)=\phi\chi$ considered in \cite{Almeida:0901} and
clarify some small issues. By applying the previous method, we
further investigate the number and the lifetime of the resonances
in the two-scalar generated Bloch brane with the coupling
$F(\phi,\chi)=\phi^{k}\chi$, $k=3,5,7,9$. Finally, we give all the
mass spectrum of the models.

\subsection{Case I: $F(\phi,\chi)=\phi^{k}\chi$ with $k=1$}

In Ref. \cite{Almeida:0901}, the simplest Yukawa coupling
$\overline{\Psi}\phi\chi\Psi$ between two scalar fields and a spinor
field was investigated for a two-scalar generated Bloch brane model
\cite{Bazeia:04a}. The fermionic resonance states for both
chiralities were discussed, and their appearance is related to the
internal structure of the brane. Here, in order to facilitate the
discussion, we recast the kink-fermion coupling
$\overline{\Psi}\phi\chi\Psi$ into $\overline{\Psi}F(\phi,\chi)\Psi$
with $F(\phi,\chi)=\phi^{k}\chi$ and $k=1$.

\begin{figure}[htb]
\begin{center}
 \subfigure[$m=0.897851$]{\label{fig_fbloch_Probability1a}
  \includegraphics[width=4.5cm,height=3.5cm]{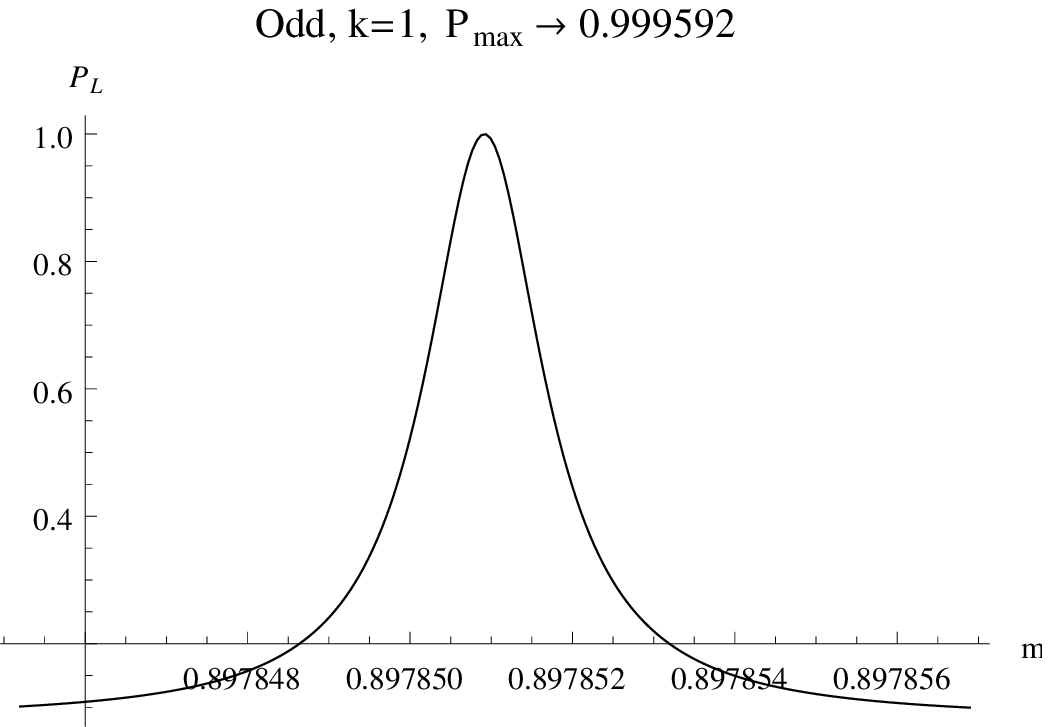}}
 \subfigure[$m=1.23208$]  {\label{fig_fbloch_Probability1b}
  \includegraphics[width=4.5cm,height=3.5cm]{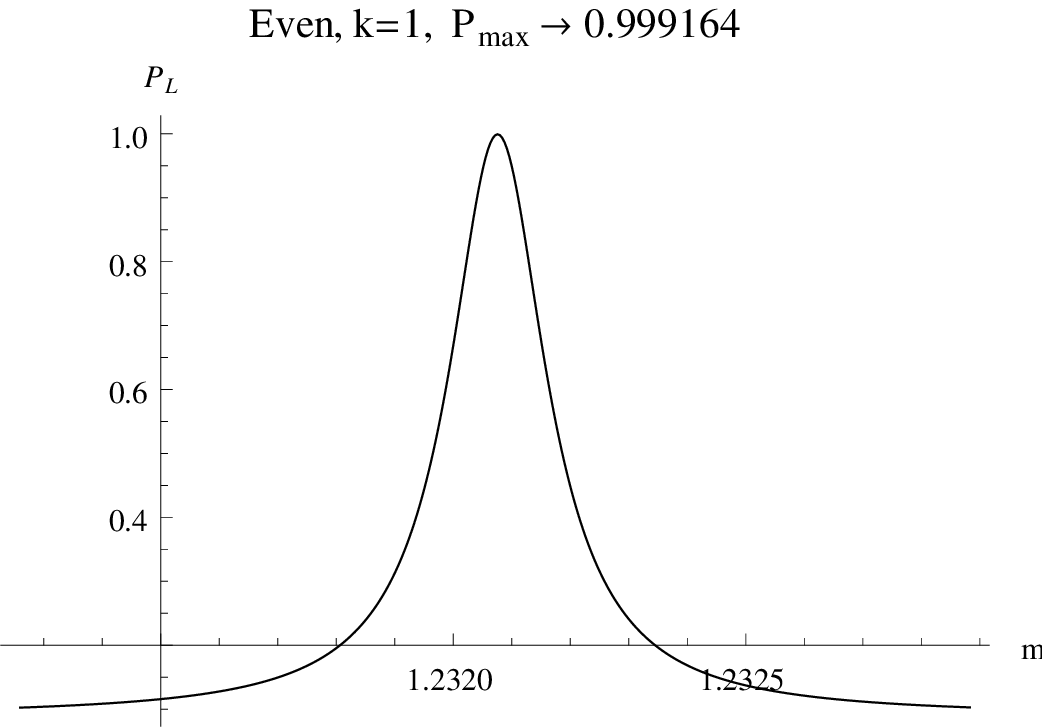}}
 \subfigure[$m=1.44997$]  {\label{fig_fbloch_Probability1c}
  \includegraphics[width=4.5cm,height=3.5cm]{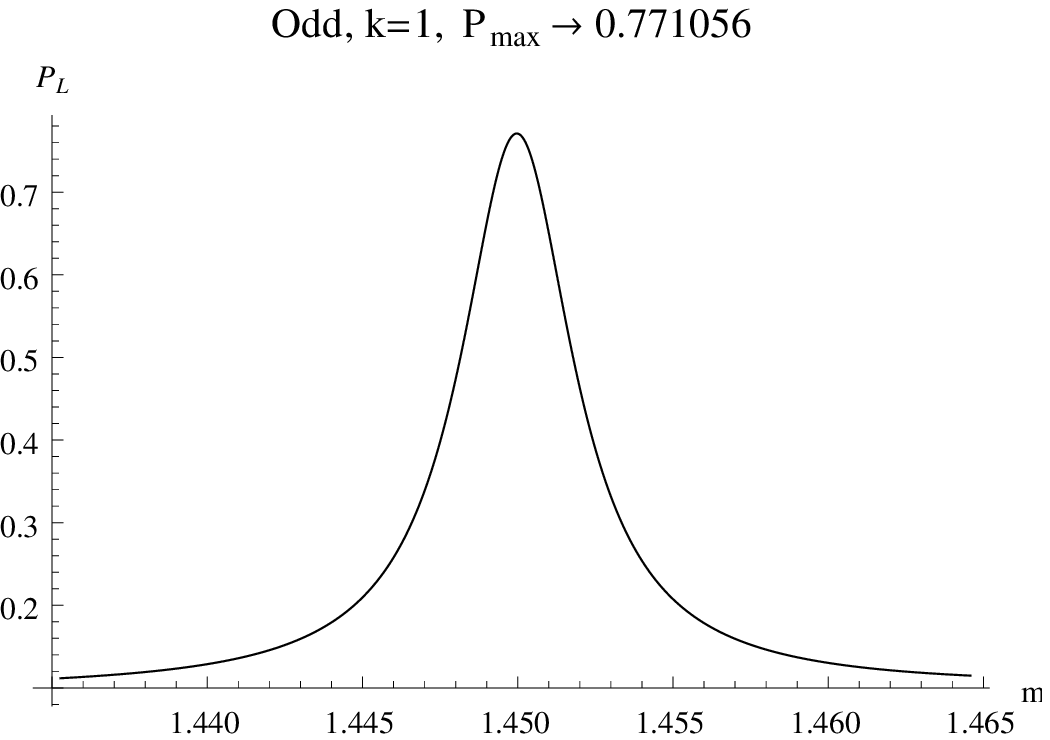}}
   \subfigure[$m=0.898066$]{\label{fig_fbloch_Probability1d}
  \includegraphics[width=4.5cm,height=3.5cm]{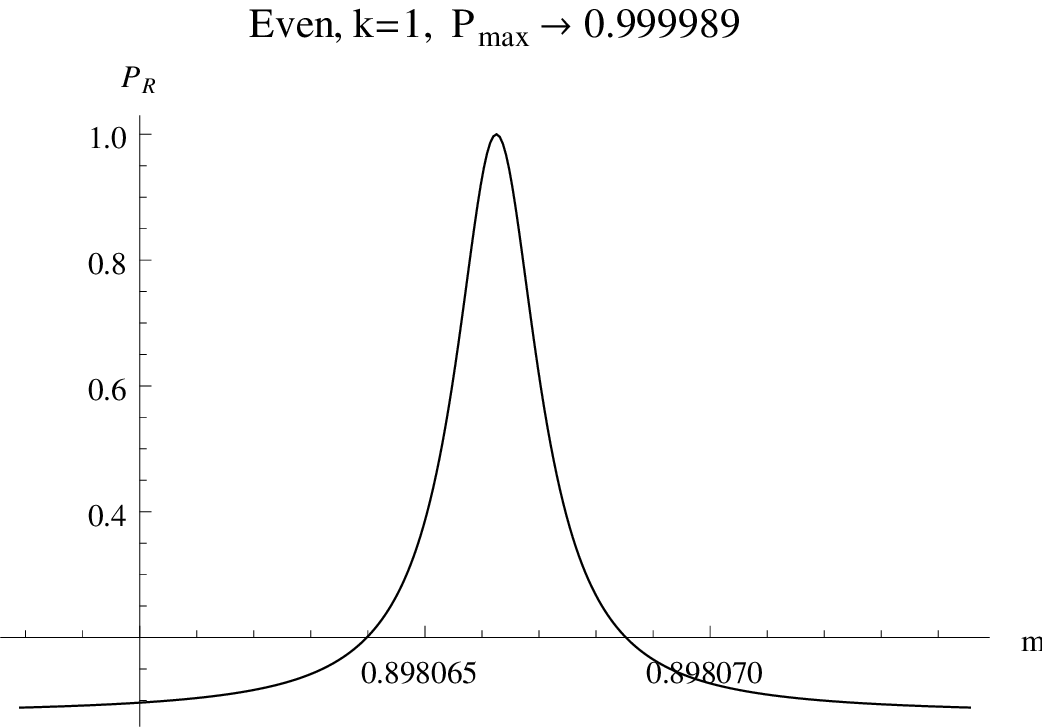}}
 \subfigure[$m=1.23244$]  {\label{fig_fbloch_Probability1f}
  \includegraphics[width=4.5cm,height=3.5cm]{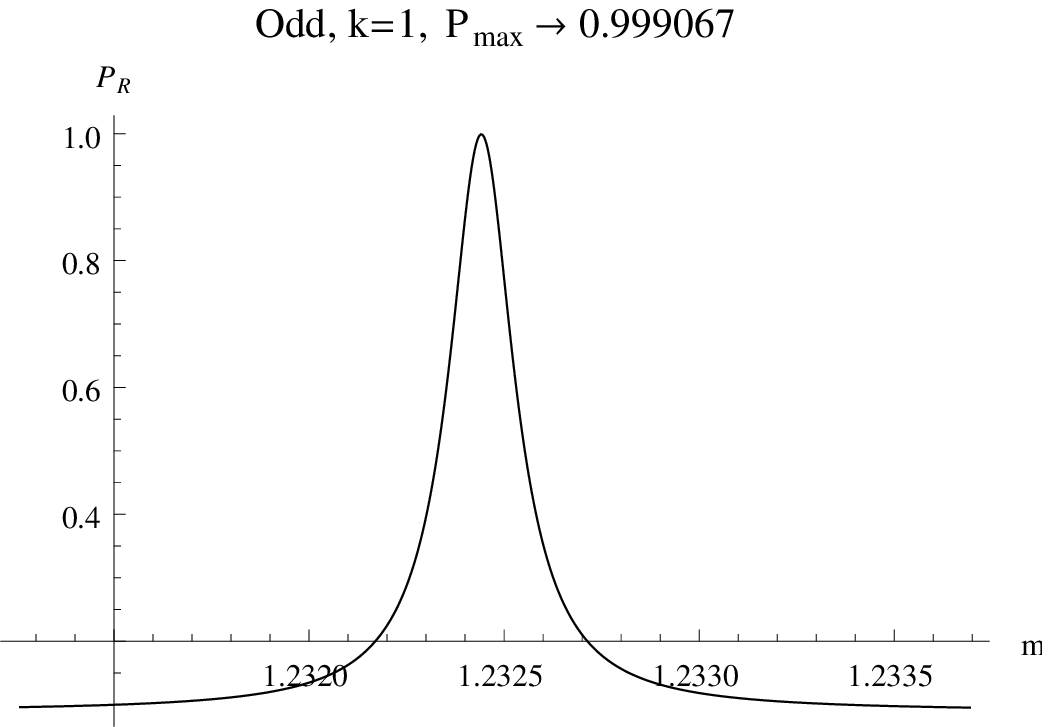}}
 \subfigure[$m=1.45041$]  {\label{fig_fbloch_Probability1f}
  \includegraphics[width=4.5cm,height=3.5cm]{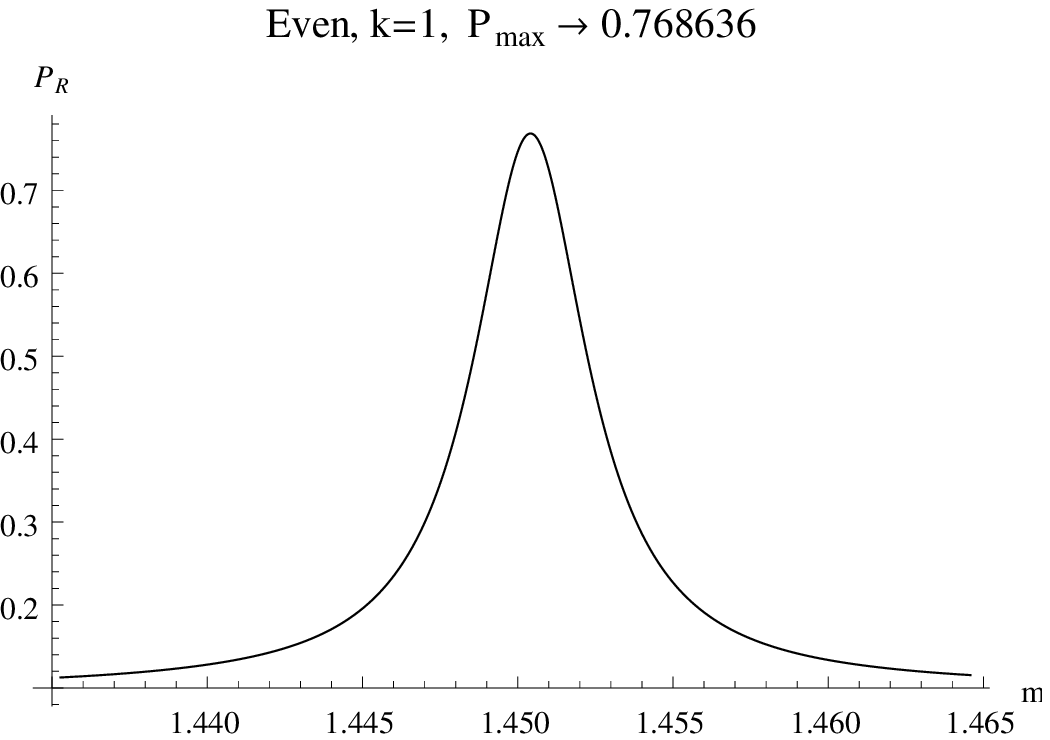}}
 \end{center}  \vskip -5mm
\caption{The probability $P_{L,R}$ (as a function of $m$) for
finding the even-parity and odd-parity massive KK modes of the left-
and right-chiral fermions around the Bloch brane \cite{Bazeia:04a}
location for the coupling $F(\phi,\chi)=\phi\chi$. Part of the
results have been given in Ref. \cite{Almeida:0901}. The parameters
are set to $z_{max}=200$, $a=0.05$, $\eta=1$.}
 \label{fig_fbloch_Probability1}
\end{figure}

By using the Numerov method, we solve the Schr\"{o}dinger equation
for the massive KK modes and obtain a series of resonances. For
$a=0.05$ and $\eta=1$, we choose $z_{max}=200$. The probability
for finding the even parity and odd parity massive KK modes of the
left- and right-chiral fermions around the Bloch brane location
are shown in Fig.~\ref{fig_fbloch_Probability1}. We found a total
of three resonant peaks and their eigenvalues $m^2$ are in the
potential well below the maximum. It is worth noting that the
number of the resonance states we discussed here is a physical
quantity which has not been well defined. The number of the
resonance states we are talking about here are the sum of the
resonance states whose energy eigenvalues are lower than the
highest point of the potential wells. The massive KK modes whose
eigenvalues $m^2$ are very close to or higher than the highest
point of the potential wells may also have a a little resonance,
but because the probability of these resonance states are
generally very small and they have quite a little lifetime, so we
do not take into account them.

\begin{figure}[htb]
\begin{center}
 \subfigure[$m^{2}=0.806136$]{\label{fig_fbloch_Eigenvalue13a}
  \includegraphics[width=4.5cm,height=3.5cm]{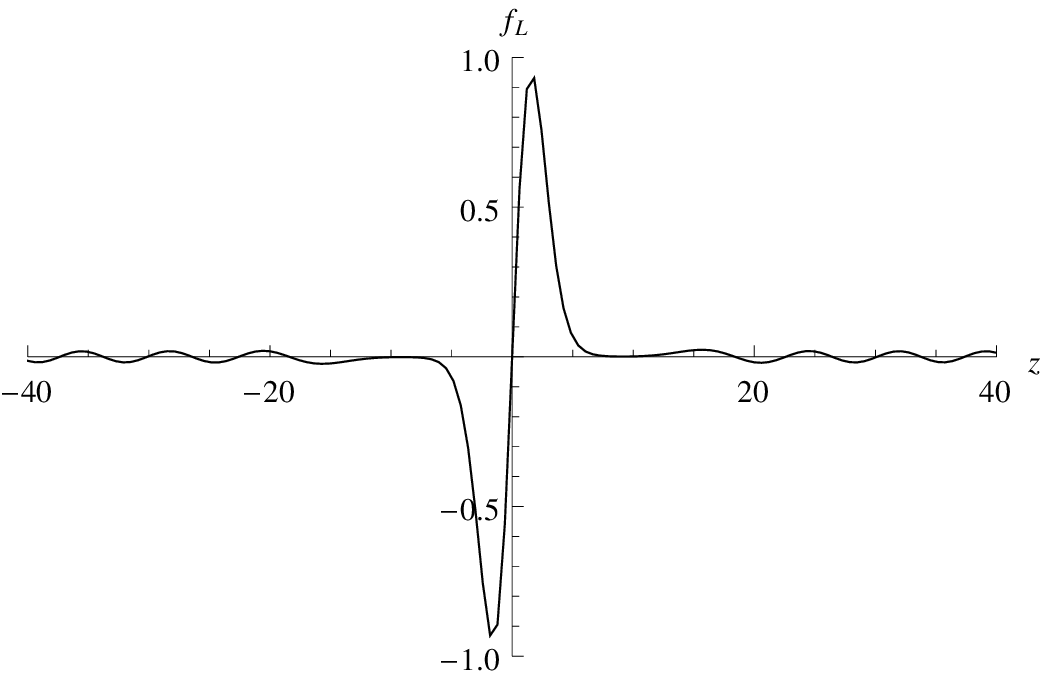}}
 \subfigure[$m^{2}=1.51801$]  {\label{fig_fbloch_Eigenvalue13b}
  \includegraphics[width=4.5cm,height=3.5cm]{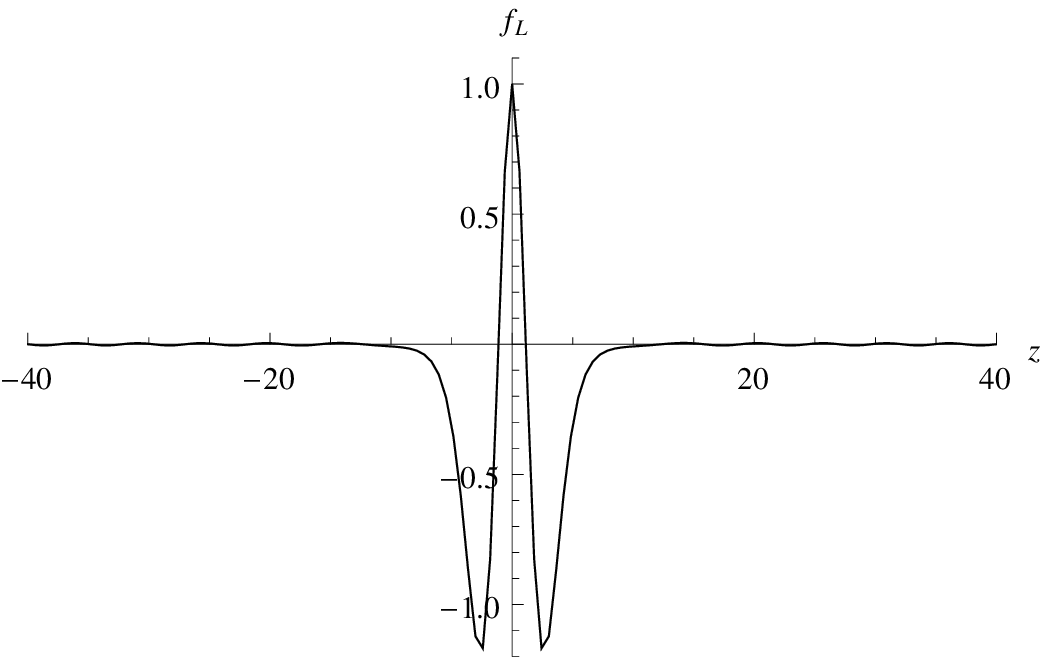}}
 \subfigure[$m^{2}=2.1024$]  {\label{fig_fbloch_Eigenvalue13c}
  \includegraphics[width=4.5cm,height=3.5cm]{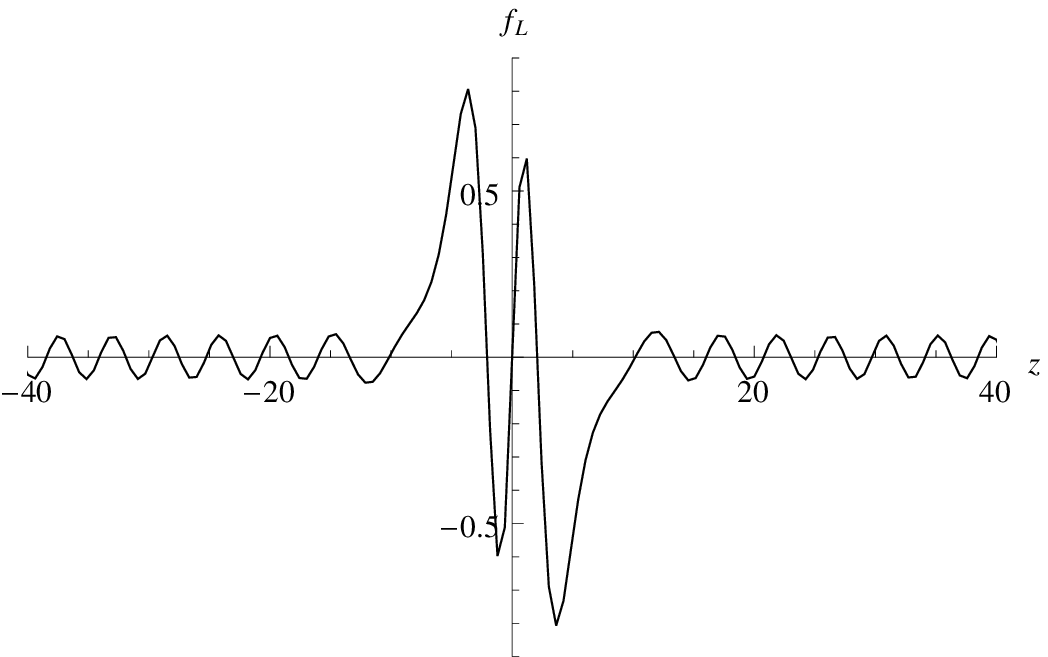}}
   \subfigure[$m^{2}=0.806523$]{\label{fig_fbloch_Eigenvalue13d}
  \includegraphics[width=4.5cm,height=3.5cm]{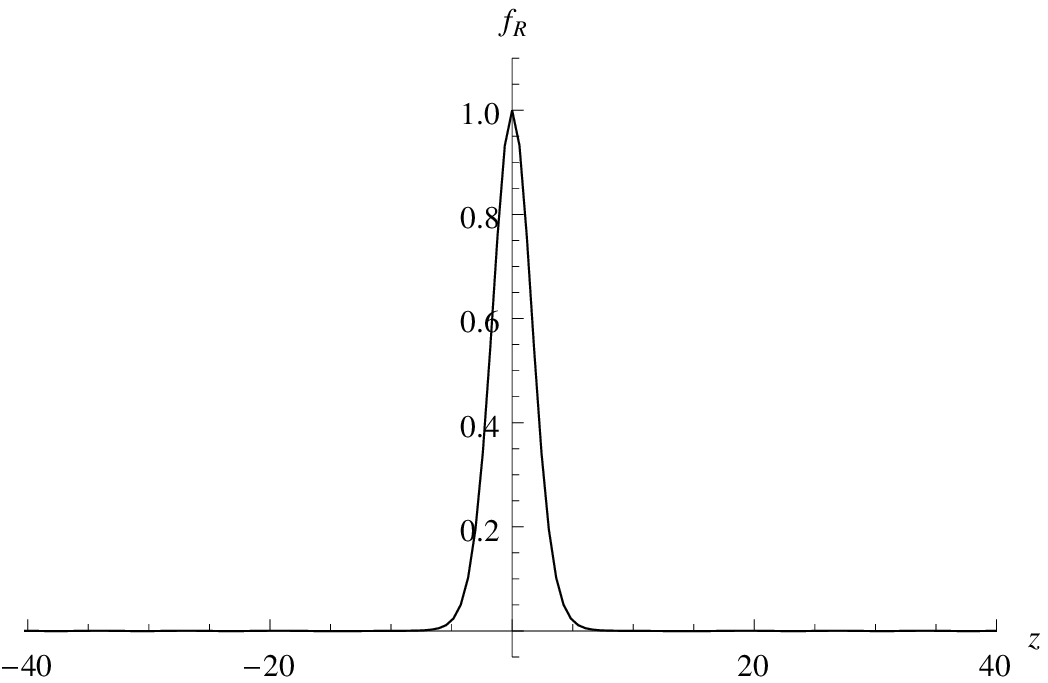}}
 \subfigure[$m^{2}=1.51891$]  {\label{fig_fbloch_Eigenvalue13e}
  \includegraphics[width=4.5cm,height=3.5cm]{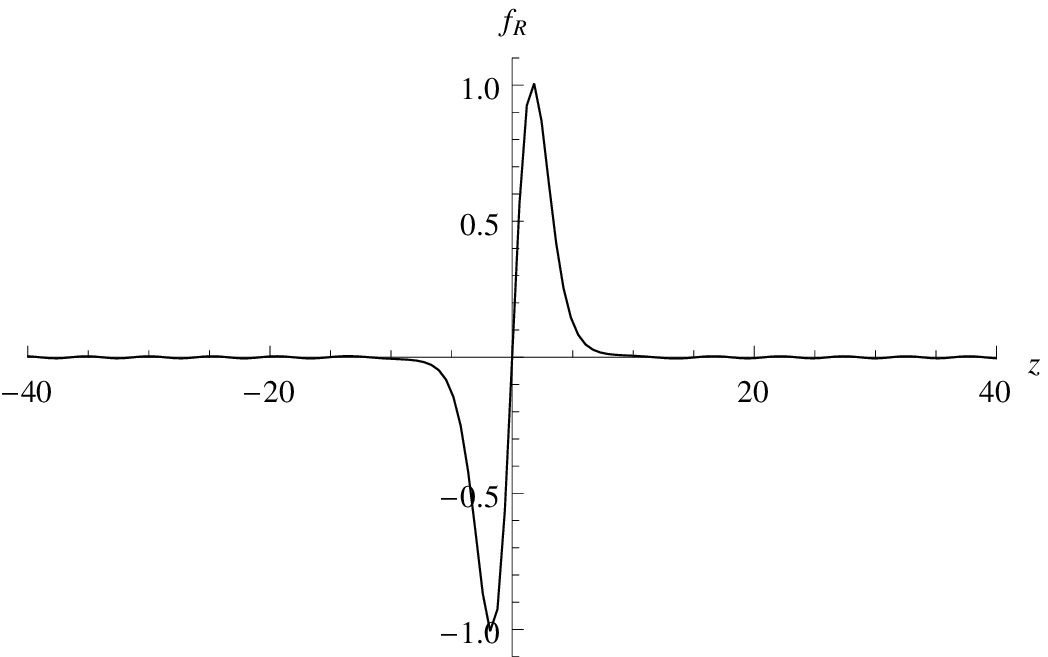}}
 \subfigure[$m^{2}=2.1037$]  {\label{fig_fbloch_Eigenvalue13f}
  \includegraphics[width=4.5cm,height=3.5cm]{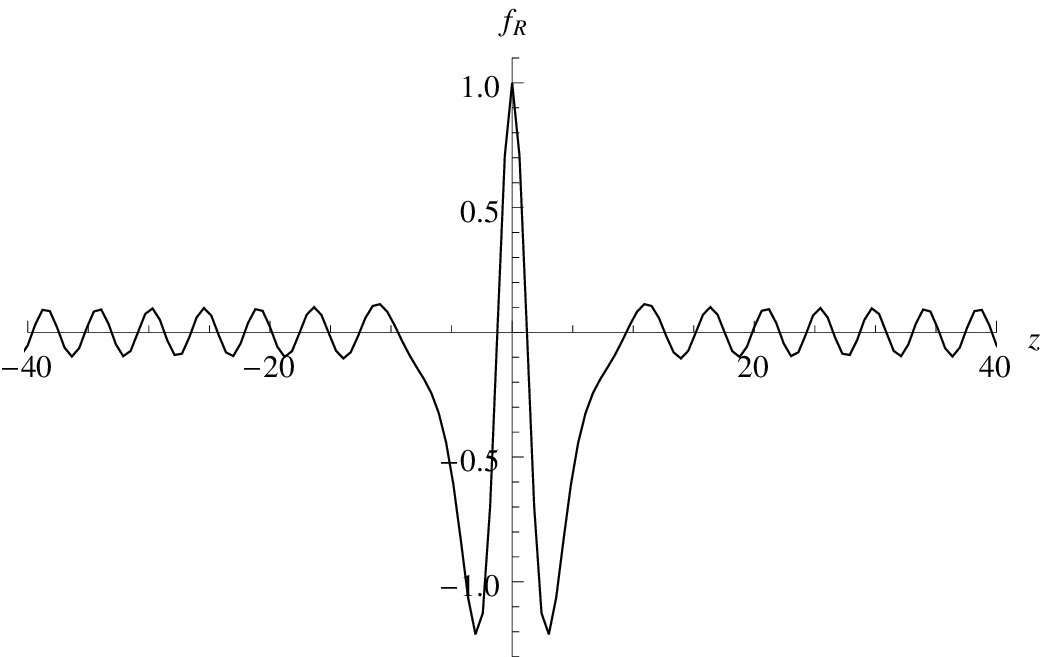}}
\end{center} \vskip -5mm
\caption{The shapes of resonances on the Bloch brane for the
coupling $F(\phi,\chi)=\phi\chi$. The parameters are set to
$z_{max}=200$, $a=0.05$, $\eta=1$.} \label{fig_fbloch_Eigenvalue13}
\end{figure}

The shapes of the corresponding massive KK modes of left- and
right-handed fermions with the even parity and odd parity for Bloch
brane with different $m^{2}$ are shown in
Fig.~\ref{fig_fbloch_Eigenvalue13}. There are a total of three
odd-parity solutions and three even-parity solutions, on the other
hand, three left-handed solutions and three right-handed solutions.
This comply with the parity-chiral decomposition
expression~(\ref{the general chiral decomposition explicit}).

\subsection{Case II: $F(\phi,\chi)=\phi^{k}\chi$ with odd $k>1$}

\begin{figure}[htb]
\begin{center}
\includegraphics[width=12cm,height=9cm]{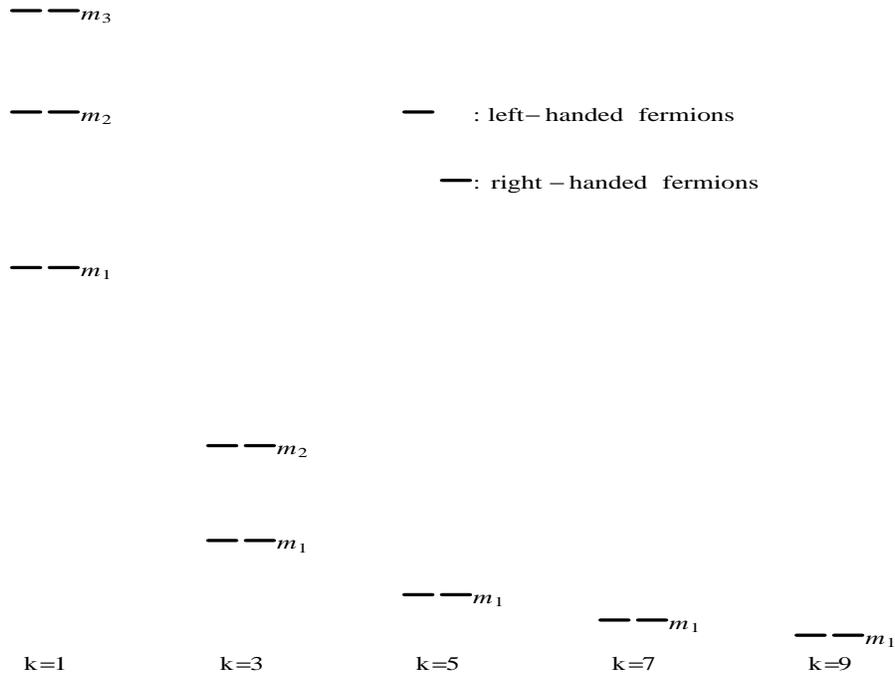}
\end{center} \vskip -5mm
\caption{Mass spectrum of resonances for the coupling
$F(\phi,\chi,\rho)=\phi^{k}\chi$. The parameters are $k=\{1$, $3$,
$5$, $7$, $9\}$, $a=0.05$ and $\eta=1$.}
\label{Blochbranemassspectrum}
\end{figure}

In this subsection, we extend the analysis to the situation $k>1$.
Use the same method, for $k=3$, $5$, $7$, $9$, the eigenvalues
$m^2_{n}$, mass $m_{n}$, width $\Gamma$ and lifetime $\tau$ for
the resonances are listed in Table.~\ref{TableSpectra2Fields}. For
given $a$ and $\eta$, as $k$ increases, the number of resonance
states reduce to zero quickly. The parity-chiral decomposition
expression~(\ref{the general chiral decomposition explicit}) holds
good for every $k$. The graphic of the mass spectrum of resonances
for the coupling $F(\phi,\chi,\rho)=\phi^{k}\chi$ ($1\leq k\leq9$)
is depicted in Fig.~\ref{Blochbranemassspectrum}.

\begin{table}[h]
\begin{center}
\begin{tabular}{||c|c|c|c|c|c|c|c||}
 \hline
 $k$ & $\mathcal {C}$ & $\mathcal {P}$ & $m^{2}_{n}$ & $m_{n}$ & $\Gamma$& $\tau$ &  $P_{max}$   \\
 \hline\hline

 & & odd & 0.806136      & 0.897851    & $1.92489$$\times$$10^{-6}$   & 519510  & 0.999592  \\ \cline{3-8}
 & $\mathcal{L}$ &  even    & 1.51801         & 1.23208      & 0.000221484    & 4515  &  0.999164  \\ \cline{3-8}
 & & odd  & 2.1024 & 1.44997 & 0.00512333 & 195.186 & 0.771056 \\
  \cline{2-8}\cline{2-8}
  \raisebox{2.3ex}[0pt]{1}& & even & 0.806523 & 0.898066 & $1.92954$$\times$$10^{-6}$ & 518260 & 0.999989
  \\ \cline{3-8} \cline{3-8}
  & $\mathcal{R}$ &odd  & 1.51891      & 1.23244     & 0.000223108  & 4482.13   &  0.999067 \\ \cline{3-8}
  &  &even      & 2.1037  & 1.45041      & 0.00517988     & 193.055   & 0.768636  \\ \hline\hline

  & & odd & 0.096964      & 0.31139    & 0.0000873266   & 11451.3  & 0.999515 \\ \cline{3-8}
 & \raisebox{2.3ex}[0pt]{$\mathcal{L}$} &  even    & 0.26526         & 0.515034      & 0.0048577    & 205.859  &  0.646296  \\  \cline{2-8}
  \raisebox{2.3ex}[0pt]{3}&  & even & 0.096982 & 0.311419 & 0.0000884991 & 11299.5 & 0.999304
  \\ \cline{3-8} \cline{3-8}
  & \raisebox{2.3ex}[0pt]{$\mathcal{R}$}  &odd  & 0.2653      & 0.515073     & 0.00487788   & 205.007 & 0.646222 \\ \hline\hline

 & $\mathcal{L}$ & odd & 0.03794      & 0.194782    & 0.000837101   & 1194.6 & 0.929206 \\ \cline{2-8}
  \raisebox{2.3ex}[0pt]{5}& $\mathcal{R}$ & even & 0.03795 & 0.194808 & 0.000833316 & 1200.03 & 0.929414 \\ \hline\hline

 & $\mathcal{L}$ & odd & 0.01956      & 0.139857    & 0.00383873   & 260.502 & 0.477145 \\ \cline{2-8}
  \raisebox{2.3ex}[0pt]{7}& $\mathcal{R}$ & even & 0.01957 & 0.139893 & 0.00373923 & 267.435 & 0.479078 \\ \hline\hline

 & $\mathcal{L}$ & odd & 0.0112      & 0.10583    & 0.015039   & 66.4936 &  0.265202 \\ \cline{2-8}
  \raisebox{2.3ex}[0pt]{9}& $\mathcal{R}$ & even & 0.0112 & 0.10583 & 0.0135679 & 73.7035 & 0.270337 \\ \hline
\end{tabular}\\
 \caption{The eigenvalues $m^2$, mass, width and lifetime
 for resonances of left- and right-chiral fermions with
 odd-parity and even-parity solutions for the coupling
 $F(\phi,\chi)=\phi^{k}\chi$. $\mathcal{C}$ and $\mathcal{P}$ stand for chirality and parity, respectively.
$\mathcal{L}$ and $\mathcal{R}$ are short for left-handed and
right-handed, respectively. The parameters
 are $a=0.05$, $k=\{1$, $3$, $5$, $7$, $9\}$ and $\eta=1$.}
\label{TableSpectra2Fields}
\end{center}
\end{table}

\section{Discussion and conclusion}
\label{secConclusion}

In this paper, we first investigated the thick branes generated
from multi-scalar (especially three-scalar) fields. Then, we
studied the localization and the resonances on the three-scalar
and two-scalar generated branes. Using the Numerov method, we
solved the Schr\"{o}dinger equations for KK modes of fermions with
the numerical potentials under two types of initial value
conditions, which lead to the odd- and even-parity solutions. We
got the wave functions of the resonance states and used them to
calculate the probability and the lifetime of resonance states.

For the three-field model, we considered the generalized Yukawa
coupling $\eta\overline{\Psi}\phi^{k}\chi\rho\Psi$ on the thick
brane, where $\eta$ is an arbitrary positive coupling constant and
$k$ is an odd positive integer. For the two-field model, we
considered the coupling $\eta\overline{\Psi}\phi^{k}\chi\Psi$, where
the case of $k=1$ was considered in \cite{Almeida:0901}. In these
models, there is a real parameter $a$, whose role is regulating the
structure of the thick branes. The results show that the behavior of
the resonances completely decided by the structure of the branes and
the coupling with scalars. For a certain $k$, the coupling constant
$\eta$ and the parameter $a$ decide the shape of the potentials and
wave functions.

For $k=1$ in the three-field model, the potential of the KK modes
of left chiral fermions $V_L$ is a modified volcano type
potential. While the shape for the right one $V_R$ is decided by
$a$ and $\eta$. For small $a$ and large $\eta$, the potential
$V_R$ will have a potential well. When the depth of the potential
well is deep enough, it will be able to trap the fermions in some
sense. We obtained a series of quasibound states or the metastable
states with finite lifetimes. The eigenvalues $m^2_{n}$ of the
resonances were also given.

For $k>1$ in the two-field and three-field models, the potential
well for left-handed fermions becomes a double-well, which is
consistent with the result obtained in Ref.~\cite{Liu:09b}. However,
for given $a$ and $\eta$, the number of the resonant states in these
two models decreases with $k$. This is opposite to the result given
in Ref. \cite{Liu:09b} for the single-scalar generated brane case.
The reason is that the absolute value of the kink $\phi$ in current
models is less than $1$, and so, when $k$ increases, the strength of
the coupling decreases.

We find that the $n$-th resonance of left-handed fermions with odd
parity and the $n$-th resonance of right-handed fermions with even
parity have the same mass and lifetime, and vice versa. This
demonstrates that it is possible to compose a Dirac fermion from
the left and right KK modes~\cite{Liu:09b}. However, in both
chiral fermions, the parity is opposite. We call this phenomenon
as the parity-chiral decomposition, and it is the explicit form of
the general KK chiral decomposition of the Dirac fermions.

In our numerical calculations, for those KK modes with $m^{2}$
much larger than the maximum of the corresponding potential, they
would be approximately plane waves and the probabilities for them
would trend to $0.1$, where we chose $z_{max}=10z_{b}$ to
calculate the relative probabilities. For the KK modes with
$m_{2}$ close to but larger than the maximum of the potential, the
probabilities for finding them around the brane $P$ is quite
small. For the sake of clarity, we give the definition of the
number of the resonant states: the number of these resonant states
with eigenvalue $m^2$ lower than the highest point of the
potential. We also took into account the impact of the numerical
precision. Besides, the probabilities $P$ for the resonances are
affected by $z_{b}$. So $z_{b}$ cannot be too small. The
resonances in other thick brane models with different kink-fermion
couplings could be considered, we will propose the corresponding
work on this subject in a near future.

\section{Acknowledgement}

This work was supported by the Program for New Century Excellent
Talents in University, the National Natural Science Foundation of
China(NSFC)(No. 10705013), the Doctoral Program Foundation of
Institutions of Higher Education of China (No. 20070730055), the Key
Project of Chinese Ministry of Education (No. 109153).


\end{document}